\documentclass [11pt]{article}
\usepackage{amsmath,amsthm,amsfonts,amscd,eucal,latexsym,amssymb}
\usepackage{epsfig}   
\input xy
\xyoption{all}
\def\sp{\hskip -5pt} 
\def\spa{\hskip -3pt}

\newsymbol\bt 1202           %%% \boxtimes 

\def\cH{{\ca H}}

\def\cE{{\ca E}}

\def\cS{{\ca S}}
\def\cA{{\ca A}}

\def\cW{{\ca W}}

\def\cR{{\ca R}}

\def\bC{{\mathbb C}}           %%%  complex numbers and so on 
 
\def\bI{{\mathbb I}}

\def\bN{{\mathbb N}} 
\def\bM{{\mathbb M}} 
 
\def\bR{{\mathbb R}}

\def\bS{{\mathbb S}}

\def\bZ{{\mathbb Z}} 
 
\newsymbol \rest 1316         %%% restriction symbol 
 
\def\gF{{\mathfrak F}}

\def\beq{\begin{eqnarray}}
\def\eeq{\end{eqnarray}}

\def\al{\langle}
\def\cl{\rangle}
\newcommand{\ca}[1]{{\cal #1}}         %%  calligraphic

\def\z{\zeta}
\def\bz{\overline{\zeta}}

\def\bSf{{\mathbb S}^2}  %2-sphere
\def\scri{\Im^+}         %null future boundary
\def\scrip{\Im^-}         %null past boundary
\def\tg{\tilde{g}}
\def\tM{\tilde{M}}
\def\tphi{\tilde{\phi}}
\def\tV{\tilde{V}}
\def\MV{M_{\tilde{V}}}

\def\lie{\pounds}
\def\bms{G_{BMS}}
\def\gbms{G_{BMS}}
\def\tgbms{\widetilde{{G}_{BMS}}}
\def\cU{{\cal U}}
\def\bx{{\bf x}}

\def\bp{{\bf p}}
\def\vpsi{\vec{\psi}}
\def\vPsi{\vec{\Psi}}

\def\iso{ISO(3,1)}

\newcounter{proposition}[section]
\newcounter{theorem}[section]
\newcounter{lemma}[section]
\newcounter{definition}[section]
\newcounter{remark}[section]

\def\theproposition{\thesection.\arabic{proposition}}
\def\thetheorem{\thesection.\arabic{theorem}}
\def\thelemma{\thesection.\arabic{lemma}}
\def\thedefinition{\thesection.\arabic{definition}}
\def\theremark{\thesection.\arabic{remark}}

\def\s #1 {\section{#1}}

\def\ssa #1 {\ifhmode{\par}\fi\refstepcounter{subsection}
  \noindent {\bf\thesubsection}. {\em #1}.\quad
  \addcontentsline{toc}{subsection}{\protect\numberline{\thesubsection} #1}%
  }

\def\ssb #1 {\ifhmode{\par}\fi\refstepcounter{subsection}
  \noindent {\bf\thesubsection.} {\em #1.}\quad
  \addcontentsline{toc}{subsection}{\protect\numberline{\thesubsection} #1}%
  }

\def\proposizione {\ifhmode{\par}\fi\refstepcounter{proposition}
  \noindent {\bf Proposition \theproposition}. \quad}
\def\teorema {\ifhmode{\par}\fi\refstepcounter{theorem}
  \noindent {\bf Theorem \thetheorem}. \quad}
\def\lemma {\ifhmode{\par}\fi\refstepcounter{lemma}
  \noindent {\bf Lemma \thelemma}. \quad}
\def\definizione {\ifhmode{\par}\fi\refstepcounter{definition}
  \noindent {\bf Definition \thedefinition}. \quad}
\def\remark {\ifhmode{\par}\fi\refstepcounter{remark}
  \noindent {\bf Remark \theremark}. \quad}

\begin{document} 
 
\hfill{\sl June 2005, Preprint  FNT/T-2005/05 and UTM  683} \\

%%%%%%%%%%%%%   Title %%%%%%%%%%%%%%%%%%%%%%%%%% 
 
\par 
\LARGE 
\noindent 
{\bf Rigorous steps towards holography in asymptotically flat spacetimes} \\
\par 
\normalsize 
 
%%%%%%%%%%%%%%%%%%%%%%%%%%%%%%%%%%%%%%%%%%%%% 
 
%%%%%%%%%%%% Authors %%%%%%%%%%%%%%%%%%%%%%%%%%% 

\noindent {\bf Claudio Dappiaggi$^{1,a}$},
{\bf Valter Moretti$^{2,b}$}, {\bf Nicola Pinamonti$^{2,c}$} \\
\par
\small
\noindent $^1$ Dipartimento di Fisica Nucleare e Teorica, Universit\`a di Pavia and  Istituto Nazionale di Fisica Nucleare  
Sezione di Pavia, via A.Bassi 6 I-27100 Pavia, Italy.\smallskip

\noindent$^2$ Dipartimento di Matematica, Universit\`a di Trento, 
 and Istituto Nazionale di Alta Matematica ``F.Severi''  unit\`a locale  di Trento,
 and  Istituto Nazionale di Fisica Nucleare  Gruppo Collegato di Trento, via Sommarive 14  
I-38050 Povo (TN), Italy. \smallskip

\noindent $^a$  claudio.dappiaggi@pv.infn.it,
$^b$  moretti@science.unitn.it,  $^c$  pinamont@science.unitn.it\\ 
 \normalsize

\small 
\noindent {\bf Abstract}. {
\noindent Scalar QFT on the boundary $\scri$ at future null
infinity of a general asymptotically flat $4D$ spacetime is 
constructed using the algebraic
approach based on Weyl algebra associated to a BMS-invariant  symplectic form. 
The constructed theory turns out to be invariant under a suitable 
strongly-continuous unitary representation of the BMS group 
with manifest meaning when the fields are interpreted as suitable
 extensions to $\scri$ 
 of massless minimally coupled fields
propagating in the bulk. The group theoretical analysis of the found unitary 
BMS representation proves that 
such a field on $\scri$
coincides with the natural
wave function constructed out of the unitary BMS irreducible representation
induced from the little group $\Delta$, the semidirect
product between $SO(2)$ and the 
two-dimensional translations group.
This wave function 
is massless with respect to the notion of mass for BMS representation theory.
The presented result
proposes a natural criterion to solve the long standing problem of the topology 
of BMS group. Indeed the found natural
correspondence of quantum field theories holds only if the BMS group
is equipped with the nuclear topology rejecting instead the Hilbert one.
Eventually some theorems towards a holographic description on $\scri$ of QFT in 
the bulk are established at level of 
$C^*$ algebras of fields for asymptotically flat at null infinity spacetimes. 
It is proved  that preservation of a certain symplectic form implies the existence
of an injective $*$-homomorphism from the Weyl algebra of  fields of the  bulk into
that associated with the boundary $\scri$.
Those results are, in particular,
applied to  $4D$ Minkowski spacetime where a nice interplay between Poincar\'e 
invariance in the bulk and BMS invariance  on the boundary 
at null infinity is established at level of QFT. It arises that,
in this case, the $*$-homomorphism admits unitary implementation and Minkowski vacuum is 
mapped into the BMS invariant vacuum on $\scri$.}

\normalsize
\newpage 
\tableofcontents

\s{Introduction}
\ssa{Holography in asymptotically flat spacetimes}
One of the key obstacles in the current, apparently never-ending, quest to combine in a 
unique framework general relativity and 
 quantum mechanics consists in a deep-rooted lack 
of comprehension of the role and the number of quantum degrees of freedom of gravity. 
Within this respect, a new insight has been gained from the work of G. 't Hooft who suggested to address this 
problem from a completely new perspective which is now referred to as {\em the holographic principle} \cite{hooft}.
This principle states, from the most general point of view,  that {\em physical information in spacetime is fully 
encoded on the boundary of the region under consideration}. 't Hooft paper represented a cornerstone for innumerable research papers
which led to an extension of celebrated Bekenstein-Hawking results 
about black hole entropy to a wider class of 
spacetime regions (see in particular the {\em covariant entropy conjecture} in \cite{Bousso}).
Furthermore a broader version of the holographic principle arisen from the above-cited developments 
according to which {\em any quantum field theory} - 
gravity included - living on a $D$-dimensional spacetime can be fully described by means of a second 
theory living on a suitable submanifold, with codimension $1$, which is not necessary (part of the) 
boundary of the former.
However the holographic principle lacks any 
general prescription on how to concretely construct a holographic counterpart 
of a given quantum field theory. In high energy physics in the 
past years  the attempt to fill this gap succeeded achieving some remarkable results. 
The most notable is the so-called 
$AdS/CFT$ correspondence \cite{Aharony} or Maldacena conjecture, 
the key remark being the existence of the equivalence 
between the bulk and the boundary partition function once asymptotically $AdS$ 
boundary conditions have been imposed on the physical fields.
Without entering into  details 
(see \cite{deBoer} for a recent review), it suffice 
to say that in the low energy limit a supergravity theory living on a 
$AdS_D\times X^{10-D}$ manifold is (dual to) a $SU(N)$ conformal super 
Yang-Mills field theory living on the boundary at spatial infinity of $AdS_D$.
Other remarkable versions of holographic principle for $AdS$-like spacetime
are due to  Rehren \cite{rehr00a,rehr00b} who proved rigorously several holographic 
results for local quantum fields in a $AdS$ 
background, establishing a correspondence between bulk and boundary observables 
without employing string machinery. 

 It is rather 
natural to address the question whether  similar holographic correspondences 
hold whenever a different class of spacetimes is considered. 
In this paper 
we will deal with the specific case of  asymptotically flat spacetimes and we consider fields 
interacting, in the bulk, only with the gravitational field. 
 The quest to construct a holographic correspondence in this scenario started only 
recently and a few different approaches have been proposed \cite{Arcioni, deBoer2, 
Goursat}.  In particular, in \cite{Arcioni}, in order to implement the holographic 
principle in a {\em four-dimensional} asymptotically flat spacetimes $(M,g)$, it has been 
proposed to construct a bulk to 
boundary correspondence between a theory living on $M$ and a quantum field 
theory living at future (or past) null infinity $\scri$ of $M$.
A key point is that the theory on $\scri$  is further assumed  to be invariant 
under the action of the asymptotic symmetry group of this class of spacetimes: the 
so called Bondi-Metzner-Sachs (BMS) group. The analysis performed along the lines 
of Wigner approach to Poincar\'e invariant free quantum field theory has led to
construct the
full spectrum, the equations of motion and the Hamiltonians for free quantum 
field theory enjoying BMS invariance \cite{Arcioni, Arcioni2}. 
A first and apparently surprising 
conclusion which has been drawn from these papers is that, in a BMS invariant 
field theory, there is a natural plethora of different kinds of admissible BMS-invariant fields.
As a consequence the one-to-one 
correspondence between the bulk and boundary particle spectrum, proper of the 
Maldacena conjecture, does not hold in this context or needs further information to be constructed. 
 Nevertheless such a conclusion should not be seen as a 
 setback, since it represents the symptom of a key feature 
 proper only of asymptotically flat spacetimes. This is the universality of the boundary data,  i.e. 
 as  explained  in more detail in the next section, the structure at future and past null 
 infinity of any asymptotically flat spacetime is the same. Thus, from a holographic perspective,
 a BMS-invariant field theory on $\scri$
  should encode the information from all possible asymptotically 
  flat bulk manifolds. 
  Consequently,  it is not surprising if there is such a huge 
  number of admissible BMS-invariant free fields. The main question now consists on 
  finding a procedure allowing one to single out information on 
  a specific bulk from the boundary theory.

The aim of this paper is develop part of this programme using the theory 
of unitary representations of BMS group as well as tools proper of 
algebraic local quantum field theory. 
In particular, using the approach introduced in \cite{mopi3, mopi4,mopi6} and 
fully developed in \cite{mopi6}, we define quantum field theory on the 
null surface $\scri$ using the algebraic framework based on a suitable 
representation of Weyl $C^*$ algebra of fields.
 Then we investigate the interplay of 
that theory and quantum field theory of a free scalar field in the bulk finding several 
interesting results.
There is a GNS (Fock space) representation of the field theory on $\scri$,
based on a certain algebraic quasifree state,
 which
admits an irreducible strongly-continuous unitary representation of the BMS group 
which leaves invariant the vacuum state. The algebra of fields transforms 
covariantly with respect to that unitary representation. In other words
the fields on $\scri$ and the above-mentioned unitary action of BMS group 
have manifest geometrical meaning 
when the fields on $\scri$ are interpreted as suitable extensions of massless minimally coupled fields
propagating in the bulk. 
Furthermore, the group theoretical analysis of the BMS representation proves 
that the bulk massless field ``restricted'' on $\scri$ coincides with the natural
wave function constructed out of the unitary BMS irreducible representation
induced from the little group $\Delta$: the semidirect
product between $SO(2)$ and the two dimensional translations. This wave-function 
is massless with respect to a known notion of mass in BMS 
representation theory.
In this context the found extent provides the solution of a long-standing 
problem concerning the natural topology of BMS group. 
In fact,  the found unitary representation of GNS group takes place only if the BMS group is equipped with the nuclear 
topology. In this sense the widely considered  Hilbert topology must be rejected.\\
Eventually some theorems towards a holographic description on $\scri$ of QFT in 
the bulk are established at level of 
Weyl $C^*$ algebras of fields for spacetimes which are  asymptotically flat at null infinity. 
It is shown that, if a symplectic form is preserved passing from 
the bulk to the boundary, the algebra of fields in the bulk can be identified 
with a subalgebra for the field observables on $\scri$ by means of an injective $*$-homomorphism.
Moreover, the BMS invariant state of quantum field theory on   $\scri$
induces a corresponding reference state in the bulk. It could be used 
to give a definition of particle based only upon asymptotic symmetries, 
no matter if the bulk admits any isometry group (see also \cite{Dappiaggi2}).
Those results are, in particular,
applied to  $4D$ Minkowski spacetime where a nice interplay between Poincar\'e 
invariance in the bulk and BMS invariance  on the boundary $\scri$ is established at level of quantum
field theories. Among other results it arises that the above-mentioned injective $*$-homomorphism
has unitary implementation such that the Minkowski vacuum is mapped into the 
BMS invariant vacuum on $\scri$.\\
 The outline of the paper is the following.

In {\em section 2} we 
review the notion of asymptotically flat space-time and of the 
Bondi-Metzner-Sachs group. Starting from these premises a  
field living at null infinity $\scri$ is defined as a suitable limit 
of a bulk scalar field and the set of fields on $\scri$ is endowed with 
a symplectic structure. Eventually the quantum field theory for an
uncharged scalar field living on $\scri$ is built up within  Weyl 
algebra approach and a preferred Fock representation is selected 
which also admits a suitable unitary representation of the BMS group.

In {\em section 3} the theory of unitary and irreducible 
representation for the BMS group is discussed and  quantum 
field theory on $\scri$ is defined along the lines of Wigner 
analysis for the Poincar\'e invariant counterpart. Furthermore 
it is shown that, at least for scalar fields, the approaches 
discussed in this and in the previous sections are essentially equivalent
provided one adopts a nuclear topology on the BMS group. 

In  {\em section 4}  the issue of an 
holographic correspondence is discussed 
for spacetimes satisfying a requirement weaker than strongly asymptotically 
predictability
given
in Proposition \ref{prop2}.
We show that preservation of a certain symplectic form implies existence
of a injective $*$-homomorphism from the Weyl algebras of the fields in the bulk into
that on $\scri$.
It is done devoting a particular 
attention to the specific scenario when the bulk is 
four-dimensional
Minkowski spacetime.  
 It arises that,
in this case the $*$-homomorphisms admits unitary implementation and Minkowski vacuum is 
mapped into the BMS invariant vacuum on $\scri$ and the standard unitary 
representation of Poincar\'e group in the bulk is transformed in 
a suitable unitary representation of a subgroup of BMS group on $\scri$ and the
correspondence has a clear  geometric interpretation.

In  {\em section 5} we present our conclusion with some comments about possible 
future developments and investigations.  
The {\em appendix} contains the proof of most of 
the statement within the paper.\\

\ssa{Basic definitions and notations} \label{notations}
In this paper {\em smooth} means $C^\infty$ and we adopt the signature
 $(-,+,+,+)$ 
for the Lorentzian metric. \\
The symbol $B\ltimes A$ will be reserved for a {\em semidirect product}
of a pair of groups $(B, \cdot ),(A, \ast )$. We recall the reader that $B\ltimes A$ is defined as the group
obtained by the assignment, on the set of pairs $B\times A$, of the group product $(b,a) \odot (b',a') = (b\cdot b', a \ast
\beta_b(a'))$
where $B\ni b \mapsto \beta_b$ is a fixed (it determining $\odot$) group representation of $B$ 
in terms of group automorphisms of $A$. $A$ turns out to be  naturally isomorphic to the normal subgroup of $B\ltimes A$
made of the pairs $(I,a)$ with $a\in A$, $I$ denoting the unit element of $B$.
The proper orthocronous Lorentz group will be denoted by $SO(3,1)\spa \uparrow$,
while $ISO(3,1) = SO(3,1)\sp \uparrow \sp \ltimes T^4$ is  the proper orthocronous Poincar\'e group 
with semidirect product structure induced by $(\Lambda, t)\odot (\Lambda', t') = (\Lambda \Lambda', t + \Lambda t')$.\\
In a manifold equipped with Lorentzian  metric $\Box := \nabla_a\nabla^a$ 
indicates d'Alembert operator referred to Levi-Civita connection $\nabla_a$,
$\lie_\xi$ denotes the Lie derivative with respect to the vector field $f$ and $f^*$ the push-forward
associated with the diffeomorphism $f$ acting on tensor fields of any fixed order. $C^\infty(M;N)$ and
$C_c^\infty(M;N)$ respectively
indicates the class of smooth functions and 
compactly supported smooth functions $f:M\to N$. We omit $N$ in the notation if $N=\bR$.
 $\lim_{\scri}f$ indicates a function on $\scri$ which is the smooth extension to $\scri$ of the
function $f$ defined in $M$. \\
A {\em spacetime} is a four-dimensional smooth (Hausdorff second countable)  manifold $M$ equipped
with a Lorentzian metric $g$ assumed to be everywhere smooth, finally $M$ is supposed to be time orientable and time oriented. 
A {\em vacuum spacetime} is a spacetime satisfying vacuum Einstein equations. 
In this paper we make use of several properties of {\em globally hyperbolic spacetimes}
as defined in Chapter 8 in \cite{Wald}, employing standard notations  of  \cite{Wald} concerning causal sets. 
We adopt the notion of {\em asymptotically flat at future null infinity} vacuum spacetime
presented in \cite{Wald}. 
A smooth spacetime $(M,g)$ is called {\em asymptotically flat vacuum spacetime at future null infinity} \cite{Wald} if 
it is a solution of vacuum Eintein equations and the following requirements are fulfilled. There is 
a second smooth spacetime  $(\tilde{M},\tilde{g})$ such that $M$ turns out to be an open
submanifold of $\tilde{M}$ with boundary $\scri \subset \tM$.  $\scri$ is an embedded 
submanifold of $\tM$ satisfying $\scri \cap \tilde{J^-}(M) = \emptyset$.
$(\tM,\tg)$ is required to be strongly causal in a neighborhood of $\scri$ and 
it must hold $\tilde{g}\spa\rest_M= \Omega^2 \spa\rest_M g\spa\rest_M$ where $\Omega \in C^\infty(\tM)$
is strictly positive on $M$. On $\scri$ one must have $\Omega =0$ and $d\Omega \neq 0$. 
Moreover, defining $n^a := \tg^{ab} \partial_b \Omega$, 
there must be a smooth function, $\omega$, defined in $\tM$ with $\omega >0$ on $M\cup \scri$, such that 
$\tilde{\nabla}_a (\omega^4 n^a)=0$ on $\Im$ and the integral lines of $\omega^{-1} n$ are complete on $\scri$.
Finally the topology of  $\scri$ must be that of $\bS^2\times \bR$. $\scri$ is called {\em future infinity} of $M$.\\
It is possible to make stronger the definition of asymptotically flat spacetime by requiring  asymptotic flatness
at both null infinity -- including the {\em past} null infinity $\scrip$ defined analogously to 
$\scri$ -- and {\em spatial} infinity, given by a special point in $\tM$ indicated by $i^0$. The complete definition 
is due to Ashtekar (see Chapter 11 in \cite{Wald} for a general discussion). We stress that the results presented in this work do not
require such a stronger definition: for the spacetimes we consider existence of $\scri$ is fully enough.
Hence, throughout this paper {\em asymptotically flat spacetime} means {\em asymptotically flat vacuum spacetime at future null infinity}.

\section{Scalar QFT on $\scri$.} 
\ssa{Asymptotic flatness, asymptotic Killing symmetries, BMS group and all that} \label{BMS}
Considering an asymptotically flat spacetime,  the metric structures of $\scri$ are affected by a {\em gauge freedom}
due the possibility of changing the metric $\tg$ in a neighborhood of $\scri$
with a factor $\omega$ smooth and strictly positive. 
It corresponds to the freedom involved in transformations $\Omega \to \omega \Omega$ in a neighborhood of $\scri$.
The topology of $\scri$ (which is that of $\bR \times \bS^2$) as well as the differentiable structure 
 are not affected 
by the gauge freedom. 
Let us stress some features of this extent.
Fixing $\Omega$, $\scri$  turns out to be
the union of  future-oriented  integral lines of the field 
$n^a :=\tilde{g}^{ab}\tilde{\nabla}_b\Omega$.
This property is, in fact, invariant under gauge transformation, but the field $n$
depends on the gauge. 
For a fixed asymptotically flat vacuum spacetime $(M,g)$,
the manifold $\scri$ together with its degenerate metric $\tilde{h}$ induced by $\tg$ and
 the field $n$ on $\scri$
form a triple which, under gauge transformations $\Omega \to \omega \Omega$, transforms as
\beq
\scri \to \scri \:,\:\:\:\:\: \tilde{h} \to \omega^2 \tilde{h} \:,\:\:\:\:\: n \to \omega^{-1} n \label{gauge}\:.
\eeq
If $C$ denotes the class containing all of the triples $(\scri,\tilde{h}, n)$  transforming as in (\ref{gauge})
for a fixed
asymptotically flat vacuum spacetime $(M,g)$,
there is no general physical principle which allows one to select a preferred element in $C$.
Conversely, $C$ is {\em universal} for all asymptotically flat vacuum spacetimes in the following sense. 
If $C_1$ and $C_2$ are the classes of 
triples associated respectively to $(M_1,g_2)$
and $(M_2,g_2)$ there is a diffeomorphism $\gamma: \scri_1 \to \scri_2$ such that for suitable $(\scri_1,\tilde{h}_1, n_1)\in C_1$
and $(\scri_2,\tilde{h}_2, n_2)\in C_2$, 
\beq
\gamma(\scri_1) = \scri_2 \:,\:\:\:\:\: \gamma^* \tilde{h}_1=\tilde{h}_2 \:,\:\:\:\:\:\gamma^* n_1=n_2 \nonumber\:.
\eeq
The proof of this statement  relies on the following nontrivial result \cite{Wald}. 
For whatever asymptotically flat 
vacuum spacetime $(M,g)$ (either $(M_1,g_1)$  and $(M_2,g_2)$ in particular) and whatever initial 
choice for $\Omega_0$,
varying the latter with a judicious choice of the gauge $\omega$, 
one can always fix $\Omega := \omega \Omega_0$ in order that the metric $\tg$ associated with $\Omega$ satisfies
\beq
\tg\spa\rest_{\scri}  = -2du \:d\Omega +  d\Sigma_{\bS^2}(x_1,x_2)\:. \label{met}
\eeq
This formula uses the fact that in a neighborhood of $\scri$, $(u, \Omega,
x_1,x_2)$ define a meaningful coordinate system.
$ d\Sigma_{\bS^2}(x_1,x_2)$ is the standard metric on a unit $2$-sphere
(referred to arbitrarily fixed coordinates $x_1,x_2$) 
and $u \in \bR$ is nothing but an affine parameter along
the {\em complete} null geodesics forming $\scri$ itself with tangent vector $n= \partial/\partial u$. In these coordinates $\scri$ is just the set of the points with
$u \in \bR$, $(x_1,x_2) \in \bS^2$ and, no-matter the initial spacetime $(M,g)$ 
(either $(M_1,g_1)$  and $(M_2,g_2)$ in particular), one has finally the triple 
$(\scri,\tilde{h}_B, n_B) := (\bR\times \bS^2, d\Sigma_{\bSf}, \partial/\partial u)$. \\

\noindent {\bf Bondi-Metzner-Sachs (BMS) group}, $G_{BMS}$ \cite{Penrose, Penrose2, Geroch, AS}, is  the group of diffeomorphisms of 
$\gamma : \scri \to \scri$ 
which preserve the  universal structure of $\scri$, i.e. $(\gamma(\scri),\gamma^*\tilde{h}, \gamma^*n)$
differs from $(\scri,\tilde{h},n)$ at most by a gauge transformation (\ref{gauge}). The following proposition holds
\cite{Wald}.\\

\proposizione\label{prop1} {\em The one-parameter group of diffeomorphisms generated by a smooth vector 
field $\xi'$ on $\scri$ is a subgroup of $G_{BMS}$ if and only if  the following holds. 
$\xi'$ can be extended smoothly to a field $\xi$ (generally not unique) 
defined in $M$ in some neighborhood of $\scri$
such that $\Omega^2 \lie_\xi g$ has a smooth extension to $\scri$
and $\Omega^2 \lie_\xi g \to 0$ approaching $\scri$. }\\

\noindent The requirement $\Omega^2 \lie_\xi g \to 0$ approaching $\scri$
is the best approximation of the Killing requirement $\lie_\xi g=0$ for a generic asymptotically
flat spacetime which does {\em not} admits proper Killing symmetries. In this sense BMS group describes
{\em asymptotic null Killing symmetries} valid for all asymptotically flat vacuum spacetimes.\\

\remark \label{fundamental}  \\ {\bf (1)} Notice that BMS group is {\em smaller} than the group of gauge transformations 
in equations (\ref{gauge}) because not all those transformations can be induced by diffeomorphisms of $\scri$. 
On the other hand the restriction of the gauge group to those transformations induced by diffeomorphisms
 permits to view BMS group
as a group of asymptotic Killing symmetries.\\ {\em Henceforth, whenever it is not explicitly stated otherwise,
we consider as admissible realizations
of the unphysical metric on $\scri$ only those metrics $\tilde{h}$ which can be reached through transformations 
of BMS group -- i.e. through asymptotic symmetries -- from a metric whose associated triple is $(\scri,\tilde{h}_B,n_B)$}.\\
 {\bf (2)} Therefore $\tilde{h}$ in general may not coincide with the initial metric induced by $\tg$ on $\scri$
 but
a further, strictly positive on $\scri$, factor $\omega$ defined in a neighborhood of $\scri$  
may take place\footnote{In case the spacetime is, more strongly, asymptotically flat at future and past null infinity
 and spatial infinity
 \cite{Wald}, $\omega\Omega$
could have singular behaviour at spatial infinity $i^0\in \tM$ which does {\em not} belong to $\scri$ by definition, 
see footnote on p.279 in \cite{Wald}
for details.}.
In this sense freedom allowed by rescaling with factors $\omega$ is larger that freedom involved
in re-defining the unphysical metric $\tg$ on the whole unphysical spacetime $\tM$. \\

\noindent To  give an explicit representation of $G_{BMS}$ we need a suitable coordinate frame 
on $\scri$.   
Having fixed the triple $(\scri,\tilde{h}_B,n_B)$ one is still free to select an arbitrary coordinate frame 
on the sphere and, using the parameter $u$ of integral curves of $n_B$ to complete the coordinate system, one is
free to fix the origin of $u$ depending on $\z,\bz$ generally.
Taking advantage of stereographic projection one may adopt complex coordinates $(\z,\bz)$ on the (Riemann) sphere,
$\z= e^{i\phi}\cot(\vartheta/2)$,  $\phi,\vartheta$ being usual spherical 
 coordinates. \\
{\em Coordinates $(u,\z,\bz)$ on $\scri$ define a  {\bf  Bondi frame} when $(\z,\bz)\in \bC\times \bC$ are 
complex stereographic coordinates on $\bS^2$,
 $u\in \bR$  (with the origin fixed arbitrarily) is the 
 parameter of the integral curves of $n$ and $(\scri,\tilde{h},n)= (\scri,\tilde{h}_B,n_B)$.}\\ 
In this frame  the set $G_{BMS}$  is nothing but $SO(3,1)\sp\uparrow \sp \times C^\infty(\bS^2)$, and
 $(\Lambda, f) \in SO(3,1)\sp\uparrow \times C^\infty(\bS^2)$ acts on $\scri$ as
 \cite{sachsa}
\begin{eqnarray}
u &\to & u':= K_\Lambda(\z,\bz)(u + f(\z,\bz))\:,\label{u}\\
\z &\to & \z' :=\Lambda\z:= \frac{a_\Lambda\z + b_\Lambda}{c_\Lambda\z +d_\Lambda}\:, \:\:\:\:\:\:
\bz \: \to \: \bz' :=\Lambda\bz := \frac{\overline{a_\Lambda}\bz + \overline{b_\Lambda}}{\overline{c_\Lambda}\bz +\overline{d_\Lambda}}\:.
\label{z}
\end{eqnarray}
 \begin{eqnarray}
 K_\Lambda(\z,\bz) :=  \frac{(1+\z\bz)}{(a_\Lambda\z + b_\Lambda)(\overline{a_\Lambda}\bz + \overline{b_\Lambda}) +(c_\Lambda\z +d_\Lambda)(
 \overline{c_\Lambda}\bz +\overline{d_\Lambda})}
\label{K}\:\: \: \mbox{and}\:\:\:\:
 \left[
\begin{array}{cc}
  a_\Lambda & b_\Lambda\\
  c_\Lambda & d_\Lambda 
\end{array}
\right] = \Pi^{-1}(\Lambda)\:.
\end{eqnarray}
$\Pi$ is the well-known surjective covering homomorphism $SL(2,\bC) \to SO(3,1)\sp\uparrow$. 
Thus the matrix  of coefficients $a_\Lambda, b_\Lambda, c_\Lambda, d_\Lambda$
is an arbitrary element of $SL(2,\bC)$ determined by $\Lambda$ up to an overall sign. However $K_\Lambda$ and the 
right hand sides of (\ref{z}) are manifestly independent from any choice of such a sign. It is clear from
(\ref{z}) and (\ref{K}) that, 
 $G_{BMS}$ can be viewed as the semidirect product of $SO(3,1)\sp\uparrow$ and the Abelian additive
group $C^\infty(\bS^2)$, the group product depending on the used Bondi frame. 
The elements of this subgroup are called {\bf supertranslations}.
In particular, if $\odot$ denotes the  product in $G_{BMS}$, $\circ$ the composition of functions, $\cdot$
the pointwise product of scalar functions
and $\Lambda$ acts on $(\z,\bz)$ as said in the right-hand sides of (\ref{z}):
\begin{eqnarray}
K_{\Lambda'}(\Lambda(\z,\bz)) K_\Lambda(\z,\bz) &=& K_{\Lambda' \Lambda}(\z,\bz) \label{KK}\:.\\
(\Lambda',f') \odot (\Lambda,f) &=& \left(\Lambda' \Lambda,\: f + (K_{\Lambda^{-1}} \circ \Lambda)\cdot (f'\circ \Lambda)  \right)\:.
\label{product}
\end{eqnarray}\\

\remark\label{generatoremalditesta}{\em 
We underline that in the literature the factor $K_\Lambda$ does not always have
the same definition. In particular, in \cite{Mc1, Mc2, Mc5, Girardello, Mc4} 
$$K_\Lambda(\z,\bz) :=  \frac{(a_\Lambda\z + b_\Lambda)(\overline{a_\Lambda}\bz + \overline{b_\Lambda}) +(c_\Lambda\z +d_\Lambda)(
 \overline{c_\Lambda}\bz +\overline{d_\Lambda})}{(1+\z\bz)},$$
but in this paper we stick to the definition (\ref{K}) as in
\cite{sachsa, rappresentazioni_con_peso} adapting accordingly the calculations and results from the above
mentioned references.}\\

The following proposition arises from the definition of Bondi frame and the equations above.\\

\proposizione\label{exremark1} {\em Let $(u,\z,\bz)$ be a Bondi frame on $\scri$. The following holds.\\
{\bf (a)} A global coordinate frame $(u',\z',\bz')$ on $\scri$ is a Bondi frame if and only if 
\begin{eqnarray}
u &=& u'+g(\z',\bz')\:,\label{transform1}\\
\z &=& \frac{a_R\z' + b_R}{c_R\z' +d_R}\:, \:\:\:\:\:\:
\bz \:\:=\:\: \frac{\overline{a_R}\bz' + \overline{b_R}}{\overline{c_R}\bz' +\overline{d_R}}\:,\label{transform2}
\end{eqnarray}
for $g\in C^\infty(\bS^2)$ and
$R\in SO(3)$ 
referring to the canonical inclusion $SO(3)\subset SO(3,1)\sp\uparrow$
(i.e. the canonical inclusion $SU(2)\subset SL(2,\bC)$ for matrices of coefficients 
$(a_\Lambda,b_\Lambda,c_\Lambda,d_\Lambda)$ in (\ref{K}).)\\ 
{\bf (b)} The functions $K_\Lambda$ are smooth on the Riemann sphere $\bS^2$.  Furthermore
 $K_\Lambda(\z,\bz)=1$ for all $(\z,\bz)$ if and only if $\Lambda \in SO(3)$.\\
{\bf (c)} Let $(u',\z',\bz')$ be another Bondi frame as in (a).
If $\gamma \in \bms$ is represented by $(\Lambda,f)$ in $(u,\z,\bz)$, the same $\gamma$ is represented
by $(\Lambda',f')$ in $(u',\z',\bz')$ with
\beq (\Lambda',f') = (R,g)^{-1}\odot (\Lambda,f)\odot (R,g)\:.\label{indep} \eeq}

\ssa{Space of fields with BMS representations} \label{QFT1} Let us consider QFT on $\scri$ developed in the way presented in 
\cite{mopi3,mopi4,mopi6}
where QFT on null hypersurfaces was investigated in the case of Killing horizons. $\scri$ is not a Killing horizon 
but the theory can be re-adapted to this case with simple adaptations. The procedure we go to introduce is similar to that
sketched in \cite{APRL} for graviton field.\\
 First of all we fix a relation between 
scalar fields $\phi$ in $(M,g)$ and scalar fields $\psi$ on $\scri$. The idea is to consider the fields $\psi$
as re-arranged smooth restrictions to $\scri$ of fields $\phi$. Simple restrictions make no sense because 
$\scri$ does not belong to $M$. We aspect that a good definition of fields $\psi$ is a suitable smooth limit to $\scri$
of products $\Omega^\alpha \phi$ for some fixed real exponent $\alpha$. A strong suggestion  for the value of $\alpha$ 
is given by the following proposition. (Below $\tilde{\Box}$ is d'Alembert operator referred to $\tg$ and $R$ and $\tilde{R}$ are 
the scalar curvatures on $M$ and $\tM$ respectively.) \\

\proposizione\label{prop2} {\em Assume that $(M,g)$ is asymptotically flat with associated unphysical spacetime 
$(\tM,\tg)$ with $\tg\sp\rest_M = \Omega^2 g$.
Suppose that there is an open set $\tilde{V}\subset \tM$ with 
$\overline{M\cap J^-(\scri)}\subset \tilde{V}$ (the closure being referred to $\tM$) such that $(\tilde{V}, \tg)$ 
 is globally hyperbolic so that $(M\cap V,g)$ is globally hyperbolic, too.  
 If $\phi : M\cap\tilde{V}  \to \bC$ 
has compactly supported Cauchy data on some Cauchy surface of $M\cap \tilde{V}$ and satisfies 
massless conformal Klein-Gordon equation,
\beq
\Box \phi - \frac{1}{6} R \phi =0\label{cc}\:,
\eeq
{\bf(a)} the field $\tphi := \Omega^{-1}\phi$ can be  extended uniquely into a smooth solution 
in $(\tilde{V}, \tg)$
of 
\beq
\tilde{\Box} \tphi - \frac{1}{6} \tilde{R} \tphi =0\label{cc2}\:;
\eeq  
{\bf(b)} for every smooth positive factor $\omega$ defined in a neighborhood of $\scri$ used to rescale 
$\Omega \to \omega \Omega$ in such a neighborhood,
 $(\omega\Omega)^{-1} \phi$
extends to a smooth field $\psi$ on $\scri$ uniquely.\\}

\noindent We have assumed the possibility of having $R\neq 0$ in $M$ because, as noticed in \cite{Wald}, all we said in 
 section  \ref{BMS} holds true dropping the hypotheses for the spacetime $(M,g)$ to be a vacuum Einstein 
solution, but requiring  that the stress energy tensor $T$ is such that $\Omega^{-2}T$ is smooth on $\scri$.
 A simple and well-known example of the application of the theorem  is given by
  Minkowski spacetime, but also Schwarzschild spacetime fulfills these hypotheses (more precisely the hypotheses
  are satisfied for regions of the cited spacetimes in the future of a fixed suitable spacelike Cauchy surface). \\

\noindent {\em Proof}. 
 In this proof we define $M_{\tV}:= M\cap \tilde{V}$ and the symbol ``tilde'' written on a causal set indicates that the metric $\tg$
 is employed, otherwise the used metric is $g$. 
 (In the figure below, for the sake of simplicity, it has been assumed that
 $\tV\supset M$ so that $M_{\tV} = M$.)
 Notice that $\tilde{J^-}(M)\cap \scri =
 \emptyset$ so that $J^-(p;M_{\tV})= \tilde{J^-}(p; \tilde{V})$ if $p\in M_{\tV}$. 
 $(M_{\tV},g)$  is globally hyperbolic 
because it is strongly causal and the sets $J^-(p;M_{\tV})\cap J^+(q;M_{\tV})$
are compact for $p,q\in M_{\tV}$ (see sec. 8 in \cite{Wald}).
Indeed, $(\tilde{V},\tg)$ is strongly causal and thus $(M_{\tV},g)$  is 
strongly causal, moreover,  if $p,q \in M_{\tV}$, $J^-(p; M_{\tV})\cap J^+(q; M_{\tV})$ is compact because
 $J^-(p;M_{\tV})\cap J^+(q;M_{\tV}) = \tilde{J^-}(p; \tilde{V})\cap \tilde{J^+}(q; \tilde{V})$ 
 and $\tilde{J^-}(p; \tilde{V})\cap \tilde{J^+}(q; \tilde{V})$ is compact since $(\tilde{V},\tg)$
 is globally hyperbolic. As a consequence, we can use in $M_{\tV}$ (but also in $\tV$) standard results of solutions of Klein-Gordon
  equation with compactly supported Cauchy data in globally hyperbolic spacetimes \cite{Wald}. \\
(a) Let  $S$ be a spacelike Cauchy surface for $(M_{\tV}, g)$.
It is known \cite{Wald} that, in any open subset of $M$ and under the only hypothesis $\tg= \Omega^2 g$, (\ref{cc}) 
is valid for $\phi$ if and only if (\ref{cc2}) is valid for $\tilde{\phi}:= \Omega^{-1}\phi$. 
The main idea of the proof is to associate $\phi$  with  Cauchy data for $\tilde{\phi}$ on a suitable Cauchy surface
of the larger spacetime $(\tilde{V}, \tg)$, so that the unique maximal solution $\tilde{\Phi}$ of (\ref{cc2}) uniquely determined 
in  $(\tilde{V}, \tg)$ by those Cauchy data,
 on a hand is well defined on $\scri \subset \tV$, on the other hand it is a smooth extension of $\Omega^{-1}\phi$ initially defined 
in $M_{\tV}$ only.
Let $K_S$ be the compact support of Cauchy data of $\phi$ on $S$. As $\tV$ is homeomorphic to the product manifold 
$\bR \times \Sigma$, $\bR$ denoting a global time coordinate on $\tV$ and
$\Sigma$ being a spacelike Cauchy surface of $\tV$, one can fix $\Sigma$ in the past of the compact set $K_S$. 
Since $K_S$ is compact and the class of the open sets $\tilde{I^-}(p; \tilde{V})\cap \tilde{I^+}(q; \tilde{V})$ 
with $p,q\in M_{\tV}$
is a basis of the topology of $M_{\tV}$, it is possible to determine a {\em finite} number of points $p_1,\ldots p_n\in M_{\tV}$ 
in the future of $K_S$ in order that
$\cup_{i}{I^-}(p_i; M_{\tilde{V}}) \supset K_S$. In this way one also has 
$\cup_{i}{J^-}(p_i; M_{\tilde{V}}) = \cup_{i}\tilde{J^-}(p_i; \tilde{V}) \supset K_S$. On the other hand, as is well known $\cup_{i}\tilde{J^-}(p_i; \tilde{V}) 
\cap D^+(\Sigma)$
is compact and, in particular, $K_\Sigma := \cup_{i}\tilde{J^-}(p_i; \tilde{V}) \cap \Sigma = 
\cup_{i}{J^-}(p_i; M_{\tilde{V}}) \cap \Sigma$ is compact too, it being a closed subset
of a compact set. Notice that, outside $J^-(K_S; M_{\tV})\cup J^+(K_S; M_{\tV})$ the field $\phi$ vanishes in $M_{\tV}$. Thus we are
naturally lead to consider compactly supported (in $K_\Sigma$) Cauchy data on $\Sigma$ for the equation (\ref{cc2}), obtained 
by restriction of $\Omega^{-1}\phi$ and its derivatives to $\Sigma$.
Let $\tilde{\Phi}$ be the unique solution of (\ref{cc2}) in the whole globally hyperbolic spacetime
 $(\tilde{V},\tg)$, associated with those Cauchy data on $\Sigma$.
 By construction $\tilde{\Phi}$ must be an extension to $(\tilde{V},\tg)$ of $\tilde{\phi}$ defined in
$M_{\tV}$ (more precisely in  $\tilde{D^+}(\Sigma; \tV) \cap M_{\tV}=D^+(\Sigma\cap M_{\tilde V}; M_{\tilde V})$),
since they satisfy the same equation and have the same Cauchy data on $\Sigma$.
The proof concludes by noticing that $\scri \subset \tilde{V}$
and thus $\psi:= \tilde{\Phi}\spa\rest_{\scri}$ is, in fact,  a smooth extension to $\scri$ of $\tilde{\phi}$.\\
(b) The case with $\omega\neq 1$ is now a  trivial consequence of what proved above replacing 
$\Omega$ with $\omega \Omega$ in the considered neighborhood of $\scri$ where $\omega>0$.
$\Box$\\

\begin{figure}[th]
\begin{center}
\includegraphics[bb=0 0 500 270, scale=.7]{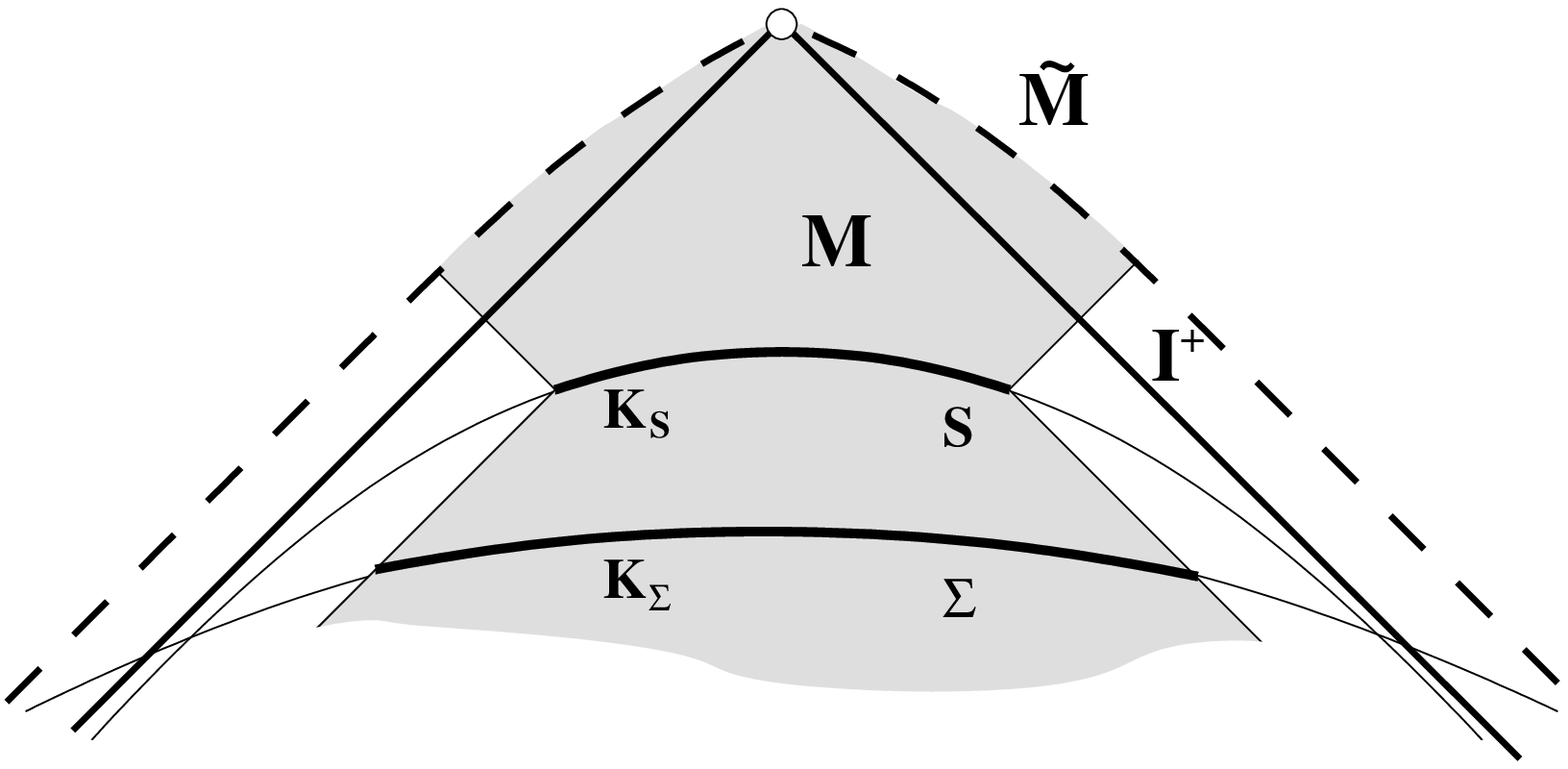}
\end{center}
%\caption{Cosa dire?}
\end{figure}

\remark We recall the reader that an asymptotically flat spacetime at null {\em and spacelike infinity} \cite{Wald}
$(M,g)$ is said to be {\em strongly asymptotically 
predictable} in the sense of \cite{Wald},
if in the unphysical associated spacetime there is an open set $\tilde{V}\subset \tM$ with 
$\overline{M\cap J^-(\scri)}\subset \tilde{V}$ (the closure being referred to $\tM$ so that $i^0 \in V$
also if, by definition, $i^0 \not \in \scri$) such that $(\tilde{V}, \tg)$ 
 is globally hyperbolic. 
Minkowski spacetime is such \cite{Wald}.  For those spacetimes in particular, the proposition above
applies.\\

\noindent 
We go to define a field theory on $\scri$ -- thought as a pure differentiable manifold -- based on smooth scalar fields $\psi$ 
and assuming $G_{BMS}$ as the natural 
symmetry group. The latter assumption is  in order to try to give some physical interpretation of the theory, 
since physical information is invariant under $G_{BMS}$ as said above. In particular, we have to handle 
the extent of a  metrical structure on $\scri$ which 
is not invariant under BMS group. The field theory should be viewed, more appropriately, as QFT on the  class
of all the triples $(\scri,\tilde{h}, n)$ connected with $(\scri, \tilde{h}_B, n_B)$ by the transformations of $\bms$.
In this way one takes asymptotic Killing symmetries into account.
Therefore we need a representation $G_{BMS} \ni \gamma \mapsto A_\gamma$ in terms of transformations  
$A_\gamma: C^\infty(\scri;\bC) \to C^\infty(\scri;\bC)$.
The naive idea is to define such an action as the push-forward on scalar fields of  
diffeomorphisms $\gamma \in G_{BMS}$, i.e.  $A_\gamma := \gamma^*$. However this is not a very satisfactory idea,
if one wants to maintain the possibility to interpret some of the fields $\psi$ as  extensions to $\scri$ of 
fields 
$(\omega\Omega)^{\alpha}\phi$ defined in the bulk.
  Proposition \ref{prop1} shows that there are
one-parameter (local) groups of diffeomorphisms $\{\gamma_t\}$ in the physical spacetime (in general not preserving 
(\ref{cc}))
which induce one-parameter subgroups of $G_{BMS}$, $\gamma'_t$.
A natural requirement on the wanted representation $A^{(\alpha)}$ is that, for a scalar field $\phi$ on $M$ such that 
$(\omega\Omega)^{\alpha}\phi$ 
admits a smooth extension $\psi$ to $\scri$
\beq 
 A^{(\alpha)}_{\gamma'_t} \psi =\lim_{\scri}(\omega\Omega)^{\alpha}\gamma^*_{t}(\phi) 
\label{req}\eeq
for every (local) one-parameter group of diffeomorphisms $\{\gamma_t\}$ generated by any vector field $\xi$ 
as in Proposition \ref{prop1}, for every value $t$ of the 
associated (local) one-parameter group of diffeomorphisms.
We have the following result whose proof is in the Appendix.\\

\proposizione\label{prop3} {\em Assume that $(M,g)$ is asymptotically flat with associated unphysical spacetime 
$(\tM,\tg)$  (with $\tg\sp \rest_M= \Omega^2 g$). Fix $\omega>0$ in a neighborhood of $\scri$ such that $\omega\tg$
is associated with the triple $(\scri, \tilde{h}_B, n_B)$.
Consider, for a fixed $\alpha \in \bR$, a representation $G_{BMS} \ni \gamma \mapsto A^{(\alpha)}_\gamma$ in terms 
of transformations  
$A_\gamma: C^\infty(\scri;\bC) \to C^\infty(\scri;\bC)$ such that $t\mapsto A^{(\alpha)}_{\gamma_t} \psi_0$ 
is smooth
for every fixed $\psi_0$ and every fixed one-parameter group of diffeomorphisms $\{\gamma_t\}$ subgroup of 
$G_{BMS}$. Finally assume that (\ref{req}) holds for 
any $\psi$ obtained as 
smooth extension to $\scri$ 
of $(\omega\Omega)^{\alpha}\phi$, $\phi\in C^\infty(M;\bC)$. 
Then, in any Bondi frame
\beq
\left(A^{(\alpha)}_{(\Lambda,f)}\psi\right)(u',\z',\bz') := K_\Lambda(\z,\bz)^{-\alpha} \psi(u,\z,\bz)\:.\label{alpha}
\eeq
for any $(\Lambda,f)\in G_{BMS}$ and referring to (\ref{u}),(\ref{z}), (\ref{K}).}\\

\noindent From (\ref{KK}), equation
(\ref{alpha}) defines, in fact, a representation of $G_{BMS}$ when assumed valid on all the fields 
$\psi \in C^\infty(\scri,\bC)$ or some BMS-invariant subspace of $C^\infty(\scri)$ as $C_c^\infty(\scri;\bC)$
or similar. 

 \noindent {\em From now on we assume that the action of $G_{BMS}$ on scalar fields $\psi\in C^\infty(\scri;\bC)$ 
is given from a representation 
$A^{(\alpha)}: G_{BMS} \ni \gamma \mapsto A^{(\alpha)}_\gamma$
 defined in (\ref{alpha})
with $\alpha$ fixed.}\\

\noindent Transformations  (\ref{alpha}) are well-known and used in the literature 
\cite{rappresentazioni_con_peso}. We stress that our interpretation of $A^{(\alpha)}_{(\Lambda,f)}$
is  {\em active} here, in particular the fields $\psi$ are scalar fields 
and thus they  transform as usual scalar fields under change of coordinates related or not 
by a BMS transformation (passive transformations). 
Using Proposition \ref{exremark1}, (\ref{K}) in particular, the reader can easily prove the following result.\\

\proposizione \label{propaggiunta}
{\em Consider two Bondi frames $B$ and $B'$ on $\scri$.  Take $\gamma\in G_{BMS}$ 
and represent it as $(\Lambda,f)$ and $(\Lambda',f')$ in $B$ and $B'$ respectively
(so that (\ref{indep}) holds).\\
Acting on a scalar fields $\psi$, $A^{(\alpha)}_{(\Lambda,f)}$ and $A^{(\alpha)}_{(\Lambda',f')}$ produce
the same transformed scalar field.}\\

\noindent The proposition says that the representation defined  in Proposition \ref{prop3} {\em does not depend} 
on the particular Bondi frame used to represent $\scri$, but it depends only on the
diffeomorphisms $\gamma\in G_{BMS}$ individuated by the pairs $(\Lambda,f)$ in the  Bondi frame
used to make explicit the representation.  In this way we are given a 
{\em unique representation} $G_{BMS} \ni \gamma \mapsto A^{(\alpha)}_\gamma$ 
not depending on the used Bondi frame which can be represented as in (\ref{alpha}) when a Bondi frame is selected.\\

\ssa{BMS-Invariant Symplectic form} \label{QFT2}
 As a second step we introduce the {\bf space of (real) wavefunctions on $\scri$},  $\cS(\scri)$.
In a fixed Bondi frame  $\cS(\scri)$ is the {\em real linear space} of the smooth functions $\psi:\scri \to \bR$ such that 
$\psi$ itself and all of its derivatives in any variable vanish as $|u|\to +\infty$, uniformly in $\z,\bz$,
faster than any  functions $|u|^{-k}$ for every natural $k$. 
It is simply proved that actually $\cS(\scri)$
 does not depend on 
the used Bondi frame (use  Proposition \ref{exremark1}  and the fact that functions $f$ are 
continuous and thus bounded on the compact $\bS^2$).
Obviously $C_c^\infty(\scri) \subset \cS(\scri)$ and it is simply proved that $\cS(\scri)$ is invariant under 
the representation $A^{(1)}$ of $G_{BMS}$ defined in the previous section.\\
One has the following result that shows that $\cS(\scri)$ can be equipped with a symplectic form 
invariant under the action of BMS group. That symplectic form was also studied in \cite{AS} and \cite{mopi6}.\\

\teorema\label{theo1}
{\em  Consider the representations $A^{(\alpha)}$ on $C^\infty(\scri;\bC)$ of $G_{BMS}$ introduced above
and the  map: $\sigma: \cS(\scri)\times \cS(\scri) \to \bR$
\beq
\sigma(\psi_1,\psi_2) := \int_{\bR\times \bS^2} 
\left(\psi_2 \frac{\partial\psi_1}{\partial u}  - 
\psi_1 \frac{\partial\psi_2}{\partial u}\right) 
du \wedge \epsilon_{\bS^2}(\z,\bz)\:, \label{sigma}
\eeq
$(u,\z,\bz)$ being a Bondi frame  on $\scri$ and
$\epsilon_{\bS^2}$ being the standard volume form of the unit $2$-sphere 
\beq
\epsilon_{\bS^2} (\z,\bz):= \frac{2d\z \wedge d\bz}{i(1+\z\bz)^2} \label{polar} \:.
\eeq
The following holds.\\
%%%%CORREZZIONE: aggiunto on $\cS(\scri)$ sotto
{\bf (a)} $\sigma$ is a nondegenerate symplectic form on $\cS(\scri)$ (i.e. it is linear, antisymmetric
%%%
and $\sigma(\psi_1,\psi_2)=0$ for all $\psi_1 \in \cS(\scri)$  implies 
$\psi_2=0$)
independently from the used Bondi frame.\\
{\bf (b)} $\cS(\scri)$ is invariant under every representation $A^{(\alpha)}$, whereas 
$\sigma$ is invariant under  $A^{(1)}$.}\\

\noindent{\em Proof}. (a) can be proved by direct inspection  using Proposition \ref{exremark1} to check 
on the independence from the used Bondi frame
and taking advantage of the fact that $\epsilon_{\bS^2} (\z,\bz)$ is invariant under three dimensional rotations. 
Invariance of $\cS(\scri)$ under $A^{(\alpha)}$ can be established immediately using the fact that the functions
$f$ in (\ref{u}) and the functions $K_\Lambda$ in (\ref{K}) and (\ref{alpha}) are bounded.
Let us prove the non trivial part of item (b). 
One has
$$\sigma(\psi'_1,\psi'_2) = 
\int_{\bR\times \bS^2} 
\left(\psi'_2 \frac{\partial\psi'_1}{\partial u'}  - 
\psi'_1 \frac{\partial\psi'_2}{\partial u'}\right) 
du' \wedge \epsilon_{\bS^2}(\z',\bz')\:.$$
Now we can use (\ref{alpha}) together with the known relation
$$\epsilon_{\bS^2} (\z',\bz') = K_{\Lambda}(\z,\bz)^{2} \epsilon_{\bS^2} (\z,\bz)$$
obtaining 
$$\sigma(\psi'_1,\psi'_2) = 
\int_{\bR\times \bS^2} 
\left(\psi_2 \frac{\partial\psi_1}{\partial u}  - 
\psi_1 \frac{\partial\psi_2}{\partial u}\right) 
du \wedge \epsilon_{\bS^2}(\z,\bz)$$
which is the thesis. $\Box$\\

\remark \label{remark2} From now on  the restriction to the invariant space
$\cS(\scri)$ of  $A_\gamma^{(1)}$
is indicated by $A_\gamma$, similarly
 $A$ denotes the representation $G_{BMS}\ni \gamma \mapsto A_\gamma$.\\

\ssa{Weyl algebraic quantization and Fock representation} \label{QFT3} As third and last step we define QFT on $\scri$ for 
uncharged scalar fields in Weyl approach giving  also a preferred
Fock space representation.  \\ 
The formulation of real scalar QFT on the degenerate manifold $\scri$ 
we present here  is an almost straightforward adaptation of 
the  theory presented  in \cite{mopi6} 
(see section \ref{bulkQFT} for the corresponding in general curved
spacetime \cite{Wald2}). 
As $\cS(\scri)$ is a real vector space equipped with a nondegenerate
symplectic form $\sigma$, 
there exists a complex $C^*$-algebra  (theorem 5.2.8 in \cite{BR2})
generated by 
nonvanishing
elements, $W(\psi)$ with $\psi\in \cS(\scri)$
satisfying, for all $\psi, \psi' \in \cS(\scri)$,
$$\mbox{(W1)} \:\:\:\:\: \:\:\:\:\:W(-\psi) = W(\psi)^*,\:\:\:\:\: \:\:\:\:\: 
\mbox{(W2)}\:\:\:\:\: \:\:\:\:\: W(\psi)W(\psi') = e^{i\sigma(\psi,\psi')/2} W(\psi+\psi')\:.$$
That $C^*$-algebra, indicated by ${\cal W}(\scri)$,  is unique up to (isometric) $^*$-isomorphisms (theorem 5.2.8 in \cite{BR2}).
As consequences of (W1) and (W2), ${\cal W}(\scri)$ admits unit $I=W(0)$, each $W(\psi)$ is unitary and, 
from the nondegenerateness of $\sigma$, 
$W(\psi)= W(\psi_1)$ if and only if $\psi=\psi_{1}$.
${\cal W}(\scri)$ is called {\bf Weyl algebra associated with} $\cS(\scri)$ and $\sigma$
whereas the $W(\psi)$ are called {\bf (abstract) Weyl operators}.
The formal interpretation of elements $W(\psi)$ is   $W(\psi) \equiv  e^{i\Psi(\psi)}$
where $\Psi(\psi)$ are 
{\em symplectically smeared field operators} as we shall see shortly.
The definition of $\sigma$ entails straightforward implementation of {\em locality principle}:
\beq
[W(\psi_1), W(\psi_2)] =0 \:\:\:\:\: \mbox{if $\:\:(supp \psi_1) \cap (supp \psi_2) = \emptyset$}\:.
\eeq
Differently from QFT in curved spacetime, but similarly to \cite{mopi6}, 
here we do not impose any equation of motion. On the other hand 
the space of wavefunctions, differently from the extent in the case of  degenerate manifolds studied 
in \cite{mopi6}, gives rise to direct implementation of locality. 
No ``causal propagator'' has to be introduced in this case.\\
A Fock representation of  ${\cal W}(\scri)$ based on a $BMS$-invariant vacuum state can be introduced as follows.
 From a physical point of view, the procedure   
resembles  quantization with respect to Killing time in a static spacetime.
Fix a Bondi  frame $(u,\z,\bz)$ on $\scri$.
Any  $\psi \in \cS(\scri)$ can be written as a  Fourier integral in the parameter
$u$ and one may extract the  {\bf positive-frequency part} (with respect to $u$):
\beq
 \psi_+(u,\z,\bz)  := \int_{\bR^+} \frac{dE}{\sqrt{4\pi E}} e^{-iE u}\widetilde{\psi_+}(E,\z,\bz) \:.\label{one}
\eeq
where $\bR^+:= [0,+\infty)$ and 
\beq\widetilde{\psi_+}(E,\z,\bz) :=  
 \sqrt{2E}\int_\bR \frac{du}{\sqrt{2\pi}} e^{+iE u}{\psi}(u,\z,\bz)\:\:\:\: \mbox{for $E \in \bR^+$}\label{two}\:.
 \eeq  
 Obviously it also holds $\psi = \psi_+ + \overline{\psi_+}$. It could seem
 that the definition of positive frequency part depend on the used Bondi frame and the coordinate $u$ in particular; actually,  by direct inspection based on Proposition \ref{exremark1},  one finds that:\\ 
 
 \proposizione \label{exremark3}
 {\em Positive-frequency parts do not depend to the Bondi frame and define scalar fields. In other words
if $\psi \in \cS(\scri)$ has positive frequency parts $\psi_+$ and $\psi'_+$ respectively
in Bondi frames $(u,\z,\bz)$ and $(u',\z',\bz')$,
it holds 
 \beq
 \psi_+(u,\z,\bz) = \psi'_+(u'(u,\z,\bz),\z'(\z,\bz),\bz'(\z,\bz))\:,\:\:\:\: \mbox{for all $u\in \bR$, $(\z,\bz) \in \bC\times \bC$.}
 \label{psi+invariant}
 \eeq}
 We are able to give a definition of one-particle Hilbert space and show that it is isomorphic to a suitable space $L^2$.
Let us denote by $\cS(\scri)^{\bC}_+$ 
 the space  made of the complex  finite linear
combinations of positive-frequency parts 
of the elements of $\cS(\scri)$.
The proof of the following result is in the appendix.\\
\teorema\label{theorem2}
{\em  With the given definition of $\cS(\scri)$, $\sigma$ and $\cS(\scri)^{\bC}_+$, the following holds.\\
{\bf (a)} The right-hand side of the definition of $\sigma$ (\ref{sigma}) is well-behaved if evaluated on functions in 
$\cS(\scri)^{\bC}_+$ and it is independent from the used Bondi frame.\\
{\bf (b)}  Using (a) and extending the definition of $\sigma$ (\ref{sigma}) to $\cS(\scri)^{\bC}_+$,
consider the complex numbers
 \beq
\al \psi_{1+},\psi_{2+}\cl  :=-i\sigma(\overline{\psi_{1+}},\psi_{2+}) 
\:,\:\:\:\mbox{for every pair $\psi_{1},\psi_{2} \in \cS(\scri)$.} \label{prodscalar}
\eeq
There is only one Hermitean scalar product $\al \cdot ,\cdot\cl$ on $\cS(\scri)^{\bC}_+$
which fulfils (\ref{prodscalar}). $\al \cdot ,\cdot\cl$ is independent from the used Bondi frame,
whereas, referring $\widetilde{\psi_{+}}$ to a given Bondi frame $(u,\z,\bz)$,
\beq
\al \psi_{1+},\psi_{2+}\cl = \int_{\bR^+\times \bS^2}\sp\sp \sp\sp\overline{\widetilde{\psi_{1+}}(E,\z,\bz)}\:
\widetilde{\psi_{2+}}(E,\z,\bz)\:dE\otimes\epsilon_{\bS^2}(\z,\bz)\:,\:\:\:
\mbox{for every pair $\psi_{1},\psi_{2} \in \cS(\scri)$.}\label{prodscalar2}
\eeq
{\bf (c)} Let $\cH$ be the Hilbert completion of $\cS(\scri)^{\bC}_+$ with respect to $\al \cdot ,\cdot\cl$.
The unique complex linear and continuous extension of the map $\psi_+ \mapsto \widetilde{\psi_+}$ 
(for $\psi \in \cS(\scri)$) 
with domain given by the whole $\cH$
is a unitary isomorphism  onto
$L^2(\bR^+\times \bS^2, dE\otimes \epsilon_{\bS^2})$.\\
%%% CORREZZIONE: aggiunto punto (d)
{\bf (d)} The map $K : \cS(\scri) \ni \psi \mapsto \psi_+ \in \cH$ has range dense in $\cH$.}\\ 
%%%

\noindent In the following $\cH$  will be called {\bf one-particle space}. Quantum field theory on $\scri$ relies on the bosonic
(i.e. symmetric) Fock space ${\gF_+}(\cH)$ built upon the  vacuum state $\Upsilon$ (we assume $||\Upsilon||=1$ explicitly). 
The  {\bf field operator symplectically smeared with} $\psi \in \cS(\scri)$ is now defined as 
\cite{Wald2}
\beq
\sigma(\psi, \Psi) :=  ia(\psi_+)-i a^\dagger(\psi_+) \label{dec}\:,
\eeq
where the operators $a^\dagger({\psi_+})$  and 
%%% CORREZZIONE $a({\psi_+})$ 
(anti-linear in ${\psi_+}$)
%%%%
respectively create and annihilate the state $\psi_+\in \cH$. 
The common invariant domain of all the involved  operators 
is the dense linear manifold $F(\cH)$ spanned by the vectors with finite number of particles.
$\Psi(\psi)$ is essentially self-adjoint on $F(\cH )$
(it is symmetric and $F(\cH )$ is dense and made of analytic vectors)
and satisfies bosonic commutation relations (CCR): 
\beq [\sigma(\psi,\Psi), \sigma(\psi',\Psi)] = -i\sigma(\psi,\psi')I \nonumber
\:.\eeq 
Since there is no possibility of misunderstandings because we will not introduce other, non symplectic, 
smearing procedures for field operators defined on $\scri$,
from now on we use the simpler notation
\beq
\Psi(\psi) :=  \sigma(\psi,\Psi)\:,
\eeq
{\em however the reader should bear in his mind that symplectic 
smearing is understood}. 
Finally the unitary operators
\beq \widehat{W}(\psi) := e^{i\overline{\Psi(\psi)}}\label{utile}
\eeq
enjoy properties (W1), (W2)  so that they define a unitary representation
$\widehat{\cal W}(\scri) $  of  ${\cal W}(\scri)$ which is also irreducible.
A  proof of these properties is contained 
in propositions  5.2.3 and 5.2.4 in \cite{BR2}
where the used field operator is
$\Phi(f)$ with $f\in {\mathfrak h}:= \cH$ and it holds
$\Psi(\psi) = \sqrt{2}\Phi(i\psi_+)$
for $\psi \in \cS(\scri)$. In particular irreducibility arises from (3) and (4) in proposition 5.2.4 using the fact that 
the real linear map $K: \cS(\scri)\ni \psi \mapsto \psi_+\in \cH$ has range is dense as stated in (d) of theorem \ref{theorem2} (notice that 
this is not obvious in the general case since, by definition of $\cH$ and (c) of the mentioned theorem, the  {\em complexified} range of $K$ is dense in $\cH$, but not necessarily 
the range itself).\footnote{With the formalism of \cite{KW} the irreducibility of the representation follows from (ii) 
in Lemma A.2 in \cite{KW} making use of (\ref{GNS}) and (d) in theorem \ref{theorem2} again.}\\
If $\Pi  : {\cal W}(\scri) \to \widehat{\cal W}(\scri)$ 
denotes the unique 
 ($\sigma$ being nondegenerate)
 $C^*$-algebra isomorphism between those two Weyl representations,  $({\gF_+}(\cH ), \Pi ,  \Upsilon )$ 
coincides, up to unitary transformations,  with  the GNS triple associated with the  algebraic pure state $\lambda$
on ${\cal W}(\scri)$ uniquely defined  by the requirement (see the appendix) 
\beq\lambda (W(\psi)) := e^{-\langle \psi_+, \psi_+\rangle /2}\label{GNS}\:.\eeq

\ssa{Unitary BMS invariance} \label{QFT4} Let us show that ${\gF}(\cH)$ admits a unitary representation of $G_{BMS}$
which is covariant with respect to an analogous representation of the group given in terms of $*$-automorphism
of $\widehat{\cal W}(\scri)$. Moreover we show that the vacuum state $\Upsilon$ (or equivalently, the associated algebraic state $\lambda$
on ${\cal W}(\scri)$) is invariant under the representation.\\ 
Consider the representation $A$ of $G_{BMS}$ in terms of transformations of fields in $\cS(\scri)$
used in sections \ref{QFT1} and \ref{QFT2}. 
As a consequence of the invariance of $\sigma$ under the action of $A_{\gamma}$, by (4) in theorem 5.2.8 of \cite{BR2} one has the 
following straightforward result concerning the $C^*$-algebra $\cW(\scri)$ constructed with $\sigma$.\\

\proposizione\label{prop4}
{\em  With the given definitions of $A$ (remark \ref{remark2}) and $\cW(\scri)$ there is a unique representation 
of $G_{BMS}$, indicated by $\alpha : G_{BMS} \ni \gamma \mapsto \alpha_\gamma$, and made of $*$-automorphisms of $\cW(\scri)$, satisfying
\beq
\alpha_\gamma(W(\psi)) = W\left(A_{\gamma} \psi\right)\:.
\eeq}
 Let us come to the main result given in the following theorem.\\

\noindent \teorema \label{theo2}
{\em Consider the representation of $\cW(\scri)$ built upon $\Upsilon$ in the Fock space ${\gF}_+(\cH)$ 
equipped with the representation of $G_{BMS}$, $\alpha$, given above. The following holds.\\
{\bf (a)} There is unique a unitary representation
$U : G_{BMS} \ni \gamma \mapsto U_\gamma$ 
such that both the requirements below are fulfilled.\\
(i) It is covariant with respect to the representation $\alpha$ , i.e.
\beq
U_\gamma \widehat{W}(\psi) U^\dagger_\gamma = \alpha_\gamma(\widehat{W}(\psi))\:, \:\:\:\:
\mbox{for all $\gamma \in G_{BMS}$ and $\psi \in \cS(\scri)$.}\label{main}
\eeq
(ii) The vacuum vector $\Upsilon$ is invariant under $U$: $U\Upsilon = \Upsilon$. \\
{\bf (b)} Any projective unitary representation\footnote{See also \cite{Mc11, Lau} for 
an earlier discussion on this issue.} $V : G_{BMS} \ni \gamma \mapsto V_\gamma$   
on ${\gF}_+(\cH)$ which is covariant with respect to $\alpha$ can be made properly 
unitary, since it must satisfy,
\beq
 e^{ig(\gamma)} V_\gamma = U_\gamma \:, \:\:\:\:\mbox{with $e^{-ig(\gamma)} = 
 \langle \Upsilon, V_\gamma \Upsilon \rangle$,  for every $\gamma \in G_{BMS}$.}\label{main2}
\eeq
{\bf (c)} The subspaces of ${\gF}_+(\cH)$ with fixed number of particles are invariant under $U$
 and $U$ itself is constructed canonically by tensorialization of $U\sp\rest_{\cH}$. The latter 
 satisfies, for every $\gamma\in G_{BMS}$ and the positive frequency part of any $\psi\in \cS(\scri)$
 \beq
  U_\gamma\psi_+ = A_\gamma^{(1)}(\psi_+) = \left(A_\gamma^{(1)}(\psi)\right)_+\:. \label{Ureducted}
 \eeq
 %%%%
 Equivalently, in a fixed Bondi frame, where $G_{BMS}\ni \gamma \equiv 
 (\Lambda, f) \in SO(3,1)\sp\uparrow \sp\ltimes C^\infty(\bS^2)$,
 \beq
  \left(U_{(\Lambda,f)}\varphi\right)(E,\z,\bz) = \frac{e^{iE K_{\Lambda}(\Lambda^{-1}(\z,\bz))f(\Lambda^{-1}(\z,\bz))}}{
  \sqrt{K_{\Lambda}(\Lambda^{-1}(\z,\bz))}}
 \varphi\left(E K_{\Lambda}\left(\Lambda^{-1}(\z,\bz)\right),\Lambda^{-1}(\z,\bz)\right) \:, \label{Ureducted2}
 \eeq
 is valid for every $\varphi \in L^2(\bR^+\times \bS^2; dE \otimes \epsilon_{\bS^2})$, $\varphi = \widetilde{\psi_+}$ in particular.}\\

\noindent {\em Proof}. (a) and (c). 
Let us assume it exists $U$ which satisfies (i) and in particular (ii). Then the 
uniqueness property
is a straightforward consequence of (b) (whose proof is
independent from (a) and (c)) since, from (\ref{main2}), $V\Upsilon=\Upsilon$ which
implies $e^{-ig(\gamma)} = \langle \Upsilon, V_\gamma \Upsilon \rangle =1$.
Let us pass to prove the existence of $U$. 
Consider the positive frequency part $\psi_+$
of $\psi\in \cS(\scri)$.  Theorem \ref{theorem2} (in the appendix) we have that
$\psi_+ \in C^\infty(\scri;\bC)$ so that $A^{(1)}_\gamma \psi_+$ is well-defined.
Furthermore $\psi_+$ with its derivatives decay as $|u|\to +\infty$ 
fast enough and uniformly in $\z,\bz$, 
so that it makes sense to apply $\sigma$ to a pair of  functions $\psi_+$. Moreover the proof of the invariance 
of $\sigma$  under the representation $A^{(1)}$ given in Theorem
%%%%CORREZZIONE: teorema citato sbagliato
 \ref{theorem2}
%%%% can be re-formulated -- 
by  changing the relevant domains simply -- when working on functions $\psi_+$ instead of functions in $\cS(\scri)$. 
Collecting all together, since $\langle \psi_{1+},\psi_{2+} \rangle := -i \sigma(\overline{\psi_{1+}}, \psi_{2+})$,
it turns out that the map $\psi_+ \mapsto A_\gamma^{(1)} \psi_+$ preserves the values of the scalar product in $\cH$
provided  any function $A^{(1)}_\gamma \psi_+$ is the positive frequency part of some 
$\psi' \in \cS(\scri)$ when $\psi\in \cS(\scri)$. Now, by direct inspection using (\ref{one}), (\ref{two})
as well as (\ref{alpha}) and (\ref{K}), and taking the positivity of $K_\Lambda$ into account, one finds, in facts, 
that $A_\gamma^{(1)}(\psi_+) = \left(A_\gamma^{(1)}(\psi)\right)_+$.
The map $L_\gamma : \psi_+ \mapsto A_\gamma^{(1)} \psi_+$ preserve the scalar product and thus it can be extended 
by $\bC$-linearity and continuity
to an isometric transformation  $S_\gamma$ from $\cH = \overline{\cS(\scri)_+^{\bC}}$ to $\cH$. That transformation is
unitary it being surjective because $S_{\gamma^{-1}}$ is its inverse. $\gamma \mapsto S_\gamma$ gives rise, in fact, to a unitary
representation of $G_{BMS}$ on $\cH$. Let us define the unitary representation $G_{BMS} \ni \gamma \mapsto U_\gamma$
on the whole space ${\gF}_+(\scri)$ by assuming $U_\gamma \Upsilon := \Upsilon$ and using  the standard 
tensorialization of $S_\gamma$ on every subspaces with finite number of particles. To conclude the proof 
of (a) and (c) it is now sufficient to establish the validity of (\ref{main}). 
(Notice that, with the given definition of $U$, in proving the validity of the 
identity $A_\gamma^{(1)}(\psi_+) = \left(A_\gamma^{(1)}(\psi)\right)_+$ one proves, 
in  fact, also (\ref{Ureducted}) and (\ref{Ureducted2})).
To prove (\ref{main}) it is sufficient to note that, in general, whenever the unitary map $V: {\gF}_+(\cH) \to {\gF}_+(\cH)$
satisfy $V\Upsilon=\Upsilon$ and it is the standard tensorialization of some unitary map $V_1 : \cH \to \cH$
then, for any $\phi \in \cH$,
$Va^\dagger(\phi)V^\dagger = a^\dagger(V_1\phi)$ 
and $Va(\overline{\phi})V^\dagger = a(\overline{V_1\phi})$.
 Since  
$\Psi(\psi) = -ia^\dagger(\psi_+)+ ia(\overline{\psi_+})$
one has
$U_\gamma \Psi(\psi)U^\dagger_\gamma = U_{\gamma} \Psi(\psi)U^\dagger_{\gamma}\Psi(A_{\gamma}\psi)$. 
Exponentiating this identity 
(using the fact that the vectors with finite number of particles are analytic vectors for $\Psi(\psi)$ \cite{BR2})
 (\ref{main}) arises.\\
(b) By hypotheses
 $U_\gamma \widehat{W}(\psi) U^\dagger_\gamma = \alpha_\gamma(\widehat{W}(\psi))= V_\gamma \widehat{W}(\psi) V^\dagger_\gamma$
so that $[V^\dagger_\gamma U_\gamma, \widehat{W}(\psi)]=0$. On the other hand the representation of
Weyl algebra $\widehat{\cW}(\scri)$
is irreducible as said above and thus, by Schur's lemma, $V^\dagger_\gamma U_\gamma= \alpha(\gamma) I$. Since 
$(V^\dagger_\gamma U_\gamma)^{-1} = (V^\dagger_\gamma U_\gamma)^\dagger= 
\overline{\alpha(\gamma)} I$,
 it must be $|\alpha(\gamma)|^2 =1$ and so $e^{ig(\gamma)} V_\gamma = U_\gamma$. Finally 
$e^{ig(\gamma)} V_\gamma = U_\gamma$ and 
(ii) imply $e^{-ig(\gamma)} = 
 \langle \Upsilon, V_\gamma \Upsilon \rangle$. $\Box$\\

\ssa{Topology on $G_{BMS}$ in view of the analysis of irreducible unitary representations and strongly continuity}
\label{nt}
Up to now we have assumed no topology on $G_{BMS}$. As the group is infinite dimensional and made of 
diffeomorphisms, a very natural topology is that induced by a suitable 
%%CORREZZIONE
countable 
%%%
class of seminorms \cite{Milnor-leshouches}
yielding the so-called {\em nuclear topology} (see below), though other choices have been made in the literature. We spend
some words on this interesting issue.
Since its original definition in \cite{sachsa, Bondi}, the BMS group has been recognized as
a semidirect product of two groups $G_{BMS}=H\ltimes N$ as it can be directly 
inferred from (\ref{product}). The group $H$ stands for the 
proper
orthocronous Lorentz group, whereas the abelian group, the space of supertranslations $N$, is 
a suitable set of sufficiently regular real functions
on the two sphere equipped with the abelian group structure induced by pointwise 
sum of functions. 
Up to now we have chosen $N= C^\infty(\bS^2)$, but there are other possibilities connected with the question about
the topology to associate to $N$ in order
to have the most physically sensible characterization for the Bondi-Metzner-Sachs group.
In Penrose construction \cite{Penrose}, where the BMS group arises as the group of exact 
conformal motions (preserving null angles) of the boundaries $\Im^\pm$ of conformally compactified asymptotically simple 
spacetimes, a specific degree of smoothness on the elements of $N$ was never imposed. 
 Nonetheless, historically the first stringent request has been
proposed by Sachs in \cite{sachsa}, i.e. each $\alpha \in N$ must be at least twice
differentiable. This choice has been abandoned by McCarthy in his study of the BMS theory of representations \cite{Mc1}, where 
he widened the possible supertranslations 
to the set of real-valued square-integrable functions $N=L^2(\bS^2; \epsilon_{\bS^2})_\bR$ equipped with \emph{Hilbert topology}. 
The underlying reasons for this proposal are two, the former concerning the great simplification of the treatment 
of induced representations in this framework\footnote{Originally it was also thought that, at a level of representation 
theory, the results were not affected by the choice of the topology of $N$ though this claim was successively falsified.}, the latter related to the conjecture that square integrable supertranslations are 
more suited to describe \emph{bounded gravitational systems} \cite{Mc5}. It is imperative to notice that, though
 such assertions may seem at a first glance reasonable (barring a problem with the interpretation of the elements of the group
in terms of diffeomorphisms), they have never been really justified besides 
purely heuristic arguments. As a matter of fact, a natural choice for $N$ and a corresponding 
 topology is, accordingly to the discussion in section \ref{QFT1},
$N=C^\infty(\bS^2)$ equipped with the  
\emph{nuclear topology}, first proposed in $\cite{Girardello}$. 
We follow \cite{Mc4} (and references therein) according to which the nuclear 
topology on  $C^\infty(\bS^2)$
is the topology such that $C^\infty(\bS^2) \supset \{f_n\}_{n\in \bN}$ turns out to converge to $f\in C^\infty(\bS^2)$
iff, for every local chart on $\bS^2$,  $\phi : U \ni p \mapsto (x(p),y(p))$  and in any compact $K\subset U$:
 $$\sup_K \left|\frac{\partial^{\alpha +\beta}}{\partial x^\alpha \partial y^\beta}
f_n\circ \phi^{-1}-\frac{\partial^{\alpha +\beta}}{\partial x^\alpha
\partial y^\beta}  f\circ \phi^{-1}\right|\longrightarrow 0\:, \:\:\:\: \mbox{as 
$n\to +\infty$,}$$ 
for every choice of $\alpha,\beta =0,1,2\cdots$. As is well-known, this topology can be induced 
by a suitable class of seminorms. Although it has been pointed out that this choice for $N$ and its 
topology should describe more accurately unbounded gravitating sources 
\cite{Mc5}, we will nonetheless find this framework more natural than the
Hilbert topology and thus we adopt the nuclear topology on  $N=C^\infty(\bS^2)$ and equip $\gbms$ with 
the consequent topology product. 
In particular we shall show in proposition \ref{finale} that,  with our choice, it is possible
to identify a field on $\scri$, which transforms with respect to $\gbms$ as said in (\ref{Ureducted2}),
with an intrinsic BMS field as introduced in the next section. After that proposition we shall remark that 
the result cannot be achieved using Hilbert topology. 

\noindent To conclude this section we state a theorem about strongly continuity of the representation  of 
$\gbms$, $U : \gbms \ni g \mapsto U_g$, defined in theorem \ref{theo2}
on ${\gF}_+(\cH)$. 
 The relevance of strongly continuity for a unitary representation, is that, through Stone's theorem, 
it implies the existence of self-adjoint generators of the representation it-self. 
The proof of the theorem 
is in the Appendix.\\

%%%CORRETTO il TESTO
\teorema \label{theo3}
{\em Make $\gbms$ a topological group adopting the product topology of the standard topology of $SO(3,1)\sp \uparrow$ 
and the nuclear topology of $C^\infty(\bS^2)$.
The unitary representation of the topological group $\gbms$ defined in theorem \ref{theo2}, $U : \gbms \ni g \mapsto U_g$,
on ${\gF}_+(\cH)$ is strongly continuous.}\\

%CORREZZIONE eliminato il commento
%\noindent \remark \label{remarkcontinuity} From the proof of the theorem it results that
%the topology on the Abelian factor of $\gbms$, $N$,
%plays no fundamental role in the proof of strong continuity. It can be replaced by Hilbert topology, 
%taking the product topology on the whole group.\\
 
\section{BMS theory of representations in nuclear topology.}

\ssa{General goals of the section}
In the previous discussions and in particular in section \ref{QFT1}, we have developed a scalar QFT on $\scri$ 
whose  kinematical 
data are fields $\psi$ which are  suitable smooth extensions/restrictions to 
$\scri$ of  fields $\phi$ living in $(M,g)$. 
Nonetheless a second candidate way to construct a consistent QFT at null infinity consists of considering as 
kinematical data, the set of wave functions invariant under a 
unitary irreducible representation of the $G_{BMS}$ group 
\cite{Arcioni}. The support of such functions is not {\em a priori} the underlying spacetime - $\Im^+$ in 
our scenario - but it is a suitable 
manifold modelled on a subgroup of $G_{BMS}$. For this reason we shall also refer to such fields as 
{\bf intrinsic} ${\bf G_{BMS}}$ {\bf fields}. \\
The rationale underlying this section is to 
demonstrate that, at least for scalar fields, both approaches are fully equivalent. In particular we shall establish that (\ref{Ureducted2}) 
is the transformation proper of  an intrinsic scalar $G_{BMS}$ field.\\

\ssa{The group $\tgbms$ and some associated spaces}\label{rep2}
To achieve our task, in the forthcoming discussion on representations of $BMS$ group we shall study the 
unitary representations of 
 the topological group $\tgbms =SL(2,\mathbb{C})\ltimes C^\infty(\bS^2)$
where the product of the group is given by suitable re-interpretations of (\ref{KK}) and (\ref{product}) and the topology
is the product of the usual topology on $SL(2,\bC)$ and that nuclear on $C^\infty(\bS^2)$ introduced in section
\ref{nt}. 
In a fixed Bondi frame, the composition of two elements $g = (A,\alpha), g' = (A',\alpha') \in \tgbms$
is defined by 
\begin{eqnarray}
(A',\alpha') \odot (A,\alpha) &=& \left(A'A,\: \alpha + (K_{A^{-1}} \circ A)\cdot (\alpha'\circ A)  \right)\:,
\label{product22}
\end{eqnarray}
\begin{gather}
A (\z,\bz)  := \left(\frac{a\z + b}{c\z +d}\:, \frac{\overline{a}\bz + \overline{b}}{\overline{c}\bz +\overline{d}}\right)\:,
\label{z2}
\end{gather}
 \begin{eqnarray}
 K_A(\z,\bz) :=  \frac{(1+\z\bz)}{(a\z + b)(\overline{a}\bz + \overline{b}) +(c\z +d)(
 \overline{c}\bz +\overline{d})}
\label{K2}\:\: \: \mbox{and}\:\:\:\:
 A:= \left[
\begin{array}{cc}
  a & b\\
  c & d
\end{array}
\right] \:.
\end{eqnarray}
In a sense, noticing that $SL(2,\bC)$ is the universal covering of $SO(3,1)\spa\uparrow$, 
$\tgbms$ could be considered as the universal covering of $\gbms$. A discussion on this point 
would be necessary if one tries to interpret  the term  ``universal covering'' literally since 
both $\gbms$ and $\tgbms$ are {\em infinite} dimensional topological groups. 
 However we limit ourselves to say that, according to \cite{Mc1, Mc11}, 
 replacing in the structure of $\gbms$ the orthocronous proper 
Lorentz group $SO(3,1)\sp\uparrow$\footnote{The orthocronous proper 
Lorentz group is called homogeneous Lorentz group in \cite{Mc1, Mc11}.}
with  its universal covering  $SL(2,\bC)$, it introduces  only further unitary irreducible representations, 
induced by the $\bZ_2$ subgroup of $SL(2,\bC)$, beyond the unitary irreducible representations of $\gbms$. 
These represent nothing but the symptom that $SL(2,\mathbb{C})$ ``covers twice''
$SO(3,1)\!\!\uparrow$ and they will be not considered in this paper: we shall pick out 
only representations of $\tgbms$ which are as well representations of $\gbms$.

 The next step consists in the following further definition \cite{Mc4, Gelfand}:\\

\definizione\label{leD}{\em If $n\in \bZ$ is fixed,
 we call $D_{(n,n)}$ the space of real functions $f$ of two complex variables $\z_1, \z_2$ 
 and their conjugate ones $\bz_1, \bz_2$
 such that:
\begin{itemize}
\item $f$ is of class $C^\infty$ in its arguments except at most the origin $(0,0,0,0)$;
\item for any $\sigma\in\bC$,
$f(\sigma\z_1,\bar{\sigma}\bz_1,\sigma\z_2,\bar{\sigma}\bz_2)=\sigma^{(n-1)}\bar{\sigma}^{(n-1)}f(\z_1,\bz_1,\z_2,\bz_2)$ for all $\z_1, \z_2,\bz_1, \bz_2$.
\end{itemize} 
Moreover $D_{(n,n)}$ is assumed to be endowed with the topology of uniform convergence on all compact 
sets not containing the origin for the functions and all their derivatives separately.}\\

\noindent The relevance of the definition above arises from the following proposition which, first of all, allows one to 
identify $C^\infty(\bS^2)$ with the space $D_{(2,2)}$ and the subsequent space $D_2$ introduced below.
These spaces will be used later.
 The relevance of the second statement will be clarified  shortly
after proposition \ref{chara2}. 
The action $\Lambda \alpha$ of  $\Lambda \in SL(2,\bC)$ on an element $\alpha$ of $C^\infty(\bS^2)$,
considered in the equation (\ref{azionesuN}) below, is that arising from the representation $SL(2,\bC)$
in terms of  $C^\infty(\bS^2)$ automorphisms
used to define the semidirect product $SL(2,\bC) \ltimes C^\infty(\bS^2)$. 
Notice that, by the natural normal subgroup identification $C^\infty(\bS^2) \ni \alpha \equiv (I, \alpha) \in \tgbms$
one also has:
\beq \label{Gaction0} (I,\alpha) \mapsto  g \odot (I,\alpha) \odot g^{-1} = (I, \Lambda \alpha)\quad 
\mbox{for any  $g=(\Lambda,\alpha^\prime)\in \tgbms$}\:,\eeq
$I$ being the unit element of $SL(2,\bC)$. 
Since $C^\infty(\bS^2)$ is Abelian, the dependence on $\alpha'$ is immaterial as the notation suggests.\\

\proposizione\label{coppia} {\em There is a one-to-one map ${\cal T}: C^\infty(\bS^2) \ni \alpha \mapsto  
f\in D_{(2,2)}$. In this way,
the action of  $\Lambda \in SL(2,\bC)$ on an element $\alpha$ of $C^\infty(\bS^2)$  
\beq\label{azionesuN}
\left(\Lambda \alpha\right)(\z,\bz)=K_{\Lambda}(\Lambda^{-1}(\z,\bz))\alpha(\Lambda^{-1}(\z,\bz))
\eeq
is equivalent to the action (defined in \cite{Gelfand}) of the same $\Lambda$ on $f$
\beq\label{Loract}
f\circ \Lambda^{-1}:=f(a\z_1+c\z_2,a\bz_1+c\bz_2,b\z_1+d\z_2,b\bz_1+d\bz_2),\;\;\;\forall \Lambda 
=\left[\begin{array}{cc}
a&b\\
c&d
\end{array}\right]^{-1}\in SL(2,\bC)\:.
\eeq
Finally ${\cal T}$ is a homeomorphism so that the topology of $D_{(2,2)}$ coincides with that on $C^\infty(\bS^2)$.\\}

\noindent The proof of this result  may be found in the appendix of \cite{Mc100} though we review some of the 
details which will be important in the forthcoming discussion. The sketch of the argument is the following: 
the homogeneity condition for the functions $f\in D_{(n,n)}$ allows us to associate to each of such 
$f$ a pair of $C^\infty$ functions 
$\xi, \hat{\xi}$  such that
\begin{gather*}
f(\z_1,\bz_1,\z_2,\bz_2)=|\z_1|^{2(n-1)} f\left(\frac{\z_2}{\z_1},\frac{\bar{\z_2}}{\bar{\z_1}}\right)=|\z_1|^{2(n-1)}\xi(\z,\bz),\\
f(\z_1,\bz_1,\z_2,\bz_2)=|\z_2|^{2(n-1)} f\left(\frac{\z_1}{\z_2},\frac{\bar{\z_1}}{\bar{\z_2}}\right)=|\z_2|^{2(n-1)}\hat{\xi}(\z,\bz),
\end{gather*}
where $\z=\frac{\z_1}{\z_2}$ and 
\beq\label{DN}
\hat{\xi}(\z,\bz)=|\z|^{2(n-1)}\xi(\z^{-1}, \bz^{-1})
\eeq
whenever $(\z_1,\z_2)\neq(0,0)$. If we call $D_n$
the set of the functions $\hat{\xi}$,
 the above discussion can be recast as the existence of a bijection between $D_{(n,n)}$ and $D_n$ which thus 
 inherits the same topology as $D_{(n,n)}$ (or {\em viceversa}). Furthermore (\ref{Loract}) becomes, with obvious
 notation,
\begin{gather*}
(\xi\circ \Lambda^{-1})(\z,\bz) =|a+c\z|^{2(n-1)}\xi\left(\frac{d+b\z}{a+c\z},
 \frac{\bar d+ \bar b\z}{\bar a+\bar c\bz}\right),\;\;\;a+c\z\neq 0,\\
(\hat{\xi}\circ \Lambda^{-1})(\z,\bz) =|d+b\z|^{2(n-1)}\xi\left(\frac{a+c\z}{d+b\z}, 
\frac{  \bar a+ \bar c\bz}{ \bar d+ \bar b\bz}\right),\;\;\;d+b\z\neq 0.\\
\end{gather*}
If we specialize to $n=2$, it is now possible to show (see \cite{Mc4, Mc100}) that the above equations correspond
 to the canonical realization of the $\tgbms$ group as $SL(2,\bC)\ltimes C^\infty(\bS^2)$ if we associate
  the supertranslation $\alpha\in C^\infty(\bS^2)$ with $\hat{\xi}$ as:
\beq\label{rel1}
\hat\xi(\z,\bz)=(1+|\z|^2)\alpha(\z,\bz).
\eeq
Within this framework and for every $\Lambda \in SL(2,\bC)$ and $\alpha\in C^\infty(\bS^2)$ 
 (\ref{azionesuN}) turns out to be equivalent to (\ref{Loract}) as one can check by direct inspection.\\

\remark Identifying the topological vector space  of supertranslations  $C^\infty(\bS^2)$
with  $D_{2}$ and equivalently with $D_{(2,2)}$, the $\tgbms$ group turns out to be  locally homeomorphic to a nuclear 
space\footnote{We recall the reader that, given a separable Hilbert space $\cH$, 
$\cE\subset\cH$ is called a nuclear space if it is the projective limit of a decreasing
sequence of Hilbert spaces $\cH_k$ such that the canonical imbedding of $\cH_k$
in $\cH_{k^\prime}$ ($k>k^\prime$) is an Hilbert-Schmidt operator.} and thus it 
is a {\bf nuclear Lie group} as defined by Gelfand and 
Vilenkin in \cite{Gelfand2}. In other words, there exists a neighborhood of the unit
element of $\tgbms$ which is homeomorphic to a neighborhood of zero in a
(separable Hilbert) nuclear space. \\

\noindent If $N$ is the real topological vector space of supertranslation $C^\infty(\bS^2)$, $N^*$ indicates 
its topological dual vector space, whose elements  are called {\bf (real) distributions} on $N$.\\

\remark\label{duale}  Since $N$ can be topologically identified as $D_{(2,2)}$, 
$N^*$ is fully equivalent to the set of continuous linear functionals 
$D_{(-2,-2)}$ which is obtained setting $n=-2$ in definition \ref{leD} with the 
prescription that all the equations should be interpreted in a distributional 
sense \cite{Mc4, Gelfand}. Consequently each $\phi\in D_{(-2,-2)}$ is a real distribution in two 
complex variables bijectively determined by a pair $\phi,\hat{\phi}\in D_{-2}$ 
of real distributions such that $\hat{\phi}=|z|^{-6}\phi$, as in (\ref{rel1}). The counterpart of (\ref{rel1}) for $N^*$ 
is the following: to each functional $\phi\in D_{(-2,-2)}$ corresponds the distribution $\beta\in N^*$
\beq\label{relduale}
\beta=(1+|\z|^2)^3\phi.
\eeq
Furthermore, if $L^2(\bS^2, \epsilon_{\bS^2})$ is the Hilbert  completion of $N$ with 
respect to the scalar product associated with $\epsilon_{\bS^2}$, 
$N\subset L^2(\bS^2, \epsilon_{\bS^2})\subset N^*$ is a {\em rigged Hilbert space}. \\

\ssa{Main ingredients to study unitary representations of $\tgbms$} \label{mi} 
The starting point to study unitary representations of BMS group consists in the detailed analysis of 
McCarthy \cite{Mc1, Mc2, Mc4}. 
 The theory of unitary 
and irreducible representations for $\tgbms$ with  {\em nuclear topology}  has been developed
in \cite{Mc4}  by means 
either of Mackey theory of induced representation \cite{Simms, Group, LLedo} applied 
to an infinite dimensional semidirect product \cite{Piard} either of 
Gelfand-Vilenkin work on nuclear groups \cite{Gelfand, Gelfand2}.
In the following we  briefly discuss  some key points. Here we introduce the main mathematical tools in order 
to construct the intrinsic wave functions. We refer to \cite{Arcioni} for a detailed analysis 
in the Hilbert topology scenario.\\

\definizione\label{charadef2}{\em If $A$ is an Abelian topological group, 
a {\bf character} (of $A$) is a continuous group homomorphism 
$\chi: A \to U(1)$, the latter being equipped with the natural topology induced by $\bC$. 
The set of  characters $A^\prime$ is an abelian group called
the {\bf dual character group} if equipped with the group product 
$$(\chi_1\chi_2)\left(\alpha\right):= \chi_1(\alpha)\chi_2(\alpha). \:\:\:\:\:\mbox{for all $\alpha\in A$}\:.$$}

\noindent A central tool concerns an explicit representation of the characters in terms of 
distributions \cite{Mc4}. The proof of the following relevant proposition is in the appendix.\\

\proposizione\label{chara2}{\em Viewing $N := C^\infty(\bS^2)$
as an additive continuous group, for every $\chi \in N'$ 
there is a distribution $\beta \in N^*$ such that 
$$\chi(\alpha) =\exp[i\left(\alpha,\beta\right)]\:, \quad \mbox{for every $\alpha\in N$}$$
where $(\alpha,\beta)$ has to be interpreted as the evaluation of the $\beta$-distribution on the test function $\alpha$.}\\

\remark\label{directintegral}  With characters one can decompose any unitary representation 
of $N = C^\infty(\bS^2)$. Indeed, a positive finitely normalizable measure $\mu_{N^*}$ on $N^*$ exists, 
which is quasi invariant under
group translations (i.e. for any measurable $X\subset N^*$, $\mu_{N^*}(X)=0$ iff $\mu_{N^*}(N+X)=0$),
and a family of Hilbert spaces $\{\mathcal{H}_\beta\}_{\beta \in N^*}$
such that, for any unitary representation of $N$, $U: \mathcal{H} \to \mathcal{H}$,  $\mathcal{H}$
being any Hilbert space, the following direct-integral decomposition holds  ({\em c.f.} chapter I and chapter IV -- theorem 5 and subsequent discussion -- in \cite{Gelfand2}):
$$\mathcal{H}=\int\limits_{N^*}\sp\oplus\:\: \mathcal{H}_{\beta}\:d\mu_{N^*}(\beta).$$
Moreover the spaces $\cH_\beta$ are invariant under $U$ and, for every $\alpha \in N$ and $\psi_\beta \in \cH_\beta$, one has $U\rest_{\cH_\beta} \psi_\beta = e^{i(\alpha,\beta)}\psi_\beta $. Here  $(\alpha,\beta)$ denotes action
of the distribution $\beta$ on the test function $\alpha$.\\

 For any $\Lambda\in SL(2,\mathbb{C})$ a natural action $\chi \mapsto \Lambda\chi$ on $N'$
induced by duality from that on $\alpha \in N$, considered above, is
\cite{Mc1, Mc4}:  
\beq\label{charaact222}
(\Lambda\chi)(\alpha) :=\chi(\Lambda^{-1}\alpha)
\eeq
whereas an action $\beta\mapsto\Lambda\beta$ on $N^*$ is intrinsically 
defined from the identity
\beq\label{charact4}
(\Lambda\beta,\alpha)=(\beta,\Lambda^{-1}\alpha).
\eeq
If we associate to the distribution $\beta$ the pair $(\phi,\hat{\phi})$ as 
discussed in remark \ref{duale}, the latter $SL(2,\bC)$ action translates as, if 
$\Lambda=\left[\begin{array}{cc}
a&b\\
c&d
\end{array}\right]^{-1}\in SL(2,\bC)$,
\begin{gather}\label{phi}
\left(\Lambda\phi\right)(\z,\bz)=|a+c\z|^{-6}\phi\left(\frac{b+d\z}{a+c\z},  \frac{\bar d+ \bar b\z}{\bar a+\bar c\bz}\right)\:,\quad
\mbox{with}\:\:  a+cz\neq 0\\
\left( \Lambda\hat{\phi}\right)(\z,\bz)=|d+b\z|^{-6}\hat\phi\left(\frac{c+a\z}{d+b\z}, \frac{\bar a+ \bar c\bz}{ \bar d+ \bar b\bz}\right)\:,
\quad \mbox{with}\:\:  d+bz\neq 0\:.\label{hatphi}
\end{gather}

\definizione\label{littlegr2}{\em Consider a semidirect group product $G=B\ltimes A$ where $A$ 
is a  topological abelian group, $B$ is any group and $\odot$ denotes the product in $G$. 
With the identification of $A$ with the normal subgroup of $G$ containing the pairs $(I,\alpha)$, $\alpha\in A$,
define the action\footnote{It coincides with the action of $B$ on $A$ in terms of $A$-group-automorphisms
 used in the definition of $\odot$.} $g\alpha$ of $g\in G$ on $\alpha \in A$:
$$(I, g\alpha) := g\odot (I,\alpha) \odot g^{-1}\:,
\quad \mbox{for all $\alpha \in A$, $g \in G$,}$$ thus extend this action on charcters, $\chi\in A'$, by duality: 
 $$\left(g\chi\right)(\alpha) := \chi(g^{-1}\alpha)\:,
\quad \mbox{for all $\chi \in A'$, $\alpha \in A$, $g \in G$.}$$
For any $\chi\in A^\prime$,  the {\bf
orbit} of $\chi$ (with respect to $G$) is the subset of $A'$
\beq
G\chi:=\left\{\chi^\prime\in A^\prime\;\left|\;\right. \exists g\in G
\;\;\textrm{such that}\;\;\chi^\prime=g\chi\right\},
\eeq
the {\bf isotropy group} of $\chi$ (with respect to $G$) is the subgroup of $G$
\beq
H_\chi:=\left\{\left.g\in G\;\right|\;g\chi=\chi\right\}\:,
\eeq
and the {\bf little group} of  $\chi$ (with respect to $G$) is the subgroup of  $H_\chi$
\beq
L_\chi:=\left\{\left.g = (L,0) \in G\;\right|\;g\chi=\chi\right\}\:.
\eeq}

\noindent Referring to $\tgbms = SL(2,\bC) \ltimes C^\infty(\bS^2)$, to (\ref{charaact222}) and to (\ref{charact4}),
 $L_\chi$ can equivalently be seen as the subgroup of 
$SL(2,\bC)$ whose elements $L$ satisfy
\beq\label{chara12}
L\bar{\beta}=\bar{\beta},
\eeq
$\bar{\beta} \in N^*$ being associated to $\chi$ according to proposition 
\ref{chara2}.\\

\remark 
A direct inspection  shows also that the $G$ action on a character
is completely independent from  $A$ due to Abelianess. Thus the most general
isotropy group has the form $$H_\chi=L_\chi\ltimes A\:.$$ 
This applies in particular to $\tgbms$ where $A= C^\infty(\bS^2)$.\\

\noindent We now discuss a last key remark concerning the \emph{mass} of a BMS 
field. First of all, define a base of {\em real spherical harmonics} 
$\{S_{lk}\}_{l=0,1,\cdots, k= 1,2,\cdots, 2l+1}$, in the real vector space 
$C^\infty(\bS^2)$ as follows:
\begin{eqnarray}
S_{l\:k} &:=& Y_{l0}\:\:\:\:\:\:\:\:\:\:\:\:\:\:\:\:\:\:\:\:\:\mbox{if $k = 2l+1$,}\\ 
S_{l\:k} &:=& \frac{Y_{l \:-k}-Y_{l\: k}}{\sqrt{2}}\:\:\:\:\mbox{if $1< k \leq l$,}\\ 
S_{l\:k} &:=& i\frac{Y_{l \:-k}+Y_{l\:k}}{\sqrt{2}}\:\:\:\mbox{if $l< k \leq 2l$,}
\end{eqnarray}
where
$Y_{lm}$ are the usual (complex) spherical harmonics with $m\in \bZ$ such that $-l \leq m\leq l$.
Now, let us consider
a generic supertranslation $\alpha\in C^\infty(\bS^2)$ and let us decompose (in the sense of $L^2(\bS^2,\epsilon_\bS^2)$)
it in real spherical harmonics
\beq \label{decsph}\alpha(\zeta,\bar{\zeta})=\sum\limits_{l=0}^1\sum\limits_{k=1}^{2l+1}a_{lm}S_{lm}(\zeta,\bar{\zeta})
+\sum\limits_{l=2}^\infty\sum_{k=1}^{2l+1}a_{lk}S_{lk}(\zeta,\bar{\zeta}),\;\;\bar{\alpha}_{lk} \in \bR\:.\eeq
The former double sum defines the {\em translational component} of $\alpha$ and the latter the {\em pure
supertranslational component} of $\alpha$. This relation allows one to split 
$C^\infty(\bS^2)$ into an orthogonal direct sum
$T^4\oplus\Sigma$ where $T^4$ is a four-dimensional real space 
invariant under $SL(2,\bC)$ viewed as the subgroup of $\tgbms$ made of elements $(A,0)$. More precisely
(see also proposition \ref{lastproposition} below):\\

\proposizione 
{\em The subset $SL(2,\bC) \ltimes T^4 \subset \tgbms$ made of the elements $(\Lambda,\alpha)$ 
with $\alpha \in T^4$ is a subgroup 
of $\tgbms$ itself which is invariant under $SL(2,\bC)$, i.e., if $g \in SL(2,\bC) \ltimes T^4 $,
$$g \odot (A,0) \:\mbox{and}\:  (A,0) \odot g \in  SL(2,\bC)\ltimes T^4\:, \:\:\:\: \mbox{for all $A \in SL(2,\bC)$.}$$}

\remark 
Defining the analogous subset $SL(2,\bC)\ltimes \Sigma$, one finds that $\Sigma$ is {\em not} $SL(2,\bC)$ invariant.
More precisely breaking 
of invariance  happens when $A\not \in SU(2)$. \\

\noindent The  decomposition (\ref{decsph}) explicitly associates to each $\alpha\in C^\infty
(\bS^2)$ the 4-vector
\beq\label{4mom2}
a_\mu \equiv - \frac{1}{2}\sqrt{\frac{3}{\pi}}\left(\frac{a_{01}}{\sqrt{3}},a_{11},a_{12},a_{13}\right)
\eeq
One has the following very useful proposition which can be proved by direct inspection and which will be used 
in several key points in the following.\\

\proposizione \label{covariance}
{\em If $\alpha_{a} \in T^4$, where $a_\mu$ is made of the first four components of $\alpha_{a}$ as in (\ref{4mom2}),
 transforming $\alpha_{a}$ under the action of $A\in SL(2,\bC)$ as in (\ref{azionesuN}) is equivalent
to transforming the 4-vector $a^\mu$ under the action of the Lorentz transformation 
associated with $A$ itself. In other words:
\beq
 K_A(\z,\bz)^{-1}\alpha_a\left(A(\z,\bz)\right) = \alpha_{\Pi(A)^{-1}a}(\z,\bz) \:,\:\:\:\:
\mbox{for all $A\in SL(2,\bC)$,} \label{notevole}
\eeq
$\Pi : SL(2,\bC) \to SO(3,1)\spa\uparrow$ being the canonical covering projection.}\\

\noindent According to the discussion in \cite{Mc4}, (\ref{4mom2}) can be translated 
to the dual space $N^*$ where we shall define the annihilator of $T^4$ as
\beq\label{annihilator} 
\left(T^4\right)^0=\left\{\left. \beta\in N^*\;\;\right|\;\;(\alpha,\beta)=0,\;\forall
\alpha\in T^4\hookrightarrow C^\infty(\bS^2)\right\}\:.
\eeq
$\hookrightarrow$ recalls the reader that $T^4$ above is seen as a 
subspace of $C^\infty(\bS^2)$ and not as the four-dimensional translation group of vectors $a^\mu$
acting in Minkowski space.\\
 From now on  $(T^4)^* \subset N^*$ denotes the subspace generated by the subset of $N^*$ 
$$\{S^*_{lk}\:|\: -l\leq m\leq l\:,\: l=0,1 \}\:,$$ where each $S^*_{lk}$ is completely defined  
by the requirement  $$(\alpha, S^*_{lm}) := a_{lm}\quad \mbox{$\forall \alpha \in N$, and $a_{lm}$ 
given in (\ref{decsph})}\:,$$  taking into account that each map $N\ni \alpha \mapsto a_{lm}$
is continuous in nuclear tolopogy and thus it belongs to $N^*$.
 It is simply proved that $(T^4)^*$ 
and $N^*/(T^4)^0$ are canonically isomorphic and the isomorphism (first introduced in
\cite{Mc4}) is invariant under $SL(2,\bC)$ transformation.
As a consequence there is a linear projection of $N^*$ onto $(T^4)^*$ (which is, in fact, the usual projection onto the quotient space 
composed with the cited isomorphism)
\beq \label{pigiusto}
\pi : N^* \to (T^4)^*\sim \frac{N^*}{(T^4)^0}\:.
\eeq
That projection enjoys  the following remarkable properties \cite{Mc4,Gelfand} which gives the first step in order to introduce 
the notion of mass for BMS representations:\\

\proposizione \label{aggiuntaallafine}{\em Let $\beta\in N^*$ and let $\phi\in D_{(-2,-2)}$ and $\hat{\phi}=|\z|^{-6}\phi$ be the  
distributions associated with $\beta$ as in remark \ref{duale}. The function 
\beq\label{projt}
\widehat{\pi(\beta)}(\z^\prime,\bz^\prime)=\frac{i}{2(1+|\zeta^\prime|^2)}\int\limits_{\mid\z\mid<1}[(\z-\z^\prime)
(\bz-\bz^\prime)\phi(\z,\bz)+(1-\z\z^\prime)(1-\bz\bz^\prime)\hat{\phi}(\z,\bz)]d\z 
d\bz,
\eeq 
is well defined for $\zeta,\bz\in \bC$ and, in fact, it  belongs to $T^{4}$. Moreover, as the notation suggests, $\widehat{\pi(\beta)}$
depends on $\pi(\beta)$ and not on the whole distribution $\beta$. That is
$\widehat{\pi(\beta)} = \widehat{\pi(\beta')}$ if $\pi(\beta)= \pi(\beta')$ for whatever $\beta,\beta'\in N^*$.\\}

\noindent The following final  proposition \cite{Mc1, Mc2, Mc4}  is, partially, a straightforward consequence of proposition 
\ref{covariance}. It produces the preannounced notion of mass similar to that used in the theory of Poincar\'e representations. \\ 

\proposizione\label{massa}{\em The space $(T^4)^*$ is invariant under the 
$SL(2,\bC)$-action on $N^*$ and, according to (\ref{rel1}), 
the supertranslation associated to
$\widehat{\pi(\beta)}$  may be expanded in spherical 
harmonics thus extracting as in (\ref{4mom2}) a 
4-vector $\widehat{{\pi(\beta)}}_\mu$.
Moreover if one defines the real bilinear form on $N^*\ni \beta_1,\beta_2$ as
\beq\label{mass2}
B\left(\beta_1,\beta_2\right) := \eta^{\mu\nu}{\widehat{\pi(\beta_1)}}_\mu {\widehat{\pi(\beta_2)}}_\nu,
\eeq
with $\eta := diag(-1,1,1,1)$ and  it turns out that $B$  is 
$SL(2,\mathbb{C})$-invariant.}\\ 

\noindent  $-B\left(\beta,\beta\right) =m^2$ is the equation for the squared-\emph{mass} $m^2$ of an intrinsic BMS field.
It is the analog of the invariant mass of a field in Wigner's approach to define Poincar\'e-invariant particles. 
Consequently we we shall refer to  $N^*$ as the {\bf supermomentum space} and its elements as the {\bf supermomenta}.\\

\ssa{Construction of unitary irreducible representations of $\tgbms$}
Consider a group $G=B\ltimes A$ where $A$ 
is a (possibly infinite dimensional) topological abelian group and $B$ a locally compact topological group and 
a suitable group operation is defined in order to make
$G$ the semidirect product of $B$ and $A$, which is a topological group with respect to the product of topologies. \\
Using the definitions and propositions given above, the procedure to build up unitary irreducible representations of $\tgbms$ goes on as follows,
starting from representations of the little groups of characters.
The next proposition has a trivial straightforward proof.\\

\proposizione\label{irreps2}{\em Take a character $\chi\in A'$ and a closed subgroup of $B$, $K$, which leaves 
invariant $\chi$.
 If  $K \ni L \mapsto  \sigma_L$ is a unitary and irreducible 
representation of $K$  acting
on a, non necessarily finite-dimensional, target Hilbert space $V$,
an associated unitary and irreducible representation $K \ltimes A \ni g \mapsto \chi\sigma_g$ of  $K \ltimes A$ 
acting on $V$ is constructed as follows:
\beq\label{irrep12}
\chi\sigma_{(\Lambda,\alpha)}\:\vpsi := \chi(\alpha)\:\sigma_\Lambda(\vpsi)\:,\:\:\:\: \mbox{for all $\vpsi \in V$.}
\eeq}

\noindent 
Furthermore let us define  the following equivalence relation in $G\times V$ equipped with the product topology:
\beq\label{rel2}
(g,v) \sim_K (g',v') \:\:\:\:\:\mbox{iff there is $g_K \in K$ such that $\:\:\:(g',v')=(gg^{-1}_K,\chi\sigma(g_K)v)$.}
\eeq
The quotient space equipped with its natural topology, will be denoted by
$$G\times_{K} V := \frac{G\times V}{\sim_K}\:.$$ 

\noindent From now on, concerning the equivalence classes associated with the equivalence relation defined above, we use
the notation $[g,v]$ instead of the more appropriate but more complicated $[(g,v)]$.\\

\remark { A natural projection map exists
$$\tau :G\times_{K} V\longrightarrow \frac{G}{K}\:,$$
which associates  $[g,v]\in G\times_{K} V$ with  $gK$. Furthermore the inverse image $\tau^{-1}(p)$ with $p=gK\in G/K$ for some $g\in G$, 
has the form $[g,v]$ where $v\in V$
is uniquely determined by $p$. Thus it exists a natural bijection from 
$\tau^{-1}(p)$ into $V$ such that, automatically, the former acquires the 
structure of a Hilbert space and this structure  does not  depend
upon the choice of $g\in G$ with $p=gK$. As a matter of fact, if $p=gK=g_1K$
with $g\neq g_1$, then the following diagram commutes:
$$\xymatrix{\tau^{-1}(p)
\ar[dd]^{id.}\ar[rr]^{\;\;\;\;\;\;\;\left[g,v\right]\mapsto v} & &
V\ar[dd]^{\sigma^{(k)}(g_1^{-1}g)} \\
 & &\\
\tau^{-1}(p)\ar[rr]^{\;\;\;\;\;\;\;\left[g_1,v\right]\mapsto v} & & V}$$
Consequently since the representation $\sigma^{(k)}(g^{-1}_1g)$ as in
(\ref{rel2}) is unitary, the above statement naturally follows \cite{Simms}.} \\

\noindent According to the above remark we can introduce the following definition\\

\definizione\label{Hilb2}{\em A triple $(X,\tau,Y)$ is called  {\bf Hilbert
bundle} if $X$ and $Y$ are topological spaces, $\tau$ 
is a continuous surjection of $X$ on $Y$ and $\tau^{-1}(p)$ (the fiber) has an 
Hilbert space structure for each $p\in Y$ (see chapter 7 in \cite{Mackey}).}\\

\noindent In the following a Hilbert bundle $(X,\tau,Y)$ will be also denoted
$$\tau : X \to Y\:.$$

\definizione{\em Let  $(X_1,\tau_1,Y_1)$ and $(X_2,\tau_2,Y_2)$ be two Hilbert bundles. A {\bf Hilbert-bundle isomorphism}
 is a pair 
of homeomorphisms $\lambda_1 : X_1 \to X_2$, $\lambda_2: Y_1 \to Y_2$, such that 
\begin{itemize}
\item $\tau_2\lambda_1=\lambda_2\tau_1$,
\item $\lambda_1$ isometrically maps the fiber $\tau^{-1}_1(p)$ into $\tau^{-1}_2(\lambda_2 p)$ for each $p\in Y_1$.\\
\end{itemize}}

\definizione\label{GHilb2}{\em Let $G$ be a topological group and $(X,\tau,Y)$ an Hilbert bundle. 
Then $(X,\tau,Y)$ is called a {\bf ${\bf G}$-Hilbert bundle} if there are two continuous actions of 
${G}$ onto $X,Y$ such that the pair $\lambda_{{1,g}}:X\to X$, with $x\mapsto {g}x$ and 
$\lambda_{{2,g}}:Y\to Y$, with $y\mapsto {g}y$, is an Hilbert bundle automorphism for each 
${g}\in {G}$. Accordingly an isomorphism between two different ${G}$-Hilbert bundles 
is an isomorphism between the two Hilbert bundles which commutes with the ${G}$-action (see chapter 9 in \cite{Mackey}).}\\ 

\proposizione{\em According to definitions \ref{Hilb2} and \ref{GHilb2}, take a representation (\ref{irrep12}) $\chi\sigma$
associated with a character $\chi \in N'$ and an irreducible  representation $\sigma$ of $L_\chi$
on the \underline{finite dimensional} Hilbert space $\cH$.\\
A  $\tgbms$-Hilbert bundle can be built up as follows,
\beq\label{Hilbertbundle2}
\tau^\sigma_\chi :\tgbms \times_{\spa H_\chi}\mathcal{H}\longrightarrow
\frac{\tgbms}{H_\chi},
\eeq
where:\\
{\bf (a)} $\tgbms\times_{\spa H_\chi}\mathcal{H}$ consists of the 
equivalence classes $[g,\vpsi]$ associated with the equivalence relation $\sim_{H_\chi}$ in $\left(\tgbms\times\mathcal{H}\right)
\times \left(\tgbms\times\mathcal{H}\right)$
$$ (g',\vpsi') \sim_{H_\chi} (g,\vpsi)\:, \:\:\:\: \mbox{if and only if $(g',\vpsi') =
(gk^{-1},\chi\sigma(k)\vpsi)$ for some $k\in H_\chi$}$$
{\bf (b)}
the group actions,
respectively on $\tgbms\times_{\spa H_\chi}\mathcal{H}$  and $\frac{\tgbms}{H_\chi}$, 
are defined as
$$g^\prime[g,\vpsi]=[g^\prime\odot g, \vpsi],\;\;\;g^\prime(g\odot H_\chi)=(g\odot g^\prime)\odot H_\chi.$$
Eventually if considering two $\tgbms$ representations
$\chi\sigma$ on the finite dimensional Hilbert space $\mathcal{H}$ and $\chi\tau$ on the finite dimensional Hilbert space $\mathcal{H}^\prime$, 
which are unitary equivalent by $U:\mathcal{H}\to\mathcal{H}^\prime$, 
then the Hilbert bundles $\tau^\sigma_\chi :\tgbms \times_{\spa H_\chi}\mathcal{H}\to
\frac{\tgbms}{H_\chi}$ and $\tau^\eta_\chi :\tgbms \times_{\spa H_\chi}\mathcal{H}^\prime\to
\frac{\tgbms}{H_\chi}$ are $\tgbms$-isomorphic under the map $[g,\vpsi]\mapsto [g,U\vpsi]$.}\\

In order to fully control the theory of $\tgbms$ unitary representations, we
also need some measure theoretical notions which will allow us to impose
integrability conditions on the set of $\tgbms$ wave functions.\\
Consider a generic topological space $X$.
 Two  Borel measures $\mu,\nu$ on $X$ are said to be lying in the same {\bf
measure class} if they assume the value zero for the same Borel sets in $X$ so that 
$\mu$ admits Radon-Nikodym derivative with respect to $\nu$ and {\em viceversa}.\\

\noindent In particular, when we deal with locally compact groups such as
$SL(2,\mathbb{C})$, the following theorem holds (see \cite{bourbaki} for the
demonstration and also section 4 in \cite{Mackey}) and it is of a great
importance for our later applications.\\

\teorema\label{measureB2}{\em For any closed subgroup $K$ of a locally compact group $G$, there 
is a unique non vanishing measure class $M$ on $\frac{G}{K}$ such that if $\mu\in M$,
$\mu_g\in M$ for every $g\in G$, where $\mu_g(E)=\mu(g^{-1}E)$ for every Borel set $E\subset \frac{G}{K}$.
$M$ is called {\bf invariant
measure class} of $\frac{G}{K}$.}\\

\noindent 
Furthermore,
according to \cite{Simms}, consider the  Borel-measurable sections of a $\tgbms$-Hilbert bundle (\ref{Hilbertbundle2}), i.e.
 Borel measurable functions  $\vpsi:\frac{\tgbms}{H_\chi}\to \tgbms \times_{\spa H_\chi}\mathcal{H}$ such that 
$\tau^\sigma_\chi \circ\vpsi=id_{\frac{\tgbms}{H_\chi}}$.
Since the orbit 
$\mathcal{O}_\chi=\frac{SL(2,\mathbb{C})\ltimes C^\infty(\bS^2)}{L_\chi\ltimes C^\infty(\bS^2)}$ is isomorphic to 
$\frac{SL(2,\mathbb{C})}{L_\chi}$, we can exploit theorem \ref{measureB2}
introducing for any orbit $\mathcal{O}_\chi$ and for a 
$\mu \in M$ the following Hilbert space:
\beq\label{config2}
\mathcal{H}_\mu=\left\{\vpsi:\mathcal{O}_\chi \to \cH\;\left|\;\int\limits_{\mathcal{O}_\chi}d\mu(p)\langle \vpsi(p),\vpsi(p)\rangle <\infty
\right.\right\}\:.
\eeq
Above, $\langle \: ,\: \rangle$ refers to the $\tgbms$-invariant  Hermitean inner product of the fiber
$\left(\tau_\chi^{\sigma}\right)^{-1}(p)$ where $p$ is an element on the orbit $\mathcal{O}_\chi$.\\
 Each 
element\footnote{We adopt the symbol $\psi$ either for the intrinsic $\tgbms$ field either for the bulk field suitably restricted on
 $\Im^+$ since they will ultimately be the same object, at least for a scalar $\tgbms$ representation.} $\vpsi$ 
in $\mathcal{H}_\mu$, usually called an {\bf ``induced wave
function''}\footnote{For an interested reader, we underline that we adopt the
most common name for the wave functions constructed from induced
representations. Nonetheless, in the literature, it exists a zoology of different
names the most notables being {\em canonical wave function} (as in \cite{Asorey})
or {\em Mackey wave function} (as in \cite{Group}).} (or BMS intrinsic free field), inherits a natural $\tgbms$ action as:
\beq\label{induced2}
(g\vpsi)(p)=\sqrt{\frac{d\mu(gp)}{d\mu(p)}}g(\vpsi(g^{-1}(p))),\;\;\;\forall g\in \tgbms
\eeq
where $\sqrt{\frac{d\mu(gp)}{d\mu(p)}}$ is the Radon-Nikodym derivative. It is worth stressing that, by construction, the scalar product 
in $\mathcal{H}_\mu$ is invariant under the above action of $\tgbms$.\\
 
 Let us fix a little group $H_\chi$ and 
 consider the set of all possible $\tgbms$-Hilbert bundles
(\ref{Hilbertbundle2})  $\zeta^\sigma=(
\frac{\tgbms}{H_\chi},\tau^\sigma_\chi, \tgbms \times_{\spa H_\chi}
\mathcal{H})$. We are entitled to directly apply Mackey's theorem (see chapter 16 of \cite{Group} and \cite{Piard, Piard2}) 
which grants us that:\\

\proposizione{\em (\ref{induced2}) individuates  a unitary strongly continuous irreducible  $\tgbms$ representation 
$T_\mu(\zeta^\sigma)$ called  {\bf induced representation} associated with the irreducible representation $\sigma$.} \\

\remark\label{measure2}{For a fixed little group $H_\chi$, if we consider two invariant measures
 $\mu,\nu\in M$, then the map which associates to each $\vpsi\in \mathcal{H}_\mu$ the element
$\sqrt{\frac{d\mu}{d\nu}}\vpsi\in\mathcal{H}_\nu$ defines an isometry between 
$\mathcal{H}_\mu$
and $\mathcal{H}_\nu$ and, at the same time, an equivalence between
$T_\mu(\zeta^\sigma)$ and $T_\nu(\zeta^\sigma)$.
Since, according to theorem \ref{measureB2}, we have chosen 
the unique invariant measure class $\mu$ of the base space $\frac{\tgbms}{H_\chi}$
on each  $\tgbms$-Hilbert bundle, we are entitled to drop the $\mu$-dependence in the induced representation
$T_\mu(\zeta^\sigma)\equiv T(\zeta^\sigma)$.} \\

Apparently the last discussion
grants us that $T(\zeta^\sigma)$ depends only upon a selected
representation of the little group $H_\chi$, but it is rather intuitive that the
existence of Hilbert bundle isomorphisms could imply that, a priori different
representations of $H_\chi$ on different $\tgbms$-Hilbert bundles, could
actually induce equivalent full $\tgbms$ representations. 
In detail, the last assertion can be justified if we notice that
(\ref{Hilbertbundle2}) depends only on the orbit $\tgbms\chi$ 
and not on the specific choice of $\chi$. Let us thus choose two different 
bundles, namely
$$
\tau^\sigma_\chi :\tgbms \times_{\spa H_\chi}\mathcal{H}\longrightarrow
\frac{\tgbms}{H_\chi},\;\;\;
\tau^{\sigma_1}_{\chi_1} :\tgbms \times_{\spa H_{\chi_1}}\mathcal{H}\longrightarrow
\frac{\tgbms}{H_{\chi_1}}
$$
such that $\tgbms\chi=\tgbms\chi_1$ for $\chi_1\neq\chi$. As a consequence, an element $g_1\in\tgbms$ exists such 
that $\chi_1=g_1\chi$ and, according to definition \ref{littlegr2}, $L_{\chi_1}=g_1 L_\chi g_1^{-1}$. 
This identity translates at a level of representation as $\sigma_1(h)=\sigma(g_1^{-1}hg_1)$ 
for each $h\in H_\chi$. Furthermore, according to definition \ref{GHilb2}, there is an isomorphism
$$ (\lambda_1, \lambda_2):  \left(\tgbms \times_{\spa H_\chi}\mathcal{H}, \tau^\sigma_\chi,
\frac{\tgbms}{H_\chi}\right) \longrightarrow\left(\tgbms 
\times_{\spa H_{\chi_1}}\mathcal{H}, \tau^\sigma_{\chi_1}, 
\frac{\tgbms}{H_{\chi_1}}\right),$$
induced by the maps $$\lambda_1: [g,\vpsi]\mapsto[g_1\odot g\odot g_1^{-1}, \vpsi] \:\:\:\mbox{and}\quad 
\lambda_2: \tgbms\chi \ni  p\mapsto g_1 p \in \tgbms\chi_1\:,$$
where $p$ stands for a generic point on the orbit.
Thus, the irreducible representations, induced either from $\sigma$ i.e.
$T(\zeta^\sigma)$ either from $\sigma_1$ i.e. $T(\zeta^{\sigma_1})$, are
 $\tgbms$-equivalent  by
construction and, consequently, they will be  considered as the 
same. A summary of this discussion lies in the following remark:\\

\remark\label{Mack2}{ The $\sigma$-dependence of
$T(\zeta^\sigma)$ is determined up to $\tgbms$-equivalence.}\\

\remark{According to the previous discussion and, in particular, according to
remarks \ref{measure2} and \ref{Mack2}, a generic $\tgbms$ (unitary) 
representation depends only upon the choice of the character $\chi$ and of the
unitary representation $\sigma$ of $H_\chi$. Consequently it will be indicated as
$T(\zeta^\sigma_\chi)$ making explicit the dependence on $\chi$.}\\

\noindent The explicit action of a generic $T(\zeta^\sigma_\chi)$ should be defined
on the induced wave function as in (\ref{induced2}). However it is more convenient
to recast  (\ref{induced2}) as\footnote{In the
literature such as \cite{Mc1, Mc2}, the argument $(\zeta^\sigma_\chi)$ is 
considered a priori fixed and thus it is not even introduced.}:
\beq\label{conditiona2}
\vpsi(gh)= T(h^{-1})\vpsi(g),\;\;\forall g\in SL(2,\mathbb{C}),\ h\in L_\chi
\eeq
where we write $ T(h^{-1})$ instead of $[T(\zeta^\sigma_\chi)](h^{-1})\vpsi(g)$
 to stress, that for a fixed $\tgbms$-Hilbert bundle and for a
fixed representation $\sigma$ of $L_\chi$, the dependence of the induced 
representation $T$ on such data is superfluous.
The $\tgbms$ action explicitly reads, for $(\Lambda, \alpha) \in \tgbms$,
\begin{gather}
(\Lambda\vpsi)(g)=\vpsi(\Lambda^{-1}g), \label{conditionb}\\
(\alpha\vpsi)(g)=\chi\left(g^{-1}\alpha\right)\vpsi(g),
\end{gather}
which is a unitary representation induced from $T(\zeta^\sigma_\chi)$ as in (\ref{conditiona2}) 
and thus, according to Mackey
theorem, it is also irreducible. From an operative point of view, an equivalent 
definition of  an induced wave function can be
constructed dropping the condition (\ref{conditionb}). In this scenario 
we introduce the set of $\mu$ square-integrable maps of $\cH_\mu$ (see theorem
\ref{measureB2})
\beq\label{maps2}
\vpsi:\mathcal{O}_\chi= \frac{SL(2,\mathbb{C})}{L_\chi}\to\mathcal{H}\:.
\eeq 
However the absence of (\ref{conditionb}) requires the
introduction of an additional datum, namely an almost everywhere continuous
section $\omega$ of the bundle
$\tau:SL(2,\mathbb{C})\to\mathcal{O}_\chi$ 
which satisfies
$\omega(p)\chi=p,\;\;\forall p\in\mathcal{O}_\chi$. Thus we can define, as an
{\bf induced wave function}, a map (\ref{maps2}) which transforms under $(\Lambda,\alpha) \in \tgbms$ as
\begin{gather}\label{inducedLor2}
(\Lambda\vpsi_\omega)(p)=\sqrt{\frac{d\mu(\Lambda p)}{d\mu(p)}}[T(\zeta^\sigma_\chi)]
\left(\omega(p)^{-1}\Lambda\omega(\Lambda^{-1}p)
\right)\vpsi_\omega(\Lambda^{-1}(p)),\;\;\;\Lambda\in SL(2,\mathbb{C})\:,\:\: p\in \mathcal{O}_\chi
\\
(\alpha\vpsi_\omega)(p)=p(\alpha)\vpsi_\omega(p),\;\;\;\alpha\in
C^\infty(\bS^2)\label{inducedsuper2}\:.
\end{gather}  
Above, $p(\alpha)$  denotes the action of the character $p\in SL(2,\bC)\chi$ on $\alpha$
 and 
the subscript $\omega$
reflects the strict dependence of the induced wave function upon the choice of 
the section itself.\\

\ssa{The scalar induced wave function}
The long explicit construction all the $\tgbms$ irreducible unitary representations
 has been completed and extensively discussed in the Hilbert topology in \cite{Mc1,Mc2} and in the 
 nuclear topology in \cite{Mc4}, thus it will not be reviewed here. 
It is anyway interesting for our purposes to stress some of the non trivial points in McCarthy analysis; 
 in particular, whereas in the Hilbert topology all the unitary representation for the BMS group can be 
 constructed as induced representations from compact little group, in the nuclear topology the scenario is 
 far more complicated and it can be summarized in the following proposition \cite{Mc4}:\\

\proposizione {\em %The following facts hold for representations of $\tgbms$.\\
%{\bf (a)}%
 If a unitary representation of $\tgbms$ is irreducible then it must arise either from a transitive $SL(2,\bC)$ action on 
$N^*$ or from a cylinder measure with respect to which the $SL(2,\bC)$ action is strictly ergodic.}\\
%{\bf (b)} All the unitary induced $\tgbms$ representations are irreducible.}\\

\noindent That proposition is the reason 
why the current classification of unitary irreducible representations of $\tgbms$ group is not complete. As a
 matter of fact the construction of representations arising from strictly
 ergodic measure is rather challenging 
 and, up to now, it has not been solved nor addressed in detail.\\
Nonetheless, for our purposes we are mainly interested in induced representations. These 
have  been fully considered in \cite{Mc4} where, starting from the analysis in \cite{Shaw}, a plethora of $\tgbms$ possible
 little groups has been identified. These can be classified in two different families, the connected subgroups of 
$SL(2,\mathbb{C})$ and the non connected compact subgroups of $SU(2)$. We shall
now concentrate on\footnote{These compact little groups are also present in the 
Hilbert topology scenario.} $SU(2)$, $SU(1,1)$, $\Gamma$ (the universal 
covering of $SO(2)$ made of  all the matrices
$diag(e^{\frac{it}{2}},e^{-\frac{it}{2}})$ with $t\in \bR$) and on
$\Delta=\Gamma\ltimes T^2$ (the 
double covering of the two dimensional Euclidean group). The analysis 
for the $SU(2)$ scenario has been already developed in \cite{Arcioni, Dappiaggi}
in the Hilbert topology, where the wave functions of the intrinsic BMS free 
fields, their kinematical and dynamical configurations have been throughout
discussed. On the opposite, we shall now focus  attention on the $\Delta$ case
-- proper only of the nuclear topology -- which will turn out to be in direct 
correspondence with scalar fields on $\Im^+$ induced from the bulk. 

\vskip .3cm

\noindent {\bf{$\bf\Delta$} orbit classification}.  This little group
is the set of matrices
$$\Lambda_{t,\upsilon} =\left[\begin{array}{cc}
e^{\frac{it}{2}}, & \upsilon\\
0 & e^{-\frac{it}{2}} 
\end{array}\right],$$ 
with $t\in \bR$ and $\upsilon\in\bC$.
Thus, according to (\ref{chara12}), a fixed point $\bar{\beta}$ (which thus 
admits $\Delta$ as little group) satisfies:
$$(\Delta \bar{\beta})=\bar{\beta}.$$
In order to solve this distributional equation  the rationale is to 
switch from $\bar{\beta}\in N^*$ to the associated pair $(\bar{\phi},
\hat{\bar{\phi}}) \in D_{-2}$ 
as in remark \ref{duale} and to use (\ref{phi}) and (\ref{hatphi}), i.e.
$$\Delta\bar{\phi}=\bar{\phi},\;\;\;\Delta\hat{\bar{\phi}}=\hat{\bar{\phi}}\:.$$
As discussed in \cite{Mc4}, the general solution to these equations is:
\begin{gather}\label{orbitdelta2}
\bar{\phi}=S,\\
\hat{\bar{\phi}}=S|\z|^{-6}+A\delta^{2,2}+C\delta,\label{orbitdelta}
\end{gather}
where $S,A,C\in\bR$ are constants and $\delta^{p,q}$ is the p-th derivative on 
the variable 
$\z$ and q-th derivative on the variable $\bz$ of $\delta=\delta(\z)\delta(\bz)$.\\

\proposizione\label{deltamassa}{\em The mass (\ref{mass2}) associated to any 
orbit of the $\Delta_\chi$ little  group is $0$.\\}

\noindent {\em Proof}. The demonstration  follows from proposition
\ref{massa} and (\ref{projt}) directly according to which 
$$\widehat{\pi(\bar{\beta})}(\z^\prime,\bz^\prime)= \frac{C}{1+|\z^\prime|^2}\not \equiv
0;$$  
thus (\ref{4mom2}) grants us that
$\widehat{\pi(\bar{\beta})}_\mu=C(1,0,0,1)$; consequently we conclude from (\ref{mass2}) 
that $m^2=\eta^{\mu\nu}\widehat{\pi(\bar{\beta})}_\mu\widehat{\pi(\bar{\beta})}_\nu=0$. $\Box$\\

\noindent Only a three 
dimensional orbit $\frac{SL(2,\bC)}{\Delta}=\bR\times \bS^2$ with a 
vanishing mass can be associated  to the $\Delta_\chi$ little group. Furthermore, 
from (\ref{orbitdelta}) we can infer that, besides the constant $C$ which plays
the role of the energy, the orbit is fully determined only if we fix the
values $A,S$ which from now on are set to $0$.

\vskip .3cm

%%%%%%%%%%%%%%%%%%%%%%%%%%%%%%%%%%%%%%%%%%%%%%%%%%%%%%%%%%%%%%%%%%%%%%%%%%%%%

\noindent We can now explicitly construct the representation and we can choose 
an $SL(2,\bC)$-invariant measure on the orbit
$\frac{SL(2,\mathbb{C})}{\Delta_\chi}$ which represent the key data  
to construct the induced wave function as in proposition 3.3 and 3.4. 
Leaving the detailed analysis for the other non connected little groups to 
\cite{Arcioni, Mc2}, we concentrate on:

\vskip .3cm

\noindent{\bf ${\bf\Delta}$ induced wave functions}. Unitary and irreducible
representations of $\Delta$ are of two types \cite{Mc4, Group}. A representation $D^{\lambda,p,q}$ of the  first
type is individuated  by a triple $p,q\in\bR\setminus \left\{0\right\}\;\;\lambda\in\frac{\bZ}{2}$ and it is defined by:
$$D^{\lambda,p,q}\left(\left[\begin{array}{cc}
e^{\frac{it}{2}}, & \upsilon\\
0 & e^{-\frac{it}{2}} 
\end{array}\right]\right)=e^{i\lambda
t}e^{i(pb+qc)},\;\;\upsilon=b+ic\:.$$ 
This acts on an infinite dimensional complex target Hilbert space
by multiplication and it induces an infinite dimensional $\tgbms$ representation.
 A representation $D^{s}$ of the second type 
is individuated  by a number $s\in \bZ/2$ and it is defined by:
\beq\label{Deltarep}D^{s}\left[\begin{array}{cc}
e^{\frac{it}{2}}, & \upsilon\\
0 & e^{-\frac{it}{2}} 
\end{array}\right]=e^{is t},
\eeq
This acts on a one-dimensional complex  target Hilbert space
by multiplication. \\

\remark{Although the above representations are well known even for a Poincar\'e
invariant theory, the second being faithful the first being unfaithful, in a
$\tgbms$ scenario they are both faithful. More generally, it has been shown in 
\cite{Mc4} that an induced representation built upon an orbit $\mathcal{O}_\chi$
is faithful iff $\widehat{\pi(\bar{\phi})}\neq 0$, $\bar{\phi}$ being  the supermomentum
associated to $\mathcal{O}_\chi$ solving (\ref{chara12}).}\\

It is possible to reinforce the result presented in theorem 3.1: 
we can exploit either theorem 1 and corollary 1 in chapter 4, section 3 of 
\cite{Group} either the unimodularity\footnote{A locally compact group $G$ is
called {\em unimodular} if its right-invariant and left-invariant Haar measure
coincide ({\em c.f.} page 69 in \cite{Group})} of $SL(2,\bC)$ and $\Delta$ to
claim that the $SL(2,\bC)$-invariant measure class $M$
contains a measure $\mu$ which is $SL(2,\bC)$-invariant. Referring to this specific measure
$\mu$ we can construct the Hilbert space of induced wave functions $\psi:\mathcal{O}_\chi\to\mathbb{C}$

\beq
\mathcal{H}_\mu=\left\{\psi:\mathcal{O}_\chi\to\mathbb{C}\;\left|
\;\int\limits_{\mathcal{O}_\chi}d\mu(p)\bar{\psi}(p)\psi(p)
<+\infty\right.\right\}.
\eeq
Now we can use remark \ref{irreps2} and the formula (\ref{inducedLor2}) and
(\ref{inducedsuper2}) to construct the explicit expression of the induced wave
function (\ref{induced2}).\\

\remark{ The $\Delta$ little group is a rather special case since no global continuous section $\omega$ of the bundle 
$\tau :SL(2,\mathbb{C})\to\frac{SL(2,\mathbb{C})}{\Delta}$
exists such for the $SU(2)$ or the $\Gamma$ little group. There are different
choices commonly used and, they being far from the aim of this paper, we refer to
\cite{Boya} for a complete discussion.}\\

\noindent According to (\ref{inducedLor2}) and (\ref{inducedsuper2}), an induced wave
function transforms, for any $g=(\Lambda,\alpha)\in \tgbms$ and 
under the $\Delta$ representation (\ref{Deltarep}), as 
\begin{gather}(g\psi)(p)=\sqrt{\frac{d\mu(\Lambda p)}{d\mu(p)}}
p(\alpha)D^s(\omega(p)^{-1}\Lambda\omega(\Lambda^{-1}p))\psi_\omega(\Lambda^{-1}p)=\notag\\
\label{indSO2} p(\alpha)e^{ist}\psi_\omega(\Lambda^{-1}p),
\end{gather}
where, with the above-said choice of $\mu$,  
the Radon-Nikodym measure is $1$ and $e^{ist}$ is the action of the 
one dimensional $\Delta$ representation associated to 
$D^s(\omega(p)^{-1}\Lambda\omega(\Lambda^{-1}p))$.
Eventually we may write the \emph{induced scalar $\Delta$ $\tgbms$ wave
function} 
(i.e. $s=0$ in (\ref{indSO2})) as
\begin{gather}
\psi:\mathcal{O}_\chi\longrightarrow\mathbb{C},\\
(g\psi_\omega)(\Lambda p)=p(\alpha)\psi_\omega(p)\:,\label{inducedfin2}\;\;\;\forall\; g\in\tgbms
\end{gather}
%%%%%%%%%%%%%%%%%%%%%%%%%%%%%%%%%COVARIANT WAVE%%%%%%%%%%%%%%%%%%%%%%%%%%%%%%%%%%%%%%%%%%%

\ssa{The covariant scalar wave function and its bulk interpretation}
Although the induced wave function transforms under a unitary 
irreducible $\tgbms$ representation, thus containing all relevant physical information, from a physical perspective,
it is rather common  to start from a different wave function. That is the {\bf  covariant wave
function(al)} (or {\bf covariant free field}) which, in a BMS setting, is 
\cite{Arcioni, Goldin}:
\beq\label{covariant22}
\vPsi : N^*  \longrightarrow \mathcal{H}^\prime
\eeq
where $\mathcal{H}^\prime$ is a suitable {\em finite dimensional} target Hilbert 
space either real or complex and $N^*$ is the space of distributions over
$\bS^2$.
Under the action of 
$g=(\Lambda,\alpha) \in \tgbms$, $\vPsi$ in (\ref{covariant22})  transforms as
\beq\label{covariant222}
\left[U^\lambda(g)\vPsi\right][\beta]=\chi_{\beta}(\Lambda\alpha)
D^\lambda(\Lambda)\vPsi[\Lambda^{-1}\beta],
\eeq
where $D^\lambda(\Lambda)$ is a unitary, {\em  but not necessary irreducible},
representation of $SL(2,\mathbb{C})$ labeled by the superscript $\lambda$.
$\chi_\beta$ is the character associated with $\beta$ as in definition \ref{charadef2}  and it acts according to remark \ref{directintegral}.

\vskip .3cm

\remark\label{remarkcovind2} At  first glance (\ref{covariant22}) and 
(\ref{induced2}) are {\em a priori} unrelated, the main striking difference consisting 
in the existence of a different induced wave function for each isotropy subgroup 
$H_\chi$, whereas the covariant wave equation is unique up to the choice of a 
unitary $SL(2,\mathbb{C})$ representation. Nonetheless it is possible to show 
that both kinds of wave functions are ultimately equivalent
if suitable constraints are imposed on the covariant wave function\footnote{We shall also
refer to the construction of the covariant wave equation and of the associated equations of motion 
as the {\bf Wigner's program} in relation to Wigner seminal 
paper \cite{Wigner} where he dealt with the Poincar\'e case.} \cite{Group,
Asorey}. In the BMS scenario \cite{Arcioni}, upon selecting a specific covariant
wave function and a representation $U^\lambda$ as in (\ref{covariant222}), the
restriction to the induced wave function associated to a fixed little group $H_\chi$
operatively corresponds to:
\begin{enumerate}
\item restrict the support of (\ref{covariant222}) from $N^*$ 
to the orbit $\frac{SL(2,\mathbb{C})}{L_\chi}\hookrightarrow N^*$
($\hookrightarrow$ denoting an embedding); 
\item act on $\vPsi$ by the linear transformation 
$U^\lambda\left[\omega^{-1}\left(\beta\right)\right]$ where $\beta$ is a point on the orbit
and where $\omega$ is the section chosen in (\ref{inducedLor2});
\item select in (\ref{covariant222}) the irreducible unitary representation  
$\sigma$ of  (\ref{induced2}) contained in $D^\lambda$.
\end{enumerate}
We discuss now in details the last step in the above construction which is rather
counterintuitive.\\
Let us start either from a generic unitary, but fixed, 
representation $D^\lambda$ of
$SL(2,\mathbb{C})$ either from a generic, but fixed, irreducible unitary representation
$\sigma^j$ of a fixed little group $L_\chi$. \\
Let us now consider the restriction
of $D^\lambda$ to $L_\chi$ which decomposes as the finite sum 
$D^\lambda\spa\rest_{L_\chi}=\bigoplus\limits_{j^\prime}C_{\lambda,j^\prime}\sigma^{j^\prime},$
where $\sigma^{j^\prime}$ is a unitary irreducible representation of $L_\chi$ and
$C_{\lambda,j^\prime}$ are suitable integers standing for the multiplicity
of the $\sigma^{j^\prime}$ in $D^\lambda$. According to theorem
16.2.1 in \cite{Group}, the above decomposition translates either to the
full $\tgbms$ representation either to the target space of (\ref{covariant222})
i.e.
\beq\label{split2}
\mathcal{H}^\prime=\bigoplus\limits_{j^\prime}C_{\lambda,j^\prime}\mathcal{H}^{j^\prime}.
\eeq
Let us now recognize a fixed $\mathcal{H}^{j^\prime}$ as the target space of 
an induced
wave function (\ref{induced2}) which transforms under the action of the unitary 
and irreducible representation $\sigma^{j^\prime}$. The selection of a fixed
representation $\sigma^{j^\prime}$ in $D^\lambda$ is now equivalent to 
constrain $\mathcal{H}^\prime$ -- the target space of (\ref{covariant22}) -- to
$\mathcal{H}^{j^\prime}$ .
 This result can be operatively achieved imposing the
subsidiary condition on the covariant wave function (with support on 
$\mathcal{O}_\chi$)
\beq\label{ortop12}
\rho\vPsi[\bar{\beta}]=\vPsi[\bar{\beta}],
\eeq 
where $\rho$ is the projector selecting $\mathcal{H}^j\subset\mathcal{H}^\prime$
and where $\bar{\beta}$ is the supermomentum associated to the fixed point on 
$\mathcal{O}_\chi$.\\
If we now remember that the following identity holds:
\beq\label{ortop22}
\vPsi[\beta]=[D^\lambda(\Lambda)\vPsi][\bar{\beta}],
\eeq
where $\beta=\Lambda^{-1}\bar{\beta}\in\frac{SL(2,\mathbb{C})}{L_\chi}$
and where $\Lambda\in SL(2,\mathbb{C})$, (\ref{ortop12}) becomes
$$D^\lambda(\Lambda)\rho
[D^\lambda(\Lambda)]^{-1}\vPsi[\beta]=\vPsi[\beta].$$
If we set\footnote{The $\beta$ dependence of $\rho$ should not be interpreted
literally: it means that $\Lambda$ in (\ref{ortop22}) is the unique value such
that $\Lambda^{-1}\bar{\beta}=\beta$.} 
$\rho[\beta]=D^\lambda(\Lambda)\rho [D^\lambda(\Lambda)]^{-1}$, (\ref{ortop22})
becomes the so-called \emph{projection equation}
\beq\label{ortotot2}
\rho[\beta]\vPsi[\beta]=\vPsi[\beta]\:.
\eeq

\remark{According to the discussion of section 1B in chapter 21 of \cite{Group},
which easily generalize to the $\tgbms$ scenario, (\ref{ortotot2}) is also a 
covariant matrix operator and, since  induced wave functions  are in
one-to-one correspondence with  pairs $\left\{D^\lambda(\Lambda),\rho\right\}$, it is customary to claim that (\ref{ortotot2}) represents the most
general covariant wave equation\footnote{Chapter 21 of
\cite{Group} contains the specific discussion for the Poincar\'e
scenario where it is shown that (\ref{ortotot2}) is simply a compact expression
for the usual Dirac, Proca equations, etc... whereas, for the $\tgbms$ counterpart in the Hilbert topology, we 
refer to \cite{Arcioni}.}.}

\vskip .3cm

\noindent Since our final goal is to show the equivalence between (\ref{Ureducted2}) and the $\tgbms$ $\Delta$ scalar 
induced wave function (\ref{inducedfin2}), let us restrict our attention to this 
specific case.\\ 

\definizione {\em A covariant scalar wave function is a map
\beq\label{covprima2}
\Psi :N^*\longrightarrow\mathbb{C},
\eeq
which transforms as 
\beq\label{covscalar2}
\left[U^\lambda(g){\Psi}\right](\beta)=\chi_{\beta}(\Lambda\alpha){\Psi}[\Lambda^{-1}\beta],
\eeq
under a scalar  $SL(2,\mathbb{C})$ unitary representation $U_g$, with
$g=(\Lambda,\alpha)\in\tgbms$. In (\ref{covscalar2})
$\Lambda^{-1}\beta$ is defined as in (\ref{charaact222}).}\\

\proposizione {\em Referring to the definition above, the constraint to impose 
to reduce (\ref{covscalar2}) to  (\ref{inducedfin2}) is
\begin{equation}\label{cons12}
\left[\beta-\frac{SL(2,\bC)}{\Delta}\bar{\beta}\right]{\Psi}[\beta]=0,\;\;
\widehat{\pi(\bar{\beta})}\neq 0
\end{equation}
where $\beta \in N^*$  and $\bar{\beta}$ is the fixed point of the
$\Delta$ orbit constructed out of (\ref{orbitdelta2}) and (\ref{orbitdelta}).}\\

\noindent The proof of this proposition is a straightforward consequence of the 
analysis in \cite{Arcioni} and of the coincidence between the scalar covariant 
$SL(2,\mathbb{C})$ representation and the scalar 
representation induced from the $\Delta$ little group (i.e. 
(\ref{ortotot2}) is identically satisfied).\\
Furthermore the mass equation which
usually appears in the Hilbert topology \cite{Arcioni}, i.e.
$$\left[\eta^{\mu\nu}\widehat{\pi(\beta)}_\mu\widehat{\pi(\beta)}_\nu\right]{\Psi}[\beta]=0,$$
is automatically satisfied by (\ref{cons12}) since the little group $\Delta$ is
associated only to a vanishing mass whenever $\widehat{\pi(\bar{\beta})}\neq 0$.\\

\noindent To conclude, we want now to establish the main result of this section, namely  that a covariant massless scalar field  which satisfies
(\ref{cons12}) is identical to (\ref{Ureducted2}). Let us remember that $N^*\sim D_{(-2,-2)}
\sim D_{-2}$ as well as $N\sim D_{(2,2)}\sim D_2$. Thus an element 
$\beta\in N^*$ is bijectively related with the pair $\hat{\phi},\phi\in D_{-2}$ 
introduced in proposition \ref{coppia}. 
Furthermore let us recall that the fixed point of
the $\Delta$ orbit is $\bar{\hat{\phi}}=S|z|^{-6}+K\delta^{2,2}+C\delta$ with 
$C\neq 0$; 
if we select the specific values $S=K=0$, then $\bar{\hat{\phi}}=C\delta$ and,
according to proposition \ref{coppia} and to remark \ref{duale}, the associated supermomentum is
$$\bar{\beta}=\frac{\bar{\hat{\phi}}}{\left(1+\mid\z\mid^2\right)^3},$$
We need now the following Lemma:\\

\lemma{\em The supermomentum $\bar{\beta}$ lies in $(T^4)^*$.}\\

\noindent{\em Proof}: consider  the isomorphism discussed about (\ref{pigiusto})
 first introduced in
\cite{Mc4}
\beq\label{iso}
\frac{N^*}{(T^4)^0}\sim (T^4)^*, 
\eeq
where both sides are preserved under $SL(2,\bC)$ transformation and the isomorphism
commute with the action of that group.

It is straightforward that $\bar{\beta}\notin (T^4)^0$ since if we
consider any supertranslation $f\in T^4\hookrightarrow C^\infty(\bS^2)$ 
such that $f(0)\neq 0$, then $(f,\bar{\beta})=Cf(0)\neq 0$. 
As a consequence we are free to choose $\bar{\beta}$ as the representative
of a conjugacy class in $\frac{N^*}{(T^4)^0}$ and, according to
(\ref{iso}),  $\bar{\beta}$ also lies in $(T^4)^*$. $\Box$\\

\noindent Furthermore, since the orbit
$\mathcal{O}_{\bar{\beta}}$ is generated as 
$\frac{SL(2,\bC)}{\Delta}\bar{\beta}$, (\ref{iso}) also grants us that
$\mathcal{O}_{\bar{\beta}}\in(T^4)^*$ i.e., according to proposition (\ref{aggiuntaallafine}), 
$\pi(\beta)=\beta$ for any
$\beta\in\mathcal{O}_{\bar{\beta}}$. 
This last remark entitles to substitute in (\ref{covscalar2}) $\beta$ with 
$\pi(\beta)$.
\begin{gather*}
\Psi:
(T^4)^*\longrightarrow\bC,\\
[U(g)\Psi](\pi(\beta))=\chi_{\pi(\beta)}(\Lambda\alpha)\Psi\left[\Lambda^{-1}
\pi(\beta)\right],\;\;\;\forall\;g\in\tgbms
\end{gather*} 
which still satisfies the orbit constraint
$\left[\pi(\beta)-\frac{SL(2,\bC)}{\Delta}\pi(\bar{\beta})\right]\Psi[\pi(\beta)]=0.$\\
The next step consists in bearing in mind that $(T^4)^*\sim T^4$ 
\cite{Mc4}, i.e., according to  proposition \ref{aggiuntaallafine} and (\ref{projt}),
 any element $\pi(\beta)\in (T^4)^*$ is in one to one 
correspondence with the element $\widehat{\pi(\beta)}\in T^4$. \\
Furthermore, according to (\ref{4mom2}) and to proposition \ref{deltamassa}, we 
can identify each $\widehat{\pi(\beta)}\in\mathcal{O}_{\widehat{\pi(
\bar{\beta})}}$ with a four-vector $p_\mu$ which satisfies the mass relation 
$\eta^{\mu\nu}p_\mu p_\nu=0$. Thus we can write 
$p_\mu \equiv (E,E{\bf n}(\z^\prime,\bz^\prime))$ where ${\bf n}(\z^\prime,\bz^\prime)$ 
is a three dimensional spatial versor spanning a two-sphere of unit radius whose coordinates are 
$\z^\prime,\bz^\prime$.\\
At this stage the reader should bear in mind that we are ultimately dealing with a
Gelfand triplet i.e. $N\subset L^2(\bS^2)\subset N^*$; thus, according to 
these last remarks we are entitled to switch from the covariant wave
function living on $(T^4)^*\subset N^*$ to a second one living on 
$T^4\subset N$ which
reads\footnote{Alternatively it is possible to interpret the covariant wave
function on $T^4$ as the one on $(T^4)^*$ where the argument $\beta\in
(T^4)^*$ has been evaluated with a fixed test function as in (\ref{projt}).}:
\begin{gather*}
\Psi:
\mathcal{O}_{\widehat{\pi(\bar{\beta})}}\hookrightarrow T^4
\mapsto \bC,\\
(U(g)\Psi)[\widehat{\pi(\beta)}]=\chi(\Lambda\alpha)\Psi\left[\Lambda^{-1}\widehat{\pi(\beta)}\right]\:.\;\;\;\forall\;
g=
(\Lambda,\alpha)\in\tgbms
\end{gather*}
Since $\widehat{\pi(\beta)}$ now lies in $C^\infty(\bS^2)$ the
net effect of an $SL(2,\bC)$ action is according to (\ref{azionesuN}) and to
(\ref{charact4})
$$\left(\Lambda^{-1}\widehat{\pi(\beta)}\right)(\z^\prime,\bz^\prime)=
K_\Lambda(\Lambda^{-1}\z^\prime,\Lambda^{-1}\bz^\prime)\widehat{\pi(\beta)}(\Lambda^{-1}\z^\prime,
\Lambda^{-1}\bz^\prime).$$
In terms of 4-vectors, $\Lambda^{-1}\widehat{\pi(\beta)}$ corresponds to the 4-vector whose components are
$$p_0 = K_\Lambda(\Lambda^{-1}\z^\prime,\Lambda^{-1}\bz^\prime)E\:, \quad {\bf p} =
K_\Lambda(\Lambda^{-1}\z^\prime,\Lambda^{-1}\bz^\prime)E{\bf n}(\Lambda^{-1}\z^\prime,
\Lambda^{-1}\bz^\prime)\:.$$
The character can be directly evaluated as 
$\chi(\Lambda\alpha)=e^{iEK^{-1}_\Lambda(\z^\prime,\bz^\prime)\alpha(
\Lambda^{-1}\z^\prime,\Lambda^{-1}\bz^\prime)}$.
Substituting these results in the scalar covariant wave equation and taking into
account that each $\widehat{\left(\pi(\beta)\right)}(\z^\prime,\bz^\prime)$ is uniquely determined by
its associated four vector $p_\mu$ which, in turn, is determined by the coordinates 
$(E,\z,\bz)$, we can eventually recast (\ref{covscalar2})
in terms of a field $\varphi(E,\z,\bz) := \Psi[\widehat{\pi(\beta)}]$ as:
$$[U(g)\varphi](E,\z^\prime,
\bz^\prime)=\frac{e^{iEK_\Lambda(\Lambda^{-1}\z^\prime,\Lambda^{-1}\bz^\prime)\alpha(
\Lambda^{-1}\z^\prime,\Lambda^{-1}\bz^\prime)}}{\sqrt{K_{\Lambda}(\Lambda^{-1}(\z,\bz))}}\varphi(K^{-1}_\Lambda(\z^\prime,
\bz^\prime)E,\Lambda^{-1}\z^\prime,\Lambda^{-1}\bz^\prime)\:.$$
The square root is due to the fact that we passed
 to the measure $dE\otimes \epsilon_{\bS^2}$ from  the invariant one $d\bp/E(\bp)$.
We have found nothing but  the  unitary representation of $\gbms$ given in (\ref{Ureducted2}). Therefore this fact 
shows also that  the representation of $\tgbms$ obtained by (\ref{cons12}) is a unitary representation of $\gbms$ as well. 
\\

\noindent We have eventually proved that:\\ 

\teorema\label{finale}{\em   A field on $\scri$ satisfying (\ref{Ureducted2})
is identical to a $\tgbms$-covariant massless scalar field  which satisfies
(\ref{cons12}). Furthermore, the representation of $\tgbms$ obtained by (\ref{cons12}) is a unitary representation of $\gbms$ as well.}\\

As a last remark we wish to clarify why the above theorem holds only when a 
suitable nuclear topology is imposed on the set of supertranslations. If we choose
$N=L^2(\bS^2,\epsilon_{\bS^2})$ 
(where the field of the Hilbert space is $\bR$), 
it is still possible to construct a massless
scalar wave function induced from the $\Gamma$ little group living on an
orbit whose fixed point has a vanishing pure supertranslational component.
Nonetheless, in this framework, according to the Riesz-Fisher theorem, a 
character $\chi(\alpha)$ can be always associated with an element $\beta\in L^2
(\bS^2,\epsilon_{\bS^2})$ \cite{Mc1} such that
$$\chi(\alpha)=e^{i\int\limits_{\bS^2}\epsilon_{\bS^2}\alpha\beta}\;,\;\;\forall\alpha\in L^2(\bS^2,\epsilon_{\bS^2}).$$
This formula represents the key obstruction to obtain (\ref{Ureducted2}) in an
Hilbert topology framework since, whenever we consider a scalar covariant wave 
function $\Psi : N^* =L^2(\bS^2,\epsilon_{\bS^2})\longrightarrow \bC$ with a 
support restricted on $T^4\subset N^*$, we are requiring that
$\beta\in N^*$ can be written as
$\beta(\z,\bz)=\sum\limits_{k=0}^1\sum\limits_{l=1}^{2k+1}\beta_{lk}S_{lk}
(\z,\bz)$ where $S_{lk}$ are the real spherical harmonics.
Accordingly a character will always be written as:
$$\chi(\alpha)=e^{i\left(\sum\limits_{k=0}^1\sum\limits_{l=1}^{2k+1}
\beta_{lk}\alpha_{lk}\right)},$$
which cannot produce a phase as that in (\ref{Ureducted2})
$e^{iEK_\Lambda(\Lambda^{-1}\z^\prime,\Lambda^{-1}\bz^\prime)\alpha(
\Lambda^{-1}\z^\prime,\Lambda^{-1}\bz^\prime)}$ whenever $\alpha$ includes components in the 
space of pure supertranslations. Thus, it is this expression
which represents the symptom that a correspondence between (\ref{Ureducted2})
and an intrinsic BMS field could be achieved only if a distributional support
for the covariant wave function is considered.

\section{A few holographic issues.}
\ssa{General goals of the section} We want to start to investigate the issue of holographic correspondence between QFT
formulated in the {\em bulk} for fields $\phi$ satisfying Klein-Gordon equation (\ref{cc}) 
as in Proposition \ref{prop2},
and QFT formulated on the {\em boundary} $\scri$ as showed in the previous section.
In this sense the bulk is the globally hyperbolic subregion near the null infinity $\scri$
of an asymptotically flat spacetime contained into a globally hyperbolic nonphysical spacetime in the sense 
of the hypotheses of Proposition \ref{prop2} (in particular it could be strongly asymptotically predictable). 
We know from Proposition \ref{prop2} that,
at level of classical fields, there is a correspondence between solutions of field 
equations $(\Box - \frac{1}{6}R)\phi=0$ and associated fields $\psi$ defined on $\scri$.
We want to investigate whether or not such a correspondence can be implemented at level
of algebras of observables associated with the relevant fields. If the correspondence can be implemented
in terms of 
an injective $*$-homomorphism,
 the algebra of the bulk can be realized as a (sub)algebra of the observables
of the boundary. {\em In this sense it would realize a sort of holographic machinery which encodes complete information of
 QFT defined in the bulk in QFT living in the boundary.} \\
To this end we have to recall some features of linear QFT in globally hyperbolic spacetime \cite{Wald2}.\\

\ssa{Linear QFT in the bulk} \label{bulkQFT}
{\em Let us assume that
the spacetime $(M,g)$ is globally hyperbolic}, $K:= \Box +P$, $P$ being any smooth real valued function
on $M$, denotes a Klein-Gordon-like
operator in that spacetime 
and $\cS_K(M)$ indicates the real space of solutions $\phi$
of $K\phi=0$ with compactly supported Cauchy
data on a (and thus every) Cauchy surface of $(M,g)$. \\
A natural nondegenerate symplectic form on $\cS_K(M)$ can be defined as
\beq
\sigma_M(\phi_1,\phi_2) := \int_S \left(\phi_2 \nabla_N \phi_1 - \phi_1 \nabla_N \phi_2\right)\: d\mu^{(S)}_g\:,
\eeq
the choice of the Cauchy surface $S$ being immaterial because the right-hand side does not depend on such a 
choice. $N$ is the unit future directed normal vector to $S$ and $d\mu_g^{(S)}$ the measure associated 
with the metric induced on $S$ by $g$.
Nondegenerateness implies that there is a unique $C^*$ algebra generated by {\bf 
(abstract) Weyl operators}
$W(\phi)$, with $\phi\in \cS_K(M)$, such that 
%%%%%CORRETTO
they are not vanishing and
%%%%%
\beq
(Wb1)\:\:\:\:\: W_M(-\phi) = W_M(\phi)^*\:,\:\:\:\: (Wb2)\:\:\:\:\: W_M(\phi_1) W_M(\phi_2) = 
e^{i\sigma(\phi_1,\phi_2)/2}\: W_M(\phi_1+\phi_2)\:. \nonumber
\eeq
That $C^*$ algebra is {\bf Weyl algebra, $\cW_K(M)$, associated with $K$ in the spacetime $(M,g)$}. \\
The formal interpretation of elements $W(\phi)$ is $e^{i\sigma_M(\phi,\Phi)}$, 
$\sigma_M(\phi,\Phi)$ being the usual  field 
operator symplectically smeared with smooth field equations with  compactly supported Cauchy data.
There is an equivalent construction of $\cW_K(M)$ which allows a straightforward representation of locality
based on the linear, real, formally anti self-adjoint operator  $E_K : C_c^\infty(M) \to C^\infty(M)$ called
 {\bf causal propagator} of $K$. It is defined as the difference of advanced and retarded fundamental solutions of 
$Kf=0$ which are known to exist globally provided the spacetime is globally hyperbolic. 
Let us focus attention on remarkable features of $E_K$ we go to list.\\
(i) $E_Kf \in \cS_K(M)$ for $f\in C^\infty_c(M)$.
(ii) $E_K$ is surjective onto $\cS_K(M)$.
(iii) $supp(E_Kf) \subset J(supp f)$.
(iv) $E_Kf=0$ if and only if $f=Kg$ for some $g\in C^\infty_c(M)$.\\ 
As consequence of those properties the identity holds \cite{Wald2}
\beq
\int_M \phi f\: d\mu_g = \sigma_M(E_Kf,\phi)\:\:\:\: \mbox{and thus}\:\:\:\:\int_M fE_Kg\: d\mu_g = \sigma_M(E_Kf,E_Kg)
\label{we}\:.
\eeq 
where $d\mu_g$ is the volume form of $M$ induced by the metric $g$.
To go on, it is convenient to define 
\beq
V_M(f):= W_M(E_Kf)\:, \:\:\:\:\: \mbox{for every $\:f\in C_c^\infty(M)$\:.} \label{usefullast}
\eeq
Taking the former of (\ref{we}) into account, the formal interpretation of elements 
$V_M(f)$ is $e^{i\Phi(f)}$, $\Phi(f) = \int_M \Phi f \: d\mu_g$ being the usual field 
operator smeared with smooth compactly supported functions. The interpretation given above makes sense 
in terms of operators whenever a regular state is fixed, by applying GNS theorem. 
 It turns out, for (iv), that 
\beq V_M(f) = V_M(g)\:,\:\:\:\:\: \mbox{if and only if $\:f-g= Kh\:$ for some $\:h\in C_c^\infty(M)$.} 
\label{motion}\eeq
This is nothing but the constraint due to field equation 
$K_K\Phi=0$ given in a distributional-like fashion, using the fact that $K$ is formally self-adjoint
and $KE_K=0$ by definition of $E_K$.
By construction, generators $V_M(f)$ generate the same $C^*$-algebra, $\cW(M)$, as $W_M(\phi)$.
The improvement is due to the fact that, now , 
property (iii) together with (Wb2) and the latter in (\ref{we}) entail
 $$[V_M(f),V_M(g)]=0\:\:	\:\:\mbox{whenever the supports of $f$
and $g$ are causally separated.}$$ 
{\em This is the natural formulation of locality in spacetime}.\\

\ssb{General holographic tools} All results and tools introduced above can be used in the globally hyperbolic spacetime
$(\tV\cap M, g)$ whenever $(M,g)$ is asymptotically flat,
in accordance with the hypotheses of Proposition \ref{prop2} equipped with Klein-Gordon operator for a conformally coupled 
massless scalar field $K:= \Box -\frac{1}{6}R$.\\
The main proposition concerning holographic relations between $\cW_K(\tV\cap M)$ and $\cW(\scri)$
consists of the following proposition. We need a preliminary definition. 
If $(M,g)$ is an asymptotically flat spacetime, satisfying hypotheses of proposition \ref{prop2} 
with respect to $\tV\subset \tM$ and $K:= \Box -\frac{1}{6}R$,
the {\bf projection map} $\Gamma_{\MV} : \cS_K(\MV) \to \cS(\scri)$ is the real linear map 
which associates every $\phi\in \cS_K(\MV)$ with the  smooth extension to 
$\scri$ of $(\omega\Omega)^{-1}\phi$ as in Proposition \ref{prop2}, where $(\omega\Omega)^2 g$
induces the triple $(\scri, \tilde{h}_B,n_B)$ on $\scri$. \\

\proposizione \label{holographicproposition}
{\em Let $(M,g)$ be a globally hyperbolic asymptotically flat spacetime satisfying 
the hypotheses of Proposition \ref{prop2} with respect to
 $\tV\subset \tM$ and 
let $E_K$ denote the causal propagator  in $\MV:= \tV\cap M$ of $K:= \Box -\frac{1}{6}R$.
Assume that both conditions below hold true for the  projection map $\Gamma_{\MV}$:\\ 
{\bf (a)} $\Gamma_{\MV}(\cS_K(\MV))\subset \cS(\scri)$,\\
{\bf (b)} symplectic forms are preserved by $\Gamma_{\MV}$, that is, for all $\phi_1,\phi_2 \in \cS(\MV)$, 
\beq\label{symplflow}
\sigma_{\MV}(\phi_1,\phi_2) = \sigma(\Gamma_{\MV}\phi_1,\Gamma_{\MV}\phi_2)\:,
\eeq 
Then $\cW(\MV)$ can be identified with a sub $C^*$-algebra of $\cW(\scri)$
by means of a $C^*$-algebra isomorphism $\imath$ uniquely determined by the requirement
\beq
\imath(W_{\MV}(\phi)) = W(\Gamma_{\MV} \phi)\:, \:\:\:\:\mbox{for all $\phi\in \cS_K(\MV)$} \label{lc}\:,
\eeq
or, equivalently,
\beq
\imath(V_{\MV}(f)) = W(\Gamma_{\MV} E_Kf)\:, \:\:\:\:\mbox{for all $f\in C_c^\infty(\MV)$} \label{lc'}\:.
\eeq}

\noindent {\em Proof}. 
For (\ref{usefullast}), the thesis can be proved referring to generators $V_{\MV}(\phi)$ only. 
We start by fixing the relevant sub $C^*$-algebra of $\cW(\scri)$ as follows.
 As a consequence of (a), it makes sense to
consider the $^*$-algebra  in $\cW(\scri)$, $\cA$, finitely generated by the elements $V_{\MV}(\Gamma_{\MV} \phi)$
for all $\phi\in \cS(\MV)$. The closure (in $\cW(\scri)$) of that $^*$-algebra, $\overline{\cA}$, is a 
sub $C^*$-algebra of $\cW(\scri)$
by construction. On the other hand, by construction and using the uniqueness of the norm of a $C^*$-algebra, 
$\overline{\cA}$ must coincide with Weyl algebra associated with the real vector space 
$\cS_0:= \Gamma_{\MV}(\cS(\MV))$ and the nondegenerate symplectic form $\sigma$ restricted to that space.
Whenever the real linear application $\Gamma: \cS(\MV) \to \cS_0$ is bijective, the validity of requirement (b) entails 
(as an immediate consequence of the main statement of theorem 5.2.8 in \cite{BR2}) that there is a 
$^*$-algebra isomorphism $\imath: \cW(\bM) \to \cW(\cS_0)\equiv 
\overline{\cA}$ uniquely determined by the requirement $\imath(V_{\MV}(\phi)) = W(\Gamma_{\MV} \phi))$,
which is nothing but  (\ref{lc}). As is well known,  $^*$-algebra isomorphisms of 
$C^*$-algebras
are $C^*$-algebra isomorphisms. Hence the thesis holds true provided the map $\Gamma_{\MV}: \cS(\MV) \to 
\cS_0$
is bijective. $\Gamma_{\MV}$ is surjective by construction. Assume that $\phi \in Ker(\Gamma)$ then, by condition (b)
and using left-argument linearity of $\sigma$ one has 
$\sigma_{\MV}(\phi,\psi)= 0$ for all $\psi\in \cS(\MV)$. Thus it must hold $\phi=0$ because $\sigma_{\MV}$
is nondegenerate. It implies that $\Gamma_{\MV}$ is also injective concluding the proof.
$\Box$\\

\remark The hypotheses in the previous proposition are, to a certain extent, rather
restrictives. In particular condition b) automatically excludes a large class of
manifolds such as asymptotically flat spacetimes with a black hole since part of the
symplectic flow of data crosses the event horizon and it does not reach $\Im^+$. Thus,
in this framework, equation (\ref{symplflow}) is never satisfied and a different
holographic mechanism must be considered. Adopting an ``Occam razor'' perspective, the
simplest road to pursue would be to consider, as the screen where holographic data are
encoded, both the event horizon and null infinity. Thus, within this perspective, the
full content of the bulk theory can be reconstructed starting from two lower
dimensional quantum field theories. Nonetheless, here, we will not deal with this issue 
in detail since it would need an extensive analysis far from the aims of this paper. \\

\noindent In case the hypotheses of Proposition \ref{holographicproposition} is fulfilled, another 
relevant consequence will take place. In that case any algebraic state 
$\nu : \cW(\scri) \to \bC$ 
can be pulled 
back on $\cS(\MV)$
through $\imath$ to the state $\nu_\imath: \cW(\MV)\to \bC$, defined as  $\nu_\imath(a) :=  \nu(\imath(a))$
for all $a\in \cW(\MV)$. In particular it happens for the BMS-invariant state $\lambda$ (corresponding to $\Upsilon$ in its GNS 
representation) of section \ref{QFT3}: the state $\lambda_\imath$ could be used to build up QFT in the bulk.
For instance, it may give a notion of particle also  if the bulk spacetime does not admit any group of isometries 
(Poincar\'e group in particular). From the fact that bulk isometries, barring pathological situations prepared
{\em ad hoc}, give rise to asymptotic symmetries and $\lambda$ is invariant under all asymptotic symmetries, we expect that 
$\lambda_\imath$ is invariant under isometries of the bulk. The formal investigation on this fact in the general case
will be performed elsewhere. Another relevant point which deserves investigation concerns the short distance behaviour 
of $n$-point functions associated with $\lambda_\imath$. In fact it is a well-established result that physically meaningful
states must have Hadamard behaviour property (see \cite{Wald2} for a general discussion on this extent). There is no 
evidence, from our construction,  that $\lambda_\imath$ is Hadamard. However all those properties can be studied in the 
particular and relevant case of Minkowski spacetime. This is the content of next section.\\

\ssa{Holographic interplay of Minkowski space and $\scri$} \label{QFTMINKOWSKI} 
 Let us consider the case of 
four dimensional Minkowski space $(\bM^4,\eta)$. That spacetime
is asymptotically flat. More precisely 
it is asymptotically flat at past null, future null and spatial infinity and it is strongly asymptotically
predictable in the sense of \cite{Wald}.\\
Starting from a fixed Minkowski frame referred to coordinates $(t,\bx)$,
the unphysical spacetime $(\tM,\tg)$ can be fixed to be Einstein static universe \cite{Wald} as follows.
One passes to spherical coordinates in the rest space of the initial Minkowski frame, obtaining coordinates
$(t,r, \vartheta,\phi)$ on $\bM^4$, and finally one adopts null coordinates $u:= t-r \in \bR$, $v:= t+r \in \bR$
obtaining global coordinates $(u,v,\vartheta,\phi)$ on $\bM^4$.
Using these initial coordinates, 
define coordinates $\vartheta=\vartheta$, $\varphi=\phi$, $T= \tan^{-1}v + \tan^{-1}u$ 
and $R= \tan^{-1}v - \tan^{-1}u$  and assume
$\Omega^2\sp\rest_{\bM^4} = 4[(1+v^2)(1+u^2)]^{-1}$. With these definitions  $\tg := \Omega^2 \eta$ reads
\beq \tg = -dT^2 + dR^2 + \sin^2 R (d\vartheta^2 + \sin^2\vartheta d\varphi^2)\:.\label{EM}\eeq
This metric makes sense in a larger spacetime $(\tM,\tg)$ obtained by assuming
$T\in \bR$, $R\in (0,\pi)$ and $\vartheta,\varphi$ varying everywhere on $\bS^2$. That is Einstein static spacetime.
(The singularities for $R\to 0,\pi$ in $\tM$ are only apparent
they being ``origins of  spherical coordinates'' and the expression of the metric (\ref{EM}) is valid throughout
all Einstein spacetime except for the two  one-dimensional submanifolds corresponding to values ``$R=0$'' and ``$R=\pi$''. To cover the
whole  manifold $\tM$
 two charts at least are necessary.)\\
With that procedure  $(\bM^4,\eta)$ turns out to be embedded into $(\tM,\tg)$  as a globally hyperbolic submanifold  and
$\scri$ is completely represented by the set of points with $T+R =\pi$, $R\in (0,\pi)$.
Actually, with the definition given at the beginning, the space $(M,g)$ which fulfills the very definition of asymptotic 
flat at future null infinity is the portion of $\bM^4$ in the future of any fixed $t$-constant Cauchy surface. As a matter of 
fact in the following we consider only this region also if we shall not stress it explicitly.\\
Rescaling $\tg$ on $\scri$ by the further regular factor $\omega^2 := (\sin R)^{-2}$ (i.e.  $\omega^2 :=1+u^2$)  and 
changing coordinates in the sector
$T,R$ one gets a metric with associated  triple
$(\scri, \tilde{h}_B, n_B)$.\\
A natural Bondi frame on $\scri$, which we say to be {\bf associated with the  Minkowski frame}
 $(t,\bx)$ in $\bM^4$,
is finally obtained as $(u,\z,\bz)$ where $u$ is just 
the (limit to $\scri$ of the) null coordinate $u$ in the reference frame initially fixed in 
Minkowski spacetime and $\zeta := e^{i \phi} \cot
 \frac{\vartheta}{2}$, also $(\vartheta,\phi)$ being angular spherical coordinates in the reference frame initially fixed in 
Minkowski spacetime. \\

\remark \label{lastperhaps} 
The metric of the $2$-sphere determined by $\tilde{h}_B$ is nothing but that of the unit 
$2$-spheres 
$4d\z d{\bz}/(1+\z{\bz})^2$
in the rest frame of initial Minkowski coordinates
$(t,\bx)$. Starting form a different initial Minkowski frame
$(t',\bx')$ connected with the initial one by means of a 
orthocronous proper 
Poincar\'e transformation
$(\Lambda, a)$,
 one would determine the same asymptotic manifold $\scri$
but he would find a different  metric $\tilde{h}'_B$ on $\scri$ itself,
 $\tilde{h}_B = 0du' + 4d\z' d{\bz}'/(1+\z'{\bz}')^2$. Notice that the non degenerate 
part is again the standard metric of the unit $2$-sphere
 but, as $\z\neq \z'$ and $\bz\neq {\bz}'$, it is not the standard metric of the unit $2$-sphere 
determined in the former case: Conversely, it is 
that of the unit 
$2$-spheres 
in the rest frame of Minkowski coordinates
$(t',\bx')$.
 However, a closer scrutiny  shows that the triples 
$(\scri, \tilde{h}_B, n_B)$ and $(\scri, \tilde{h}'_B, n'_B)$ are connected by
a transformation of BMS group $(\Lambda, f_a)$. Indeed one has the following result whose (simple) 
proof is left to the reader (see also \cite{Frittelli4, Frittelli8}).\\

\proposizione \label{lastproposition}
{\em Let $(t,\bx)= (x^0,x^1,x^2,x^3)$ and $(t,\bx)=(x'^0,x'^1,x'^2,x'^3)$ be Minkowski frames in $(\bM^4, \eta)$ such that
$x'^\mu = {\Lambda^\mu}_\nu(x^\nu +a^\nu)$ for some $(\Lambda,a) \in \iso$ and let 
$(u,\z,\bz)$ and $(u',\z',\bz')$ be  the respectively associated Bondi frames on $\scri$. The following holds.\\
{\bf(a)} The Bondi frames are connected by means of
the  BMS transformation
$$u':= K_\Lambda(\z,\bz)(u + f_a(\z,\bz))\:,\:\:\:\: (\z',\bz') = \Lambda (\z,\bz)\:,$$
where the action of $\Lambda$ on $(\z,\bz)$ is that in (\ref{z}) and
 the function $f_a$ belongs to the space $T^4$ spanned by the first four real spherical harmonics as defined in Section \ref{mi}, 
 that is\footnote{In angular spherical coordinates, we recognize in the factors below in front of $-a^1$, $-a^2$ and $-a^3$ 
 the components of the radial versor, respectively,
 $\sin \vartheta \cos \phi$, $\sin \vartheta \sin \phi$ and $\cos\vartheta$.} 
\beq
f_a:= a^0 - \frac{a^1(\z+\bz)}{\z\bz+1} -
 \frac{a^2(\z-\bz)}{i(\z\bz+1)} - \frac{a^3(\z\bz-1)}{\z\bz+1} \:.
\eeq
{\bf (b)} The set
$$ \cR := \left\{(\Lambda,f_a)\in G_{BMS}\:\left|\: f_a= a^0 - \frac{a^1(\z+\bz)}{\z\bz+1} -
 \frac{a^2(\z-\bz)}{i(\z\bz+1)} - \frac{a^3(\z\bz-1)}{\z\bz+1} \:,\:\: a \in \bR^4\right.\right\}$$
is a subgroup of $G_{BMS}$,  the map $\iso\ni (\Lambda,a) \mapsto (\Lambda,f_a) \in\cR$ 
being a continuous-group isomorphism.}\\

\noindent The second statement is a straightforward consequence of proposition \ref{covariance}.\\
The quantum version of proposition above will be  
established in Theorem \ref{lasttheorem} below. 
These results are due to the fact that Poincar\'e isometries are 
also asymptotic symmetries (see also \cite{Frittelli4}).\\

\noindent Einstein static universe is globally hyperbolic because it is static and $T$-constant 
sections are compact (see chapter 6 in \cite{Fu}). As a consequence $(\bM^4,\eta)$ 
(more precisely, the region in $(\bM^4,\eta)$  in the future of a fixed Minkowskian spacelike Cauchy surface) 
fulfills the hypotheses of 
Proposition \ref{prop2}
with respect to $\tV:=\tM$ itself. 
The part of standard free QFT in Minkowski spacetime \cite{PCT,Haag} for a massless scalar field $\phi$, 
we are interested in, can be summarized as follows in 
 Weyl quantization referred to 
Weyl algebra $\cW(\bM^4)$ with $K:= -\Box$. Standard free QFT can be viewed as  the GNS realization of $\cW(\bM^4)$
based on a preferred algebraic state $\lambda_{\bM^4}$ invariant under Poincar\'e group and individuated as we go to
describe. 
 Take a Minkowski frame with coordinates $(t,\bx)\in \bR^4$ and, for every $\phi \in \cS_K(\bM^4)$, define its positive 
 frequency part, $\phi_+$,
  \beq
 \phi_+(t,\bx)  := \int_{\bR^3} \sp\spa \frac{d\bp\: e^{i(\bp\cdot \bx - t |\bp|)}}{\sqrt{16\pi^3 |\bp|}} 
 \widetilde{\phi_+}(\bp) \:,\:\:\:\:
\widetilde{\phi_+}(\bp) :=  
 \sqrt{\frac{|\bp|}{16\pi^3}}\sp \int_{\bR^3} \sp\sp d\bx
  \spa \left(\spa\phi(0,\bx) -i \frac{(\partial_t \phi)(0,\bx)}{|\bp|}\right)\spa e^{-i\bp\cdot \bx}\label{two'}\:.
\eeq
$\phi_+$ has no compactly supported Cauchy data  and $\phi = \phi_+ + \overline{\phi_+}$. The sesquilinear form 
  \beq
\al \phi_{1+},\phi_{2+}\cl_{\bM^4}  :=-i\sigma_{\bM^4}(\overline{\phi_{1+}},\phi_{2+}) 
\:,\:\:\:\mbox{for every pair $\phi_{1},\phi_{2} \in \cS_K(\bM^4)$} \label{prodscalarM}
\eeq
is well-defined and give rise to a Hermitean scalar product on the space $\cS_K(\bM^4)^{\bC}_+$ of complex 
linear combinations of positive frequency parts and
\beq
\al \phi_{1+},\phi_{2+}\cl_{\bM^4} = \int_{\bR^3} d\bp\: \overline{\widetilde{\phi_{1+}}(\bp)}\widetilde{\phi_{2+}}(\bp)\:,\:\:\:
\mbox{for every pair $\phi_{1},\phi_{2} \in \cS_K(\bM^4)$.}\label{prodscalar2M}
\eeq
As a consequence $\cS_K(\bM^4)^{\bC}_+$ is isomorphic to a subspace of $L^2(\bR^3,d\bp)$.
Since the former is also dense in the latter\footnote{As is well known, the map (see (\ref{two'})) 
$C_c^{\infty}(\bR^3)\ni f \mapsto \int_{\bR^3} d\bp
f(\bx) e^{-i\bp\cdot \bx}$ has range dense in $L^2(\bR^3,d\bp)$
because Fourier transform is a Hilbert-space isomorphism and $C_c^{\infty}(\bR^3)$
is dense in $L^2(\bR^3,d\bp)$, therefore the range is also
 $L^2$-dense in the space $B\subset L^2(\bR^3,d\bp)$
of functions which are in $C_c^\infty(\bR^3)$ and vanish in a neighborhood of $\bp=0$. Finally $B$ is
dense in $L^2(\bR^3,d\bp)$ and it is
invariant under multiplication of its elements with  either $\sqrt{|\bp|}$ and   $1/\sqrt{|\bp|}$.
Thus, by the latter equation in (\ref{two'}), we find that, up to Hilbert-space isomorphisms, $\overline{\cS_K(\bM^4)^\bC_+} = 
L^2(\bR^3,d\bp)$.} 
by (\ref{two'}),  one finds that the  {\bf one-Minkowski-particle space} $\cH_{\bM^4}$, i.e.
 the Hilbert completion  of $\cS_K(\bM^4)^{\bC}_+$,
is isomorphic to $L^2(\bR^3,d\bp)$ itself.\\
The orthocronous proper Poincar\'e group $\iso$ acts naturally on wavefunctions  
via push-forward: $g^*: \cS_K(\bM^4)\ni \phi \mapsto \phi\circ g^{-1}$ for every $g \in \iso$. 
The symplectic form $\sigma_{\bM^4}$ is invariant under such $g^*$, $g$ being an isometry. 
Furthermore, it turns out that  there is an irreducible  strongly-continuous unitary representation 
$L^{(1)}: \iso \ni g \mapsto L^{(1)}_g$ with  $L^{(1)}_g : \cH_{\bM^4} \to \cH_{\bM^4}$ such that  
$(g^* \phi)_+ = L_g \phi_+$ for every 
$g\in \iso$ 
and every $\phi\in \cS_K(\bM^4)$. In particular this implies that the decomposition in positive and negative frequency 
parties 
as well as the scalar product, do not depend on the particular Minkowski frame used. An irreducible operator
representation $\widehat{\cW}(\bM^4)$ of Weyl algebra $\cW(\bM^4)$ is constructed on $\gF_+(\cH_{\bM^4})$ 
in terms of usual symplectically-smeared field operators and their exponentials
\beq
\sigma_{\bM^4}(\phi,\Phi) := ia({\phi_+}) -i a^\dagger(\phi_+)\:,\:\:\:\:
\widehat{W}_{\bM^4}(\psi):= e^{\overline{i\sigma_{\bM^4}(\phi,\Phi)}}\:.
\eeq
The vacuum state $\Upsilon_{\bM^4}$ of $\gF_+(\cH_{\bM^4})$ is, by definition, invariant under the unitary 
representation $L$ of $\iso$  obtained by tensorialization 
of $L^{(1)}$ and the following covariance relations hold
\beq L_g \widehat{W}_{\bM^4}(\phi) L_g^\dagger = \widehat{W}_{\bM^4}(g^*\phi)\:,\:\:\:\: \mbox{for every $\phi \in \cS_{K}(\bM^4)$
 and
$g\in \iso$}\:.\label{covarianceM}\eeq
If $\Pi_{\bM^4}  : {\cal W}(\bM^4) \to \widehat{\cal W}(\bM^4)$ 
denotes the unique 
 ($\sigma_{\bM^4}$ being nondegenerate)
 $C^*$-algebra isomorphism between those two Weyl representations,  $({\gF_+}(\cH_{\bM^4} ), \Pi_{\bM^4} ,  \Upsilon_{\bM^4} )$ 
coincides, up to unitary transformations,  with  the GNS triple associated with the  algebraic  state $\lambda_{\bM^4}$
on ${\cal W}(\bM^4)$ uniquely defined  by the requirement (see the appendix) 
\beq\lambda_{\bM^4} (W_{\bM^4}(\phi)) := e^{-\langle \phi_+, \phi_+\rangle_{\bM^4} /2}\label{GNSM}\:.\eeq
 $\lambda_{\bM^4}$ is pure as well-known, however this is also a direct consequence of (b) in theorem
\ref{teoremaolograficoMinkowski} below since $\lambda$ is pure.
We can now state and prove the main results of this section.\\

\teorema \label{teoremaolograficoMinkowski}
{\em Consider free QFT for a real scalar field $\phi$ propagating in four-dimensional
Minkowski spacetime $(\bM^4,\eta)$ and QFT for a real scalar field on $\scri$. 
Let $\cW(\bM^4)$ be the Weyl algebra 
associated with the space $\cS_K(\bM^4)$ and  the symplectic form
$\sigma_{\bM^4}$ as defined in section \ref{bulkQFT} (with $M_{\tV} := \bM^4$ and $K:= -\Box$).
The following  holds.\\
{\bf (a)} $\Gamma_{\bM^4}(\cS_K(\bM^4))\subset \cS(\scri)$ because $\Gamma_{\bM^4} \phi$ has compact support for 
 $\phi\in \cS_K(\bM^4)$, moreover $\Gamma_{\bM^4}$ preserves symplectic forms. 
As a consequence
$\cW(\bM^4)$ can be identified with a sub $C^*$-algebra of $\cW(\scri)$
by means of a $C^*$-algebra isomorphism $\imath_{\bM^4}$ uniquely determined by the requirement
\beq
\imath_{\bM^4}(W_{\bM^4}(\phi)) = W(\Gamma_{\bM^4} \phi)\:, \:\:\:\:\mbox{for all $\phi\in \cS_K(\bM^4)$} 
\label{lcMinkowski}\:.
\eeq
{\bf (b)} Consider Minkowski vacuum $\lambda_{\bM^4}$ on $\cW(\bM^4)$ and the BMS-invariant vacuum $\lambda$ on $\cW(\scri)$
and focus on the respectively associated GNS realizations
$(\gF(\cH_{\bM^4}), \Pi_{\bM^4}, \Upsilon_{\bM^4})$ and  $(\gF(\cH), \Pi, \Upsilon)$.
 The $C^*$-algebra isomorphism $\imath_{\bM^4}$ corresponds
to a unitary (i.e. isometric surjective) operator $\cU : \gF(\cH_{\bM^4}) \to \gF(\cH)$ such that:
(i) $\cU: \Upsilon_{\bM^4} \mapsto \Upsilon,$ and
 (ii) $\cU \widehat{W}_{\bM^4}(\phi) \cU^{-1}= \widehat{W}(\Gamma_{\bM^4}\phi)\:.$
 
\noindent Therefore the algebraic state induced by $\lambda$ on $\cW(\bM^4)$ through $\imath_{\bM^4}$ is Minkowski vacuum
$\lambda_{\bM^4}$.}\\

\noindent {\em Proof}. (a) Fix a Minkowski reference frame $(t,\bx)$ in $\bM^4$,
pass to spherical coordinates in the rest frame obtaining coordinates $(t,r,\z,\bz)$, next pass to null coordinates 
in the sector $t,r$ and, finally,
construct coordinates $(u,\z,\bz)$
on $\scri$ referred to a Bondi frame as described at the beginning of this section.
In Minkowski spacetime solutions of $K\phi=0$ propagate along null geodesics \cite{Friedlander}.
In other words, if $\phi = Ef$, the support of $\phi$ is included in the union of null geodesics 
originated from every point $q\in supp f$.
On the the hand the map  $u= Z(q, \z,\bz)$ that associates the unique null geodesics starting from the 
point $q\in \bM^4$
and direction $(\z,\bz)$ with the coordinate $u$ where the geodesics reaches $\scri$ 
(the remaining coordinates being $(\z,\bz)$) is well defined and smooth \cite{Kozameh, Frittelli7}.
If $\phi\in \cS_K(\bM^4)$, $\phi= Ef$ where $f$ is smooth with compact support, as a consequence
$supp \:\Gamma_{\bM^4} \phi \subset  \{Z(q,\z,\bz)\:|\: q\in supp f\:, (\z,\bz)\in \bS^2\}\times \bS^2$ is 
 compact because $Z$ is continuous and defined on a compact set.  Since $\Gamma_{\bM^4} \phi$ is smooth by
 definition, we have proved that $\Gamma_{\bM^4}(\cS_K(\bM^4))\subset \cS(\scri)$. Now we pass to prove that
 $\Gamma_{\bM^4}$ preserves the symplectic forms. To this end we notice that, if 
 $\phi,\phi'\in \cS_{\bM^4}$ then $\sigma_{\bM^4}(\phi,\phi') = i2 Re \langle \phi_+,\phi'_+\rangle_{\bM^4}$
 and the analog holds for wavefunctions $\psi,\psi'\in \cS(\scri)$ referring to the corresponding
 symplectic form $\sigma$ and scalar product $\langle\cdot,\cdot\rangle$ as in Theorem \ref{theorem2}.
 (The proof is immediate, taking into account the fact that positive frequency parts 
 satisfy $\sigma_{\bM^4}(\phi_+,\phi'_+)=0$ and the analog for the other case.) As a consequence, 
 to show that $\sigma_{\bM^4}(\phi,\phi')= \sigma(\Gamma_{\bM^4}\phi,\Gamma_{\bM^4}\phi')$, it is 
 {\em completely equivalent} to show that
 \beq \langle\phi_+,\phi'_+\rangle_{\bM^4}= \langle(\Gamma_{\bM^4}\phi)_+,(\Gamma_{\bM^4}\phi')_+\rangle
 \:,\:\:\:\:\mbox{for every pair of wavefunctions $\phi,\phi'\in \cS_{\bM^4}$.} \label{centerofproof}\eeq
 Notice that the positive frequency parts in the left-hand side are referred to Minkowski time $t$ in $\bM^4$, 
 whereas those 
 in the right-hand side are referred to coordinate $u$ in $\scri$. Proof of (\ref{centerofproof})
 is a consequence of the following lemma whose proof is quite technical and  presented in the Appendix.\\
 
 \lemma \label{lemma1} {\em In the hypotheses of theorem \ref{teoremaolograficoMinkowski},
fix a Minkowski reference frame $(t,\bx)$ in $\bM^4$,
and consider the associated Bondi frame $(u,\z,\bz)$ on $\scri$.
  If $(E,\z,\bz)$ are the spherical coordinates of $\bp$ in the rest frame 
  (where $E:= |\bp|$ in particular),
  it holds
   \beq\widetilde{(\Gamma_{\bM^4}\phi)_+}(E,\z,\bz) = -iE \widetilde{\phi_+}(\bp(E,\z,\bz))\:,\:\:\:\:
  \mbox{for all $\phi\in \cS_{\bM^4}$,}\label{eqinlemma}\eeq
  the function in the left-hand side being that of definition (\ref{two}) with 
  $\cS(\scri)\ni \psi = \Gamma_{\bM^4}\phi$.}\\
  
\noindent From (\ref{eqinlemma}) one proves (\ref{centerofproof}). Indeed, starting from  (\ref{prodscalar2M}),
passing in spherical coordinates in the integral in the right-hand side and taking (\ref{prodscalar2}) into account, 
one gets (\ref{centerofproof})
$$\langle\phi_+,\phi'_+\rangle_{\bM^4}= \int_{\bR^+\times \bS^2} dE E^2\:  \epsilon_{\bS^2}  \:
 \overline{\widetilde{\phi_+}(\bp(E,\z,\bz))} \widetilde{\phi'_+}(\bp(E,\z,\bz))$$ 
 $$=\int_{\bR^+\times \bS^2} dE \epsilon_{\bS^2}\:  
 \overline{-iE\widetilde{\phi_+}(\bp(E,\z,\bz))}\: (-i)E\widetilde{\phi'_+}(\bp(E,\z,\bz))
 =  \langle(\Gamma_{\bM^4}\phi)_+,(\Gamma_{\bM^4}\phi')_+\rangle\:.$$
 (b) Referring to lemma \ref{lemma1}, start from the $\bC$-linear isometric map 
 $h_0 : \cS_K(\bM^4)_+^\bC \to \cS(\scri)_+^\bC$ which associates the function $\widetilde{\phi_+}(\bp)$
 with the function $-iE\widetilde{\phi_+}(\bp(E,\z,\bz))= \widetilde{(\Gamma_{\bM^4}\phi)_+}(E,\z,\bz)$. 
 The domain and the range of that map are dense
 in  $\cH_{\bM^4}$ and $\cH$ respectively: In the first case it has been proved previously,
 the proof for the latter case is immediate using the density property in the former and the measures in the relevant 
$L^2$ spaces corresponding to the two Hilbert spaces. As a consequence, $h_0$ extends to a unitary map $h :\cH_{\bM^4} \to \cH$. In turn,
 this second map extends to a unitary map $\cU: \gF(\cH_{\bM^4}) \to \gF(\cH)$ by tensorialization and assuming that (i)
 $\cU \Upsilon_{\bM^4} = \Upsilon$. By construction it also holds $\cU\sigma_{\bM^4}(\phi, \Phi)\cU^{-1} = 
 \Psi(\Gamma_{\bM^4}\phi)$ working in the dense space of analytic vectors containing a finite number of particles. 
 Passing to exponentials
 one finds (ii). $\Box$\\

 \remark The result established in (a) of theorem \ref{teoremaolograficoMinkowski} straightforwardly extends  
to the case of a spacetime $M$  obtained by switching on curvature in the past of an (arbitrarily far in the future) 
smooth spacelike Cauchy surface $\Sigma$ of $\bM^4$
contained in a Cauchy surface of $\tilde{M}$ passing for $i^0$. With obvious notation,
$\Gamma_{M}(\cS_K(M))\subset \cS(\scri)$ because $\Gamma_{M} \phi$ has again compact support for 
 $\phi\in \cS_K(M)$ by construction.
Moreover $\Gamma_{M}$ preserves symplectic forms since, after $\Sigma$, the symplectic form associated with $M$
is as the same as that of $\bM^4$ and symplectic forms are preserved under time evolution in bulk spacetime. 
As a consequence
$\cW(M)$ can be identified with a sub $C^*$-algebra of $\cW(\scri)$
by means of a $C^*$-algebra isomorphism $\imath$ uniquely determined by the requirement
\beq
\imath(W_{M}(\phi)) = W(\Gamma_{M} \phi)\:, \:\:\:\:\mbox{for all $\phi\in \cS_K(M)$} 
\nonumber \:.
\eeq\\

 \noindent A second theorem concerns the interplay of orthocronous proper  Poincar\'e group $\iso$ and $G_{BMS}$. We knows that 
 in the bulk there is a strongly-continuous unitary irreducible representation  $\iso\ni g \mapsto L_g$
 satisfying (\ref{covarianceM}). Referring to the Minkowski frame $(t,\bx)$ used to build up the metric on $\scri$ and all that,
  if $g=(\Lambda, T)$ with $\Lambda \in SO(3,1)\sp\uparrow$ and $a\in \bR^4$, 
  the action of $L_g$ on a positive frequency part $\widetilde{\phi}_+$ reads
  \beq
  \left(L_{(\Lambda, a)} \widetilde{\phi}_+\right)(\bp) = \sqrt{\frac{E_{\Lambda^{-1}}}{E}}e^{-i( p |
  \Lambda a)}\widetilde{\phi}_+(\bp_{\Lambda^{-1}}) \:,\label{Lp}
  \eeq
 where $p:=(E,\bp)$,  $(E_\Lambda,\bp_\Lambda):=\Lambda p$, whereas $(a | b)$ denotes the standard product of 
 4-vectors $a$ and $b$. The question is: {\em what is the meaning of the representation $\iso\ni g \mapsto \cU L_g \cU^{-1}$
 acting on quantum states for QFT defined in $\scri$?}\\
The following theorem gives an answer to the question which is the quantum version of Proposition \ref{lastproposition}.\\

\teorema \label{lasttheorem}
{\em With the same hypotheses as in Theorem \ref{teoremaolograficoMinkowski},
represent $G_{BMS}$ as the semidirect product of 
%%%MODIFICATO prima era ISO
$SO(3,1)\sp \uparrow$ 
%%%%%%  CONTROLLARE TUTTO
and $C^\infty(\bS^2)$
in the Bondi frame on $\scri$ associated with the Minkowski frame $(t,\bx)$.
Consider the natural unitary representation of $\iso$ in QFT in $(\bM^4,\eta)$
given in (\ref{Lp}). The representation 
 on $\gF(\cH)$,  induced on QFT on $\scri$ by means of $\cU$, is
  $\iso\ni g \mapsto \cU L_g \cU^{-1}$ and it coincides with
the restriction of the representation of $G_{BMS}$, $U$, defined in Theorem \ref{theo2}, 
to the subgroup  isomorphic to $\iso$ (see proposition \ref{lastproposition})
$$\left\{(\Lambda,f_a)\in G_{BMS}\:\left|\: f_a= a^0 - \frac{a^1(\z+\bz)}{\z\bz+1} -
 \frac{a^2(\z-\bz)}{i(\z\bz+1)} - \frac{a^3(\z\bz-1)}{\z\bz+1} \:,\:\: (\Lambda, a) \in \iso\right.\right\}$$}

\noindent{\em Proof}. By lemma \ref{lemma1} one has
\beq \left(\cU\widetilde{\phi_+}\right)(\bp,\z,\bz) = -iE \widetilde{\phi_+}(\bp(E,\z,\bz))\:.\label{HH}\eeq
Representing the right-hand side of (\ref{Lp})
in complex spherical coordinates $\z,\bz$ and applying $\cU$ on the final result making use of (\ref{HH}),
a straightforward, but tedious, computation based on (\ref{notevole}) proves that, for every $\widetilde{\psi_+} \in 
\cU(\cH_{\bM^4})$,
$\cU L_{(\Lambda,a)} \cU^{-1} \widetilde{\psi_+} = U_{(\Lambda,f_a)}\widetilde{\psi_+}$
holds true whenever  $(\Lambda, a)$ is any pure translation,  any pure  rotation  and any boost along $z$. 
Hence the decomposition theorem
of Lorentz group and the structure of the group product in $\iso$ and in $G_{BMS}$
imply that the identity holds for every element $(\Lambda,a)\in \iso$.
Since $\cU(\cH_{\bM^4})\subset \cH$ is dense in $\cH$ (and $\cU$ preserve one-particle spaces), 
we have obtained that
$\cU L_{(\Lambda,a)}\sp\rest_{\cH}  \cU^{-1} = U_{(\Lambda,f_a)}\sp\rest_{\cH}$. Finally, since 
$\cU: \gF(\cH_{\bM^4})\to \gF(\cH)$,
$L_{(\Lambda,a)}: \gF(\cH_{\bM^4})\to \gF(\cH_{\bM^4})$ and $U_{(\Lambda,f_a)}: \gF(\cH)\to \gF(\cH)$
are all obtained by tensorialization procedure, it must hold 
$\cU L_{(\Lambda,a)} \cU^{-1} = U_{(\Lambda,f_a)}$. $\Box$

\section{Conclusions}
The main purpose underlying this paper has been to show that, at least in the
scalar case, it is possible to start from a scalar free field $\phi$ living in the bulk of an
asymptotically flat four-dimensional spacetime $M$ and to relate it by means of a
suitable extension/restriction procedure with a second field $\psi$ living on 
$\Im^+$, the boundary of $M$ at future null infinity.
Under suitable hypotheses (preservation of a symplectic form), 
this relation preserves information at level of quantum field theories
when passing from the bulk to the boundary thus implementing the holographic principle.\\
However it is worth stressing that the notion of bulk to boundary correspondence that we 
have envisaged in this paper is
to all purposes rather different from the more common one proper of the
AdS/CFT scenario since we deal only with the reconstruction of test fields living in a
bulk with a fixed background metric whereas, up to now, the reconstruction of geometric data is
not addressed within this approach.\\
The main statements of this paper have been proved at level of Weyl $C^*$ algebras 
associated with the fields. 
Within this framework, $\psi$ is interpreted as a kinematical 
datum of a quantum field theory intrinsically defined on
$\Im^+$ and invariant under the action of the BMS group as discussed in
section 1. 
We have shown that such physical intuitive idea can be made
rigorously precise identifying $\psi$ with an intrinsic BMS field constructed
out of the induced unitary irreducible representations. Such result has been
achieved by means of a technology similar to celebrated Wigner's one used to
classify and construct explicitly all possible Poincar\'e-invariant wavefunctions. 
Universality of such an approach and the
techniques handled in section 2 and 3 suggest that our results, achieved for massless
fields, may be extended far beyond the case of vanishing ``spin''. Furthermore it 
would be interesting to investigate the interplay of these results with the 
{\em asymptotic quantization} 
procedure proposed by Ashtekar \cite{APRL} where the
main variable is played by BMS-invariant {\em gauge} massless fields living on
$\Im^\pm$. To this end it is worth noticing that, in \cite{APRL}
and in most of the paper concerning applications of the BMS group, the peculiar
role played by the unitary BMS irreducible representation induced from the subgroup $\Gamma$
(instead of our $\Delta$), suggests that it has been always implicitly assigned an Hilbert 
topology to the set of supertranslations $N$. This is in apparent contrast with our results
and the issue deserves future investigation.
This is because the results presented in section 3 indicates 
that, in order to ``relate'' a bulk field with the boundary BMS-invariant 
counterpart, it is necessary to adopt a nuclear
topology on $N$. The relevance of this result does not only lie in the realm of a
rigorous mathematical analysis of the BMS group, but it mainly affects the 
physical kinematical configuration of the field theory living on $\Im^+$ since, 
as discussed in \cite{Mc4} and partly in section 3, in the ``nuclear'' scenario, 
it arises a plethora of possible free fields (or equivalently little groups) which 
are {\em not} present in the Hilbert topology.\\
A further key requirement within our approach consists of considering specifically
four dimensional spacetimes. Beside the natural physical relevance, this scenario is
the lone where $SL(2,\mathbb{C})\ltimes C^\infty(\bS^2)$ plays the role of the (asymptotic)
symmetry group. Nonetheless it is natural to wonder whether a possible extension of
our results to higher (lower) dimensional spacetimes could be envisaged. Unfortunately
a straightforward attempt in this direction runs into two serious obstructions, the
first referring to the impossibility to coherently perform Penrose construction in odd
$d$-dimensional manifolds with $d>4$. As proved by Ishibashi and Hollands in
\cite{Hollands, Hollands2},
the definition itself of null infinity adopted in this paper is at stake and this
seems to force us to choose either a different codimension one submanifold where to
encode bulk data or a different projection map since the one, introduced in section 2
for the massless scalar field conformally coupled to gravity, strongly relies on the
(geometry of) Penrose compactification. 
Furthermore, although we consider even $d$-dimensional asymptotically flat spacetimes
with $d>4$, we cannot slavishly transfer our results since, as
proved in \cite{Hollands, Hollands2},  in these scenarios there is no notion of supertranslations and
thus the asymptotic symmetry group at null infinity is neither the BMS nor
a BMS-like group. Thus, although one could project bulk fields to $\Im^\pm$, in order
to interpret them as intrinsic boundary fields one is forced either to study, case by case, the
theory of unitary and irreducible representation for the new asymptotic symmetry
group either to repeat the Wigner programme. The final result of such approach would 
consists on a full construction of the kinematical and the dynamical spectrum of the 
boundary free field theory which should be compared with the projected bulk fields as 
it has been done in section 3 for the four dimensional scenario\footnote{A similar
conclusion holds also for $d=3$ where the counterpart of the BMS group is
$Diff(S^1)\ltimes C^\infty(S^1)$ \cite{Ashtekar2}. Nonetheless, in this scenario, there is a 
further key difficulty since the role of $SL(2,\bC)$, a finite dimensional Lie group,
is traded by $Diff(S^1)$, an infinite dimensional group. Thus Mackey imprimitivity
theorem may not hold and the inducing technique may not grant us an exhaustive
reconstruction of unitary and irreducible representations.}.  \\
A complete survey of the bulk to boundary relation for free fields should also
comprise the rather elusive case of massive fields. Within this specific
framework, the extension/restriction procedure proposed in section 2 fails mainly
due to the presence of an intrinsic scale length represented by the mass.
Nonetheless we believe that an ``holographic investigation'' along the lines  
proposed in \cite{Dappiaggi} is still possible and it is currently under
investigation.\\
Other key results of this paper appear in section 4 where the
holographic interplay between a bulk theory living on a 
spacetime satisfying a weaker requirement as in Proposition \ref{prop2}
(in particular a strongly  asymptotically predictable spacetime  in the sense of \cite{Wald})
and the BMS boundary theory has been discussed within the
framework of $C^*$ algebras of field-observables and their isomorphisms. In particular, 
in the specific scenario
of Minkowski spacetime, a key achievement consists on establishing an unitary 
correspondence between the bulk vacuum and the BMS counterpart on $\Im^+$, though
the uniqueness of the latter has not been proved and it should be analyzed in
detail. 
The uniqueness problem of a BMS-invariant quasifree (algebraic) state $\lambda$ on $\scri$ has relevance in the 
issue of the notion of particle in the absence of Poincar\'e group.
If the BMS-invariant quasifree state is uniquely determined,
it could be used to give a definition of particle for spacetime which does not admit
a group of isometries but are asymptotically flat and the algebra of the field in the bulk
can be identified with a subalgebra of the fields on $\scri$ by means of an injective
$*$-homomorphism $\imath$ as in proposition \ref{holographicproposition}.
In this case, $\lambda$ induces a quasifree state $\lambda_\imath$ for the algebra of fields 
in the bulk with an associated definition of particle. We have established in theorems 
\ref{teoremaolograficoMinkowski} and \ref{lasttheorem} that such a notion of particle,
whenever available, must agree with the usual one in four dimensional Minkowski spacetime
since $\lambda_\imath$, in that case, is just Minkowski vacuum.
Another relevant point which deserves investigation concerns the short distance behaviour 
of $n$-point functions associated with $\lambda_\imath$. There is no 
evidence, from our construction,  that $\lambda_\imath$ is Hadamard also if it happens 
in Minkowski spacetime trivially.

To conclude, we wish also to pinpoint that, within this paper, we have completely
discarded the role of interactions. Nonetheless, in
order to construct a full holographic bulk to $\Im^\pm$ correspondence it is
imperative to understand how to couple the boundary free field 
either with self/external interactions (barring gravitational field) either with gauge degrees 
of freedom. A complete and concrete solution of this challenging issue would possibly rule out
whether it is really possible or not to define a full asymptotically flat/BMS
correspondence and, thus, we believe it is worth to be deeply analyzed.

\appendix
\section{Appendix}

\ssb{GNS reconstruction}
The interplay of the  Fock representation presented in section \ref{QFT3}  and GNS theorem \cite{Haag,BR} is simply sketched.
(The same extent holds for QFT in Minkowski spacetime presented in section \ref{QFTMINKOWSKI} if replacing  ${\cal W}(\scri)$ 
with ${\cal W}(\bM^4)$, $W(\psi)$
with $W_{\bM^4}(\phi)$, $\Pi$ with $\Pi_{\bM^4}$,
$\Psi(\psi)$ with $\sigma_{\bM^4}(\phi,\Phi)$ and
$\lambda$ with $\lambda_{\bM^4}$.)
Using notation introduced in section \ref{QFT3}, if $\Pi  : {\cal W}(\scri) \to \widehat{\cal W}(\scri) $ denotes the unique 
 ($\sigma$ being nondegenerate)
 $C^*$-algebra isomorphism between those two Weyl representations,
it turns out that $({\gF}_+(\cH ), \Pi ,  \Upsilon )$ 
is the GNS triple associated with a particular pure algebraic state $\lambda$
({\em quasifree} \cite{BR}  and invariant under
the automorphism group associated with  $G_{BMS}$)
on ${\cal W}(\scri)$ we go to introduce.
Define $$\lambda (W(\psi)) := e^{-\langle \psi_+, \psi_+\rangle /2}$$
then extend $\lambda$ to the $^*$-algebra finitely generated by all the elements
$W(\psi)$ with $\psi \in \cS(\scri)$, by linearity  and using (W1), (W2). It is simply proved that, $\lambda(\bI)=1$ and
$\lambda(a^*a)\geq 0$ for every element $a$ of that $^*$-algebra so that $\lambda$
is a state. As the map $\bR\ni  t\mapsto\lambda(W(t\psi))$ is continuous,  
known theorems \cite{Lewis} imply that $\lambda$ extends uniquely to a state $\lambda$
on the complete Weyl algebra ${\cal W}(\scri)$.
On the other hand, by direct computation, one finds that $\lambda(W(\psi)) = \left\langle \Upsilon ,  \widehat{W}(\psi)  \Upsilon  \right\rangle$.
 Since a state on a $C^*$ algebra is continuous, this relation can be extended to the whole algebras 
by linearity and continuity
and using  (W1), (W2) so that a  general GNS relation is verified: 
%%%%ATTENZIONE CORREZZIONE, MANCAVA IL SEGNO = SOTTO%%%%  
\beq\lambda(a)  = \left\langle \Upsilon ,  \Pi (a) \Upsilon  \right\rangle
\:\:\:\:\:\mbox{for all $a\in {\cal W}(\scri)$}\:. \label{aa} 
\eeq 
To conclude, it is sufficient to show that  $ \Upsilon $
is cyclic with respect to $\Pi$. Let us show it.
If $\widehat{\cal F} (\scri)$ denotes the $^*$-algebra
generated by field operators $\Psi(\psi)$, $\psi\in \cS(\scri)$, defined on $F(\cH)$,
$\widehat{\cal F} (\scri) \Upsilon $
is dense in the Fock space (see proposition 5.2.3 in \cite{BR2}).  Let   $\Phi \in \gF_+(\cH)$  
be a vector orthogonal to both $\Upsilon$ and to all the vectors
$\widehat{W}(t_1\psi_1)\cdots \widehat{W}(t_n\psi_n)\Upsilon$ for  $n=1,2,\ldots$ and $t_i\in \bR$ and $\psi_i\in \cS(\scri)$.
Using Stone theorem to differentiate in $t_i$ for $t_i=0$, starting from $i=n$ and proceeding backwardly 
up to $i=1$, one finds that  $\Phi$ must also be  orthogonal to all of the vectors
$\Psi(\psi_1)\cdots \Psi(\psi_n)\Upsilon$
and thus vanishes because $\widehat{\cal F} (\scri) \Upsilon$
is dense.
This result means that  $\Pi ({\cal W}(\scri)) \Upsilon $
is dense in the Fock space too, i.e. $\Upsilon$ is cyclic
with respect to $\Pi$.  Since $\Upsilon$ satisfies also (\ref{aa}),  the uniqueness of the GNS triple 
 proves that the triple 
$({\gF_+}(\cH ), \Pi ,  \Upsilon )$ 
is just (up to unitary transformations) the GNS triple  associated with 
$\lambda$. Since the 
%%%CORREZZIONE: frase modificata
GNS representation is irreducible (see discussion after theorem \ref{theo2}) $\lambda$ is pure.  \\
%%%

\ssb{Proof of some propositions}

\noindent{\em Proof of  Proposition \ref{prop3}}.
In the following we assume that  $\Omega$ includes the further factor $\omega$.
Referring to the expression of BMS group in a Bondi frame, we prove the thesis for any element $(\Lambda,f)$
of BMS group 
of the form either $(\Lambda(t),0)$ or $(I,tf)$ where 
$f\in C^\infty(\bS^2)$ and $t\to \Lambda(t)$ is a one-parameter subgroup of $SO(3,1)\sp\uparrow$.
Notice that the subgroups $t \mapsto (\Lambda(t),0)$ and $t \mapsto (I,tf)$
 are also one-parameter group of diffeomorphisms of $\scri$
generated by a smooth vector fields $\xi'$ on $\scri$ as in Proposition \ref{prop1} as one may
 check by direct inspection.
 From decomposition theorem of Lorentz group, 
it is simply proved that every element of $G_{BMS}$ is a finite product of those elements
$(\Lambda(t),0)$ and $(I,tf)$.
Hence, using the property (\ref{KK}),
 the thesis turns out to be valid for a generic element of BMS group.\\
Assume that $(A,f)$ is an element of the one-parameter group of $\scri$-diffeomorphisms
 $\{\gamma'_t\}$  generated by $\xi'$  and let $\xi$ be a smooth
extension of $\xi$ to $M$ (i.e. $\tM$) as in Proposition \ref{prop1}
generating $\{\gamma_t\}$. In coordinates $(\Omega,u,\z,\bz)$ about $\scri$, (\ref{req}) can be written down
\beq (A_{\gamma'_t} \psi)(\Omega_t,u_t,\z_t,\bz_t) = \lim_{\Omega_t\to 0} \Omega_t^\alpha 
\phi_t(\Omega_t,u_t,\z_t,\bz_t),\label{bastard}\eeq
 where $\gamma_t : (\Omega,u,\z,\bz) \to (\Omega_t,u_t,\z_t,\bz_t)$ and $\phi_t := \gamma_t^* \phi$ so that
 $$\phi_t(\Omega_t,u_t,\z_t,\bz_t) = \phi(\Omega,u,\z,\bz)\:.$$
 (\ref{bastard}) can be re-written
 \beq (A_{\gamma'_t} \psi)(u_t,\z_t,\bz_t) = \lim_{\Omega_t\to 0} \frac{\Omega_t^\alpha}{\Omega^\alpha} 
 \:\Omega^\alpha \phi(\Omega,u,\z,\bz) \:,\nonumber\eeq
that is, since on $\scri$ $\gamma_t$ coincides with $\gamma'_t$ which preserves $\scri$ itself,
\beq (A_{\gamma'_t} \psi)(u_t,\z_t,\bz_t) = \left(\lim_{\Omega \to 0} \frac{\Omega_t}{\Omega} \right)^\alpha
 \psi(u,\z,\bz) \:.
 \label{bastard2}\eeq
 Using H\^opital rule
 \beq (A_{\gamma'_t} \psi)(u_t,\z_t,\bz_t) = \left( \frac{\partial \Omega_t}{\partial \Omega}|_{\Omega=0} \right)^\alpha
  \psi(u,\z,\bz) \:.
 \label{bastard3}\eeq
 Our task is computing the derivative in the right-hand side of (\ref{bastard3}). By definition of $\xi$ one finds 
 \beq
 \frac{d}{dt}\left( \frac{\partial \Omega_t}{\partial \Omega}|_{(\Omega=0, u,\z,\bz)} \right) = 
 \frac{\partial \xi^\Omega(\gamma_t(\Omega,u,\z,\bz))}{\partial
 \Omega} |_{\Omega=0}\label{bastard4}\:.
 \eeq
 Now making explicit the condition that $(\Omega^2 \lie_\xi g)_{\alpha\beta}$  extends smoothly to a vanishing field  approaching
$\scri$  (Proposition \ref{prop1}) in the considered coordinates, one easily finds for components  $\alpha =\Omega,
\beta= u$:
$$\frac{\partial \xi^\Omega(\Omega,u,\z,\bz)}{\partial \Omega}|_{\Omega=0}= 
-\frac{\partial \xi'^u(u,\z,\bz)}{\partial u}\:,$$
 where we also used $\xi'=\xi$ on $\scri$. Finally, from (\ref{bastard4})
 \beq
 \frac{d}{dt}\ln\left|\frac{\partial \Omega_t}{\partial \Omega}|_{(\Omega=0, u,\z,\bz)} \right| = 
- \frac{\partial \xi'^u(u_t,\z_t,\bz_t)}{\partial
 u_t}|_{(u_t,\z_t,\bz_t) = \gamma'_t(u,\z,\bz)} \label{bastard5}\:.
 \eeq
 Let us solve this equation in the relevant cases. By direct inspection one finds that the right hand side vanishes 
 when the one-parameter subgroup $\{\gamma'_t\}$
generated by $\xi'$
has the form $t\mapsto (I,tf)$ and so $ \frac{\partial \Omega_t}{\partial \Omega}|_{(\Omega=0, u,\z,\bz)}=$ constant in this case.
Since $\frac{\partial \Omega_0}{\partial \Omega}|_{(\Omega=0, u,\z,\bz)}=1$, (\ref{bastard3}) produces
$$(A_{\gamma'_t} \psi)(u_t,\z_t,\bz_t) = 
  \psi(u,\z,\bz)\:,$$
which is just the thesis in the considered case. Let us consider the other case with $\gamma'_t$ having
 the form $t\mapsto (\Lambda_t,0)$. In this case one gets
 $$\xi^u(u_t,\z_t,\bz_t) = \frac{u_t}{K_{\Lambda_t}(\Lambda^{-1}_t (\z_t,\bz_t))} 
 \left(\frac{dK_{\Lambda_t}(\z,\bz)}{dt}\right)|_{(\z,\bz)= \Lambda^{-1}_t (\z_t,\bz_t)}
 %%%% CORREZZIONE mancav il segno = e l'indicie t a u_t
  =
  u_t\frac{d\ln |K_{\Lambda_t}(\z,\bz)|}{dt}|_{(\z,\bz)= \Lambda^{-1}_t (\z_t,\bz_t)}
 \:.$$
 %%%
 From (\ref{bastard5}), using the fact that $\frac{\partial \Omega_0}{\partial \Omega}|_{(\Omega=0, u,\z,\bz)}=1$,
 one finds at the end via (\ref{bastard3}):
 $$(A_{\gamma'_t} \psi)(u_t,\z_t,\bz_t) = K_{\Lambda_t}(\z,\bz)^{-\alpha} \psi(u,\z,\bz)\:,$$
 which is the thesis in the considered case.
 $\Box$\\

\noindent{\em Proof Theorem \ref{theorem2}}. (a) and (b). Take $\psi \in \cS(\scri)$.
 Using integration by parts in (\ref{two}) and  standard theorem 
 (Lesbegue's dominate convergence)
 to interchange the symbol of derivative  with that of integral,
 it is simply proved that, if
  $\psi \in \cS(\scri)$, $(E,\z,\bz) \mapsto \widetilde{\psi_+}(E,\z,\bz)/\sqrt{E}$ belongs 
  to $C^\infty( \bR^+\times \bS^2; \bC)$ and, 
  as $E \to +\infty$,
  it  decays, uniformly in   $\z,\bz$ with all derivatives in any variable, faster than any negative power
   of $E$. Using the same procedure in (\ref{one}),
  one finds straightforwardly that $\z,\bz$ uniform estimates hold for $\psi_+$:
  \beq \left|\frac{\partial^k}{\partial u^k} \frac{\partial^c}{\partial\zeta^c} 
   \frac{\partial^d}{\partial\bz^d}\psi_+(u,\z,\bz)\right|
  \leq \frac{C_{k,c,d}}{ 1+|u|^{k+1}}\label{stime}\eeq
   for nonnegative constants $C_{k,c,d}$ depending on $k,c,d = 0,1,2,\ldots$. Therefore it make sense to apply
   $\sigma$ defined in (\ref{sigma}) to a pair of positive frequency parts $\psi_{1+}$, $\psi_{2+}$ when
   $\psi_1,\psi_2\in \cS(\scri)$. 
   The independence from the used Bondi frame can be proved by direct inspection using (\ref{psi+invariant}),
    Proposition \ref{exremark1}  to check 
on the independence from the used Bondi frame
and taking advantage of the fact that $\epsilon_{\bS^2} (\z,\bz)$ is invariant under three dimensional rotations.\\ 
Let us prove the item (b). 
In the following we use the notation $\psi'(u,\z,\bz):= (A_{(\Lambda,f)}\psi)(u,\z,\bz)$.
   Finally  (\ref{prodscalar2}) is a straightforward application of Fubini-Tonelli theorem
   in the explicit expression for $\Omega(\overline{\psi_{1+}}, \psi_{2+})$,
   the hypotheses being fulfilled due to the decaying estimates said above, using (\ref{one})
   (take into account that actually the apparent singularity due to the factor $E^{-1/2}$ does not exist
   because of (\ref{two}) where the integral produces a smooth function in $E$). The remaining part of (b)
   is an immediate consequence of (\ref{prodscalar2}). Let us prove (c). First of all notice that 
   the map $\psi_+ \mapsto \widetilde{\psi_+}$ for $\psi \in \cS(\scri)$ is well-defined because 
   the map $\widetilde{\psi_+} \mapsto \psi_+$ is injective. The proof
   follows straightforwardly from injectivity of Fourier transformation in Schwartz space
   referring to  Fourier transform involved 
   in (\ref{two}) and using the fact that  $\psi$ is real.
   By (\ref{prodscalar}) the complex linear extension   of $\psi_+ \mapsto \widetilde{\psi_+}$
   %%%CORREZZIONE TESTO MODIFICATO
   is bounded and isometric and thus, it being defined in a dense subspace, it admits a unique bounded isometric extension
$U$
   from the completion of $\cS(\scri)^\bC_+$, $\cH$, to 
   a closed subspace of $L^2(\bR^+\times \bS^2, dE\otimes \epsilon_{\bS^2})$.
    To prove the thesis, that is that $U$ is a Hilbert space isomorphism, 
%%%%%
it is sufficient to show
   that the subspace includes $C_c^\infty((0,+\infty)\times \bS^2;\bC)$ because the latter is dense in 
   $L^2(\bR^+\times \bS^2, dE\otimes \epsilon_{\bS^2})$. To this end, take 
   $\phi \in C_c^\infty((0,+\infty)\times \bS^2;\bC)$ and define $\psi$ as:
   $$ \psi(u,\z,\bz)  := \int_{\bR^+} \frac{dE}{\sqrt{4\pi E}} e^{-iE u}\phi(E,\z,\bz) +
   \int_{\bR^+}\frac{dE}{\sqrt{4\pi E}}\overline{e^{-iE u}\phi(E,\z,\bz)}\:.$$
   Notice that the singularity of $E^{-1/2}$ at $E=0$  is harmless since 
   the support of $\phi$ does 
   not include that point and thus the whole integrand is smooth and compactly supported.
   Finally, by direct inspection, one finds that $\psi \in \cS(\scri)$ and 
   $\widetilde{\psi_+} = \phi$ so that, as wanted,  $\phi=  U\psi_+$ for some $\psi \in \cS(\scri)$.
%%%CORREZZIONE: pezzo aggiunto
The last argument actually proves that  the range of $K: \cS(\scri)\ni \psi \mapsto \psi_+$  includes the space   
$U^{-1}C_c^\infty((0,+\infty)\times \bS^2;\bC)$
which is dense in $\cH= U^{-1} L^2(\bR^+\times \bS^2, dE\otimes \epsilon_{\bS^2})$
and thus it proves  also (d).
%%%
This concludes the proof. $\Box$\\

 \noindent {\em Proof of Theorem \ref{theo3}}.  
As is well-known working with group representations, to prove the thesis it is sufficient to show that strong 
continuity holds for $g\to I$  (the unit element of $\gbms$).
Let us to prove strong continuity as $g\to I$ 
for the restriction of the representation $U$ to $\cH$. 
To this end  we prove, as the first step, the strong continuity of $U$ when it works on
one-particle states represented by smooth compactly supported functions $\tilde{\phi}(E,\z,\bz)$.
(In the following, for sake of simplicity, we write $\z,\bz$ concerning coordinates on $\bS^2$, but actually one needs 
at least two charts to cover the compact smooth manifold $\bS^2$. The use of two charts removes the apparent singularity 
of the coordinates $\z,\bz$ on the point $\infty$ of the Riemann sphere.)
Using the fact that every $U_g$ is unitary, one sees that $||U_g \tilde{\phi}- \tilde{\phi}|| \to 0$
as $g\to I$ is equivalent to $(\tilde{\phi}, U_g \tilde{\phi}) \to (\tilde{\phi},\tilde{\phi})$
as $g\to I$. With an explicit representation (by means of (\ref{Ureducted2})) we have to prove that, as $g\to I$ and for a smooth
compactly supported $\tilde{\phi}$,
$$\lim_{(\Lambda,f) \to (I,0)}\int_{\bR^+\times \bS^2} \:\sqrt{K_\Lambda(\z,\bz)}\: e^{iE f(\z,\bz)} 
\overline{\widetilde{\psi}\left(\frac{E}{K_\Lambda(\z\,\bz)},\Lambda(\z,\bz)\right)}\:
\widetilde{\psi}(E,\z,\bz )\: dE \otimes \epsilon_{\bS^2}(\z,\bz)$$
\beq = \int_{\bR^+\times \bS^2}\:\overline{\widetilde{\psi}(E,\z,\bz)}\widetilde{\psi}(E,\z,\bz ) \:
dE \otimes\epsilon_{\bS^2}(\z\,\bz) \label{end}\:.\eeq 
Taking $\Lambda$ in a relatively compact neighborhood $B$ of the unit element of $SO(3,1)\sp\uparrow$, (for any fixed $f$)
the smooth  compactly supported  map $$(\Lambda, E, \z\,\bz) \mapsto 
\left|e^{iE f(\z,\bz)} K_\Lambda(\z\,\bz)\:\overline{\widetilde{\psi}\left(\frac{E}{K_\Lambda(\z,\bz)},\Lambda(\z,\bz)\right)}\:
\widetilde{\psi}(E,\z,\bz )\right|$$
is bounded by construction by some constant $K$ not depending on $f$ (which does not give contribution to
the considered functions since it is real valued). On the other hand, there is a compact $C\subset \bR^+\times \bS^2$
containing all the supports of the maps $$(E, \z\,\bz) \mapsto 
\left|e^{iE f(\z,\bz)} K_\Lambda(\z\,\bz)\:\overline{\widetilde{\psi}\left(\frac{E}{K_\Lambda(\z,\bz)},\Lambda(\z,\bz)\right)}\:
\widetilde{\psi}(E,\z,\bz )\right|,$$ for all $\Lambda \in B$ and all $f\in C^\infty(\bS^2)$. 
As a consequence all those maps are $(\Lambda,f)$-uniformly
 bounded by a smooth compactly supported function on $\bR^+\times \bS^2$ which assumes the value $K$ in $C$.
 Thus we can use Lebesgue's dominate convergence theorem in the right-hand of (\ref{end}) establishing the validity
 of (\ref{end}) itself. We have proved strong continuity on smooth compactly supported functions in $\cH$.
 As the space of those functions is dense in $\cH$, it implies strong continuity on the whole $\cH$. Indeed,
 if $\phi\in \cH$ and for any fixed smooth compactly supported $\phi_n\in H$,  triangular inequality entails
%CORREZZIONI VARIE
 $$ ||\phi- U_g \phi || \leq ||\phi - \phi_n|| + ||\phi_n- U_g \phi_n || + 
 ||U_g(\phi_n -\phi)||  =
  2||\phi - \phi_n|| + ||\phi_n- U_g \phi_n ||\:.\label{added}$$
 Let $\{V_m\}_{m\in \bN}$ be a fundamental system of neighborhoods of $I$ -- one can always choose 
 $m\in \bN$ the topology being induced by
 a countable class of seminorms -- such that
 $V_m\supset V_{m+1}$ and $\cap_m V_m = \{I\}$. From the inequality above
 and using   $\lim_{m\to +\infty} \sup_{g\in V_m}||\phi_n- U_g \phi_n || = 0$ which is a straightforward consequence of
  $\lim_{g\to I} ||\phi_n- U_g \phi_n || = 0$, one has
 $$0\leq  \lim_{m\to +\infty} 
 \sup_{g\in V_m}||\phi- U_g \phi ||
  \leq 2||\phi - \phi_n|| \:.$$ 
Taking a sequence of $\phi_n$  with $\phi_n \to \phi$ for $n\to +\infty$,
one obtains $\lim_{m\to +\infty} 
 \sup_{g\in V_m}||\phi- U_g \phi ||=0$ which entails
$\lim_{g\to I}||\phi- U_g \phi ||=0$, i.e. strong continuity holds for $U\sp\rest_{\cH}$.\\
%%%%%
To conclude the proof we show that the strong continuity in $\cH$ implies strong continuity in the whole Fock space.
By construction, if $V_g:= U\sp\rest_{\cH}$, on the $U$ invariant subspace $\cH^N\subset {\gF}_+(\cH)$ 
 containing $N$ particles
one has $V_g^{(N)}:= U\sp\rest_{\cH^N} = V_g\otimes \cdots \otimes V_g$ where the number of factors is $N$.
(Obviously $V_g^{(0)}:=I$.)
As a consequence $g\mapsto V_g^{(N)}$ is strongly continuous. Now consider a generic element of ${\gF}_+(\cH)$
which can be viewed as a sequence $\Phi=\{\Psi_N\}_{N=0,1,\ldots}$ with $\Psi_N \in \cH^N$. Let us show that
$(\Phi,V_g \Phi) \to ||\Phi||^2$ as $g\to I$. (Using either the fact that $V_g$ is unitary either the group representation structure, 
 this is equivalent to $||V_{g'} \Phi -V_{h}\Phi||^2 \to 0$ as $g'\to h$).
Spaces $\cH^N$ are invariant,  pairwisely orthogonal and $V^{(N)}_g$ are isometric; as a consequence  
one has $$(\Phi,V_g \Phi) = \sum_{N=0}^{+\infty}(\Psi^{(N)},V^{(N)}_g \Psi^{(N)})\:,$$
where $|(\Psi^{(N)},V^{(N)}_g \Psi^{(N)})|\leq ||\Psi^{(N)}||||V^{(N)}_g \Psi^{(N)}||=||\Psi^{(N)}||^2$
and thus
$$\sum_{N=0}^{+\infty}|(\Psi^{(N)},V^{(N)}_g \Psi^{(N)})| \leq \sum_{N=0}^{+\infty}||\Psi^{(N)}||^2 = ||\Phi||^2\:.$$
This $g$-uniform bound (essentially via Lebesgue dominate convergence theorem) allows one to interchange 
symbols of summation and limit:
$$\lim_{g\to I}(\Phi,V_g \Phi)  = 
\sum_{N=0}^{+\infty}\lim_{g\to I}(\Psi^{(N)},V^{(N)}_g \Psi^{(N)}) = \sum_{N=0}^{+\infty}(\Psi^{(N)},V^{(N)}_I \Psi^{(N)}) ||\Phi||^2\:,$$  
where we have used strong continuity of each representation  $V^{(N)}$.
This is what we wanted to prove. $\Box$\\ 
 
\noindent {\em Proof of Proposition \ref{chara2}}.
It is sufficient to show that each $\chi \in N^\prime$ admits a corresponding function $\beta: N \to \bR$ continuous and linear  
such that $\chi(\alpha) = e^{i\beta(\alpha)}$ for every $\alpha \in N$. (In fact, a continuous linear functional 
$\beta: N= C^\infty(\bS^2) \to \bR$ is a distribution by definition and thus one can write
$(\alpha, \beta)$ instead of $\beta(\alpha)$.) Let us prove it. Actually,  the following proof holds true in the more general hypothesis
on $N$ to be a topological vector space.\\
 Fix $\chi \in N^\prime$. First of all we identify $U(1)$ with $\bS^1$ and, in turn, we identify $\bS^1$ with $(-\pi,\pi]$ 
 where $\pi\equiv +\pi$.
In this picture, for our fixed $\chi \in N^\prime$, there is a continuous map $f: N \to  (-\pi,\pi]$ such that
$\chi(\alpha) = e^{if(\alpha)}$  for all $\alpha \in N$. From continuity there is an open set $B_0 \subset N$ such that
$B_0 = f^{-1}((-\pi,\pi))$. $B_0$ is a neighborhood of the zero vector of $N$. Indeed $e^{if(0)}= \chi(0) =1$ since $\chi$ is a homomorphism.
We have found that $f(0)= 2k\pi$ for some $k\in \bZ$. 
On the other hand, because $f(0) \in (-\pi,\pi]$ by hypotheses, it must be $f(0) = 0$. In particular $f(0) \in (-\pi,\pi)$ hence 
$0\in B_0$ and thus, as we said, $B_0$ is a open neighborhood of $0$. As $N$ is a topological vector space, there is an open 
balanced (also said star-shaped) neighborhood of $0$, $B \subset B_0$.\\
In general the function $f$ does not satisfy $f(u) + f(v) = f(u+v)$ because $f(u)+ f(v)$ may not belong to $B_0$ also if 
$f(u), f(v)$ do. Nevertheless we define the map $\beta : N \to \bR$ such that:
\beq \beta(v) := n_vf\left(\frac{1}{n_v} v\right)\:, \:\:\: \: \mbox{for all $v\in N$, $n_v >0 $
 being the first natural with $(1/n_v) v \in B$\:.}\label{bastarda}\eeq
We have the following results.

(a) For every $\alpha \in N$ it holds

$$e^{i\beta(\alpha)} = \chi(\alpha)\:.$$
Indeed,  using $\chi(v)^m = \chi(mv)$  valid for every natural $m>0$ and  $e^{if(\alpha/n_\alpha)} = \chi(\alpha/n_\alpha)$, 
one has  $e^{i\beta(\alpha)}= e^{in_\alpha f(\alpha/n_\alpha)} = 
(\chi(\alpha/n_\alpha))^{n_\alpha} = \chi(n_\alpha (\alpha/n_\alpha)) = \chi(\alpha)$.

(b) If $\beta$ is continuous it is additive as well, i.e. $$\beta(u+v)= \beta(u) + \beta(v)\:, \:\:\:\:\: \mbox{for all $u,v\in N$}\:.$$

\noindent Indeed, from $\chi(u)\chi(v)= \chi(u+v)$ and (a), one obtains
$e^{i(\beta(u)+\beta(v))} = e^{i\beta(u+v)}$. Fix $u, v\in N$ and let $t$ range in $[0,1]$. The function
$g: t\mapsto \beta(u) + \beta(tv) - \beta(u+tv)$ must be continuous  because straightforward composition of continuous functions.
On the other hand, since $e^{i(\beta(u)+\beta(tv))} = e^{i\beta(u+tv)}$, $g$ must take 
values in the non connected and discrete  set $2\pi\bZ$. Since continuous functions transforms connected sets to connected sets, 
$g$ must take a constant value in $2\pi \bZ$. 
As $g(0)=0$, we conclude that 
$\beta(u) + \beta(tv) - \beta(u+tv)=0$ for $t\in [0,1]$, in particular $\beta(u)+\beta(v) = \beta(u+v)$.

(c) If $\beta$ is continuous it is linear as well. 

\noindent Indeed from (b) one has $m\beta(v) = \beta(mv)$ for every natural $m>0$ and $v\in N$. As a consequence, defining $u:= nv$, one obtains
 $\beta(u/n) = (1/n)\beta(u)$ valid for every natural $n>0$ and $u\in N$. Both these results entail that
 $r \beta(w) = \beta(rw)$ for every rational $r>0$ and $w\in N$. 
 By continuity one finds  $r \beta(w) = \beta(rw)$ for every real $r>0$ and $w\in N$.
 Finally
(b) implies also that $\beta(0u) = 0\beta(u)=0$ and $\beta(-u) = -\beta(u)$ for every $u\in N$.
Putting all together one obtains that $r \beta(w) = \beta(rw)$ for every  $r \in \bR$ and $w\in N$. 
 Taking additivity into account we have proved that $\beta$ is linear.

\noindent To conclude, the proof it is sufficient to show that $\beta$ defined in (\ref{bastarda}) is continuous. Let us demonstrate 
it proving that $\beta$ is continuous at each point $\alpha\in N$. 
The difficult point to handle in the proof is that $n_\alpha$ in (\ref{bastarda}) is a function of $\alpha$ itself in spite of $f$
being continuous.
If $\alpha \in N$, by definition of $n_\alpha$ one has $\alpha/n_\alpha \in B$, but $\alpha/(n_\alpha -1) \not \in B$.
If $B^c$ denotes $N\setminus B$, there 
are now two possibilities concerning the requirement $\alpha/(n_\alpha -1) \not \in B$:
(1) $\alpha/(n_\alpha -1) \in int(B^c)$ or (2) $\alpha/(n_\alpha -1) \in \partial B$.

\noindent Suppose that (1) holds, i.e. $\alpha/(n_\alpha -1) \in int B^c$, together with 
$\alpha/n_\alpha \in B$. In this case  $\alpha \in n_\alpha B$
as well as
 $\alpha \in int (n_\alpha -1)B^c$. These sets are  open by construction. As a consequence, there is 
 an open neighborhood $V$ of $\alpha$
such that, if $\alpha' \in V$, $\alpha'/(n_\alpha -1) \in int B^c$ -- so $\alpha'/(n_\alpha -1) \not \in B$
 -- and furthermore $\alpha'/n_\alpha \in B$. In other words 
 $n_{\alpha'} = n_\alpha$.
In this case, there is a constant $C = n_\alpha>0$ such that,  if $\alpha'$ lies in a neighborhood $V$ of $\alpha$, 
$\beta(\alpha') = C f(\alpha'/C)$. Since $f$ is continuous, $\beta$ is such in $V$ and thus 
$\beta$ is continuous at $\alpha$.

\noindent To conclude, suppose that (2) is valid, that is $\alpha/(n_\alpha -1) \in \partial B$, together with 
$\alpha/n_\alpha \in B$.
Consider a sufficiently small open neighborhood $V$ of such a $\alpha$. If $\alpha'\in V$ there are two possibilities:
$\alpha'\in (n_\alpha -1) B^c$ or $\alpha'\in (n_\alpha -1) B$.\\
If $\alpha'\in (n_\alpha -1) B^c$ one has $\alpha'/(n_\alpha-1) \not \in B$, but $\alpha'/n_\alpha \in B$ so that
 $n_{\alpha'} = n_\alpha$ and thus
\beq\beta(\alpha') = n_\alpha f(\alpha'/n_\alpha)\:.\label{bastarda1}\eeq
Conversely, if $\alpha'\in (n_\alpha -1) B$, it must hold $\alpha'/(n_\alpha -1)\in  B$ so that $n_\alpha$
is not the first positive natural $n_{\alpha'}$ such that  $\alpha'/n_{\alpha'}\in  B$. In this case $n_{\alpha'} < n_\alpha$ and thus
\beq\beta(\alpha') = n_{\alpha'} f(\alpha'/n_{\alpha'})\:,\:\:\:\: \mbox{where $n_{\alpha'}< n_{\alpha}$}\label{bastarda1'}\:.\eeq
Let us prove that in this second case, actually,
\beq
 n_{\alpha'} f(\alpha'/n_{\alpha'}) =  n_{\alpha} f(\alpha'/n_{\alpha}) \label{bastarda3}\:,
\eeq
holds true anyway so that $\beta(\alpha') = n_{\alpha} f(\alpha'/n_{\alpha})$ and (\ref{bastarda1}) is valid in every case.
Defining $\gamma := n_{\alpha'} \alpha'$ (notice that  $\gamma\in B$ by hypotheses)
and $m = n_{\alpha}-n_{\alpha'}$ (notice that $0< m < n_{\alpha}$ by construction), 
(\ref{bastarda3}) is equivalent to
\beq
n_{\alpha} f(\gamma) - m f(\gamma) = n_\alpha f\left( \gamma - \frac{m}{n_\alpha} \gamma\right)
\label{bastarda4}\:.\eeq
To prove (\ref{bastarda4}) notice that, from $\chi(\alpha) = e^{if(\alpha)}$ one gets (use the fact that $\chi$ is a
homomorphism and $\bN \ni n_{\alpha},m>0$),
$$n_{\alpha} f(\gamma) - m f(t\gamma) - n_\alpha f\left( \gamma - \frac{m}{n_\alpha} \gamma\right)  \in 2\pi \bZ\:.$$
Finally consider the map, with $\gamma \in B$ fixed,
$$[0,1] \ni t \mapsto h(t) := n_{\alpha} f(t\gamma) - m f(t\gamma) - n_\alpha f\left( t\gamma - \frac{m}{n_\alpha} t\gamma\right).$$
This map is continuous because $f$ is continuous on $B$, $t\gamma \in B$ and $t\gamma - \frac{m}{n_\alpha} t\gamma \in B$
for $t\in [0,1]$ since $\gamma\in B$, $B$ is balanced and $0\leq 1 - m/n_{\alpha} <1$. As $2\pi \bZ$ is not connected and discrete 
but $[0,1]$ is connected, it must be $h(t)=$ constant. On the other hand $h(0)=0$, thus $h(t)=0$ for $t\in [0,1]$ and (\ref{bastarda4}) must hold true.\\
 We have proved once again that there is a constant $C = n_\alpha>0$ such that,  if $\alpha'$ lies in a neighborhood $V$ of $\alpha$, 
$\beta(\alpha') = C f(\alpha'/C)$. Since $f$ is continuous, $\beta$ is such in $V$ and thus 
$\beta$ is continuous at $\alpha$. $\Box$\\

\noindent {\em Proof of Lemma \ref{lemma1}}.
 From  the decomposition in the former formula in (\ref{two'}), passing in 
spherical coordinates one gets, 
for $\phi\in \cS_K(\bM^4)$ (remind that $\epsilon_{\bS^2}= \sin\vartheta \: d \vartheta \wedge d\varphi$ is the standard volume form of the unit $2$-sphere),
$$\phi(t,r,\vartheta',\varphi') = \frac{1}{(2\pi)^{3/2}} \int_{\bR^+} \frac{dE \:E^2}{\sqrt{2E}}\int_{\bS^2} \epsilon_{\bS^2}(\vartheta,\phi)
e^{i E(r \cos\alpha(\vartheta,\varphi,\vartheta',\varphi') -t)} \widetilde{\phi_+}(E,\vartheta,\varphi) + c.c.$$
where $\alpha=\alpha(\vartheta,\varphi,\vartheta',\varphi')$ is the angle between vectors $\bx = (r\sin\vartheta' \cos \varphi',
r \sin\vartheta'\sin\varphi',r\cos\vartheta')$ and $\bp = (E\sin\vartheta \cos \varphi,
r \sin\vartheta\sin\varphi, r\cos\vartheta)$. Passing to null coordinates $u:= t-r$, $v:= t+r$ and using the function
$\widetilde{\phi'_+}(E,\vartheta,\varphi) := \sqrt{E}\widetilde{\phi_+}(E,\vartheta,\varphi)$ which, by the second formula in (\ref{two'}), turns out to be bounded,
smooth and $\vartheta,\varphi$-uniformly rapidly decaying as $E \to +\infty$ by construction (to prove it use the latter in (\ref{two'})
taking into account that Cauchy surfaces are smooth and compactly supported and Fourier transform maps such functions 
into Schwartz functions), the equation above can be rearranged as
$$\phi(t,r,\vartheta',\varphi') = \frac{1}{4\pi^{3/2}} \int_{\bR^+} dE \int_{\bS^2} \epsilon_{\bS^2}(\vartheta,\varphi)
e^{i E v(\cos\alpha -1)/2} e^{-i E u(\cos\alpha +1)/2}E \widetilde{\phi'_+}(E,\vartheta,\varphi) + c.c.$$
By definition of $\Gamma_{\bM^4}$ and using the fact that $\omega^2\Omega^2\sp\rest_{\bM^4} = 4(1+v^2)^{-1}$
(see the beginning of section \ref{QFTMINKOWSKI}), one has
$$(\Gamma_{\bM^4} \phi)(u,\vartheta',\varphi') = \lim_{v\to +\infty} \frac{\sqrt{1+v^2}}{2}\phi(u,v,\vartheta',\varphi')$$
Since we know that this limit does exist by Proposition \ref{prop2} and the factor in front of $\phi$ diverges, we conclude 
that $\phi$ itself must vanish as $v\to +\infty$. As a consequence, expanding  $\sqrt{1+v^2}$ in powers of $v^{-1}$,
we conclude that it must also hold 
%%%CORREZZIONE: Mancava =
$$ (\Gamma_{\bM^4} \phi)(u,\vartheta',\varphi') = \lim_{v\to +\infty} \frac{v}{2}\phi(u,v,\vartheta',\varphi') \:.$$
 In other words
 %%%CORREZZIONE: Mancava =
 \beq (\Gamma_{\bM^4} \phi)(u,\vartheta',\varphi') = \lim_{v\to +\infty}  \frac{1}{8\pi^{3/2}} \int_{\bR^+} \sp\sp dE \sp \int_{\bS^2} \sp\sp \epsilon_{\bS^2}(\vartheta,\varphi)
e^{\frac{iE v(\cos\alpha -1)}{2}} e^{\frac{-iE u(\cos\alpha +1)}{2}}vE \widetilde{\phi'_+}(E,\vartheta,\varphi) + c.c. \label{Afirst} \eeq
%%%%%
Notice that the former exponential in the integrand, essentially due to Riemann-Lebesgue's lemma, makes vanishing the integral 
except for the case $\cos\alpha -1= 0$, that is when $(\vartheta,\varphi) = (\vartheta',\varphi')$;
on the other hand the factor $v$ blows up in this point giving rise to a Dirac $\delta$. Indeed the limit can be computed using
standard Dirac-$\delta$-regularization procedures of distributional calculus  obtaining (see below)
\beq (\Gamma_{\bM^4} \phi)(u,\vartheta,\varphi)= \frac{-i}{\sqrt{4\pi}}\int_{\bR^+} dE\: 
\widetilde{\phi'_+}(E,\vartheta,\varphi) e^{iE u} + c.c.\label{Alast}\eeq
We have found out that
$$(\Gamma_{\bM^4} \phi)(u,\vartheta,\varphi)= \int_{\bR^+}  dE \:
\frac{(-i) E \widetilde{\phi_+}(E,\vartheta,\varphi)e^{iE u} }{\sqrt{4\pi E}} + \int_{\bR^+}  dE \:
\overline{\frac{(-i) E \widetilde{\phi_+}(E,\vartheta,\varphi)e^{iE u} }{\sqrt{4\pi E}}}\:.$$
 From that expression for $(\Gamma_{\bM^4} \phi)(u,\vartheta,\varphi)$, applying the definition (\ref{two}) and standard properties of 
Fourier transform for $L^1$ functions, one straightforwardly gets 
$$\widetilde{(\Gamma_{\bM^4} \phi)}_+(E,\vartheta,\varphi) = (-i) E \widetilde{\phi_+}(E,\vartheta,\varphi)\:,$$
which is the thesis we wanted to prove. To conclude let us prove (\ref{Alast}). Without loss of generality 
we can rotate the used Cartesian frame to have $\bp$ with the direction of the positive axis $z$. In this case
(\ref{Afirst}) reads, if $c:=\cos\vartheta$,
%%%CORREZZIONE mancava = sotto
$$ (\Gamma_{\bM^4} \phi)(u,0,\varphi') = \lim_{v\to +\infty}  \frac{1}{8\pi^{3/2}} \int_{\bR^+}\spa \sp\sp dE \sp \int_0^{2\pi} \sp\sp\spa d\varphi \int_{-1}^1 \sp\sp dc \:
e^{\frac{iE v(c -1)}{2}} e^{\frac{-iE u(c +1)}{2}}vE \widetilde{\phi'_+}(E,\vartheta,\varphi) + c.c. $$
$$ =  \lim_{v\to +\infty}  \frac{-2i}{8\pi^{3/2}} \int_{\bR^+}\spa \sp\sp dE \sp \int_0^{2\pi} \sp\sp\spa d\varphi \int_{-1}^1 \sp\sp dc \:
\frac{d}{dc}\left(e^{\frac{iE v(c -1)}{2}}\right) e^{\frac{-iE u(c +1)}{2}} \widetilde{\phi'_+}(E,\vartheta,\varphi) + c.c. $$
Integration by parts gives (noticing that the dependence from $\varphi$ vanishes for $\vartheta= 0,\pi$, i.e. $c=1,-1$, and, thus, integration in $d\varphi$ trivially produces a factor $2\pi$)
%%%CORREZZIONE mancava = sotto
$$ (\Gamma_{\bM^4} \phi)(u,0,\varphi') = \lim_{v\to +\infty} \frac{-i 4\pi}{8\pi^{3/2}} \int_{\bR^+}\spa \sp\sp dE e^{-iEu} \widetilde{\phi'_+}(E,0,\varphi)
 -\lim_{v\to +\infty} \frac{-i 4\pi}{8\pi^{3/2}} \int_{\bR^+}\spa \sp\sp dE e^{-iEv} \widetilde{\phi'_+}(E,0,\varphi) + c.c.$$
 $$+ \lim_{v\to +\infty}  \frac{-2i}{8\pi^{3/2}}  \int_{\bR^+}\spa \sp\sp dE \sp \int_0^{2\pi} \sp\sp\spa d\varphi \int_{-1}^1 \sp\sp dc \:
e^{\frac{iE v(c -1)}{2}} e^{\frac{-iE u(c +1)}{2}}(-i)Eu \widetilde{\phi'_+}(E,\vartheta,\varphi) + c.c. $$
In other words
$$(\Gamma_{\bM^4} \phi)(u,0,\varphi') = \frac{-i}{\sqrt{4\pi}}\int_{\bR^+} dE\: \widetilde{\phi'_+}(E,0,\varphi) e^{-iE u}  +\lim_{v\to +\infty}\frac{i}{\sqrt{4\pi}}\int_{\bR^+}\spa \sp\sp dE e^{-iEv} \widetilde{\phi'_+}(E,0,\varphi) + c.c.$$
$$+ \lim_{v\to +\infty}  \frac{-1}{4\pi^{3/2}}  \int_{\bR^+}\spa \sp\sp dE \sp \int_0^{2\pi} \sp\sp\spa d\varphi \int_{-1}^1 \sp\sp dc \:
e^{\frac{iE v(c -1)}{2}} e^{\frac{-iE u(c +1)}{2}}Eu \widetilde{\phi'_+}(E,\vartheta,\varphi) + c.c. $$
As the map  $ E\mapsto \widetilde{\phi'_+}(E,0,\varphi)$ (with $\varphi$ constant) is smooth and rapidly decaying, Riemann-Lebesgue's lemma
implies that the limit in the former line vanishes. Let us focus on the last limit. As the integrand is $L^1$, we can interchange the order of
integration via Fubini-Tonelli theorem obtaining in particular that the considered limit can be re-written (up to an overall constant)
\beq\lim_{v\to +\infty} \int_{[0,2\pi]\times [-1,1]} \sp\sp\sp \sp \sp\sp\sp d\varphi dc\:\: \left\{ \int_{\bR^+} dE
e^{\frac{iE v(c -1)}{2}} e^{\frac{-iE u(c +1)}{2}}Eu \widetilde{\phi'_+}(E,\vartheta,\varphi)\right\} + c.c. \label{double}\eeq
By Riemann-Lebesgue's lemma, the integral in brackets vanishes, as $v\to +\infty$,
almost everywhere in $(c,\varphi)$. On the other hand, since the following $c,\varphi$-uniform bound holds
$$ \left| \int_{\bR^+} dE
e^{\frac{iE v(c -1)}{2}} e^{\frac{-iE u(c +1)}{2}}Eu \widetilde{\phi'_+}(E,\vartheta,\varphi)\right|
\leq \int_{\bR^+} dE
 \left| Eu \widetilde{\phi'_+}(E,\vartheta,\varphi)\right|= M<+\infty,$$
 and the domain of integration of the external integral in (\ref{double}) has measure finite, we can use Lebesgue's dominate theorem
getting:
$$\lim_{v\to +\infty} \int_{[0,2\pi]\times [-1,1]} \sp\sp\sp \sp \sp\sp\sp d\varphi dc\:\: \left\{ \int_{\bR^+} dE
e^{\frac{iE v(c -1)}{2}} e^{\frac{-iE u(c +1)}{2}}Eu \widetilde{\phi'_+}(E,\vartheta,\varphi)\right\} + c.c. $$
$$=  \int_{[0,2\pi]\times [-1,1]} \sp\sp\sp \sp \sp\sp\sp d\varphi dc\:\:\: \lim_{v\to +\infty}\:\: \left\{ \int_{\bR^+} dE
e^{\frac{iE v(c -1)}{2}} e^{\frac{-iE u(c +1)}{2}}Eu \widetilde{\phi'_+}(E,\vartheta,\varphi)\right\} + c.c.$$
$$=  \int_{[0,2\pi]\times [-1,1]} \sp\sp\sp \sp \sp\sp\sp d\varphi dc\:\:\:0 + c.c. =0\:.$$
We conclude that 
$$(\Gamma_{\bM^4} \phi)(u,0,\varphi') = \frac{-i}{\sqrt{4\pi}}\int_{\bR^+} dE\: \widetilde{\phi'_+}(E,0,\varphi) e^{-iE u} + c.c.$$
Notice that the values $\varphi$ and $\varphi'$ are arbitrary only because of the singularity of spherical coordinates 
for $\vartheta=0$ (the problem is harmless here because the singular set has measure zero). What is relevant in the expression above
is that, barring the problem with coordinates, it says that the versor ${\bf n'}$ on $\bS^2$ in the argument of the function in the left-hand 
side,
$(\Gamma_{\bM^4} \phi)(u,{\bf n'})$,
coincides with the analog, ${\bf n}$, in the argument of the integrated function $\widetilde{\phi'_+}(E,{\bf n})$.
Rotating back the used Cartesian frame to work with a generic value of $\vartheta$
the equation above transforms into:
$$(\Gamma_{\bM^4} \phi)(u,\vartheta,\varphi) = \frac{-i}{\sqrt{4\pi}}\int_{\bR^+} dE\: \widetilde{\phi'_+}(E,\vartheta,\varphi) e^{-iE u} + c.c.$$
where we have identified the angles $\varphi$ and $\varphi'$ as it is due working for $\vartheta \neq 0,\pi$ because ${\bf n'}={\bf n}$.
This equation is (\ref{Alast}). 
$\Box$

\end{document}